\documentclass[journal]{new-aiaa} 
\usepackage[utf8]{inputenc}
\usepackage{textcomp}
\usepackage{graphicx}
\usepackage{amsmath}
\usepackage[version=4]{mhchem}
\usepackage{siunitx}
\usepackage{longtable,tabularx}
\usepackage{booktabs}
\usepackage{multirow}
\title{
Autonomous Navigation and Station-Keeping on Near-Rectilinear Halo Orbits
}

\author{
    Yuri Shimane\footnote{Assistant Professor, Department of Mechanical and Aerospace Engineering; Member AIAA; yshimane@uci.edu (Corresponding Author); Y. Shimane was with MERL at the time of this research.}
}
\affil{University of California, Irvine, CA 92617}
\author{
    Karl Berntorp\footnote{Chewy Robotics; K. Berntorp was with MERL at the time of this research.}, 
    Stefano Di Cairano\footnote{Distinguished Research Scientist.}, and
    Avishai Weiss\footnote{Senior Principal Research Scientist; Member AIAA; weiss@merl.com (Corresponding Author).}
}
\affil{Mitsubishi Electric Research Laboratories (MERL), Cambridge, Massachusetts, 02139}

\usepackage{todonotes}
\usepackage{wasysym}     
\usepackage{siunitx}
\usepackage[ruled,linesnumbered]{algorithm2e}
\DontPrintSemicolon
\SetKwComment{comment}{\color{gray}\# }{}

\usepackage{algpseudocode}
\usepackage{todonotes}
\usepackage{caption}
\usepackage{subcaption}
\newcommand{\Frame}{\mathcal{F}}
\newcommand{\TCtoP}{\boldsymbol{T}^{\rm C}_{\rm P}}
\newcommand{\TPtoC}{\boldsymbol{T}^{\rm P}_{\rm C}}
\newcommand{\yvec}{\boldsymbol{y}}
\newcommand{\abold}{\boldsymbol{a}}

\newcommand{\dbold}{\boldsymbol{d}}
\newcommand{\Pbold}{\boldsymbol{P}}
\newcommand{\ubold}{\boldsymbol{u}}

\newcommand{\Rbold}{\boldsymbol{R}}
\newcommand{\rbold}{\boldsymbol{r}}
\newcommand{\vbold}{\boldsymbol{v}}
\newcommand{\xbold}{\boldsymbol{x}}
\newcommand{\zbold}{\boldsymbol{z}}
\newcommand{\fbold}{\boldsymbol{f}}
\newcommand{\Tbold}{\boldsymbol{T}}
\newcommand{\zerobold}{\boldsymbol{0}}
\newcommand{\xibold}{\boldsymbol{\xi}}
\newcommand{\state}{\boldsymbol{x}}
\newcommand{\psibold}{\boldsymbol{\psi}}

\newcommand{\Sun}{\mathrm{Sun}}
\newcommand{\Earth}{\mathrm{Earth}}
\newcommand{\review}[1]{\textcolor{black}{#1}}  
\newcommand{\rreview}[1]{\textcolor{black}{#1}} 

\newcommand{\rrreview}[1]{\textcolor{black}{#1}}  

\begin{document}

\maketitle

\begin{abstract}
This article develops an optical navigation (OPNAV) and station-keeping pipeline for the near-rectilinear halo orbit (NRHO) in high-fidelity ephemeris model dynamics, \review{using synthetic images \rreview{of the Moon} in a non-iterative horizon-based OPNAV algorithm, applying the result in a navigation filter, and using the obtained estimates in a station-keeping control scheme that keeps the spacecraft in the vicinity of a reference orbit.}
We study differential correction-based and minimization-based implementations of the \review{so-called} $x$-axis crossing control scheme, and propose an improved targeting prediction scheme by incorporating the filter's state covariance with an unscented transform. We also introduce a hysteresis mechanism, which improves station-keeping cost and provides insight into the difference in performance between the differential correction-based and minimization-based approaches. We perform Monte Carlo experiments to assess the pipeline's tracking and $\Delta V$ performances. We report several key findings, including the variability of the filter performance with the sensor field of view and measurement locations, station-keeping cost reduction achieved by the unscented transform-based prediction and hysteresis, as well as the variability of the cumulative $\Delta V$ as a function of maneuver location due to the periodic structure in the OPNAV-based filter's estimation accuracy.
\end{abstract}

\section*{Nomenclature}


{\renewcommand\arraystretch{1.0}
\noindent\begin{longtable*}{@{}l @{\quad=\quad} l@{}}
$A/m$           & solar radiation pressure area-to-mass ratio, \SI{}{m^2/kg} \\
$C_r$           & solar radiation pressure coefficient \\
$\mathcal{F}_{\rm EM}$ & Earth-Moon rotating frame centered at the Moon\\
$\mathcal{F}_{\rm I}$ & inertial frame centered at the Moon \\
$\mathcal{F}_{\rm P}$ & Moon's principal axes frame \\
$\fbold$        & spacecraft dynamics \\
$f$             & camera focal length, \SI{}{mm}  \\ 
$\boldsymbol{K}$& filter Kalman gain  \\
$m$             & number of points lying on the lit limb \\
$\boldsymbol{Q}$ & filter process noise \\
$\Rbold_{\yvec}$ & measurement covariance  \\
$\rbold$        & spacecraft inertial position vector, \SI{}{km}  \\
$\boldsymbol{T}^{\mathrm{A}}_{\mathrm{B}}$
                & transformation matrix from frame $\rm A$ to frame $\rm B$ \\ 
$t$             & time, \SI{}{s} \\
$\vbold$        & spacecraft inertial velocity vector, \SI{}{km/s}  \\
$v_{x,\mathrm{tol}}$   & $v_x$ targeting tolerance, \SI{}{m/s} \\
$v_{x,\mathrm{trig}}$  & $v_x$ trigger threshold, \SI{}{m/s}   \\
$W^{(\ell)}_m$           & $\ell^{\rm th}$ sigma point weight for the mean \\
$W^{(\ell)}_c$           & $\ell^{\rm th}$ sigma point weight for the covariance \\  
$\mathcal{X}^{(\ell)}$   & $\ell^{\rm th}$ sigma point \\
$\state$    & spacecraft inertial state vector \\
$\yvec$        & measurement vector \\
$\boldsymbol{\kappa}$& camera calibration matrix  \\
$\boldsymbol{\Sigma}$ & filter state covariance \\
$\theta$        & true anomaly, \SI{}{deg}         \\
$\Phi$          & state transition matrix          \\
$\boldsymbol{\psi}$ & targeted subset of state
\end{longtable*}}
\setcounter{table}{0}

\section{Introduction}
A key challenge of operating space assets in cislunar space is the vast distance from the Earth.
The distance makes frequent communication and navigational updates to cislunar spacecraft a challenge due to the scarcity and congestion of ground-based infrastructures, necessitating autonomous capability for guidance, navigation, and control (GN\&C).
A particular orbit of interest is the 9:2 resonant near-rectilinear halo orbit (NRHO) about the Earth-Moon L2 point~\cite{Lee2019, Zimovan-Spreen2020, Zimovan-Spreen2022}, due to its planned use for the upcoming lunar space station, or \emph{Gateway}, humanity's next outpost in space. 

Regular station-keeping (SK) maneuvers, also referred to as orbit maintenance maneuvers, are required for spacecraft flying along the NRHO due to the orbit's unstable modes.
SK is required at an appropriate cadence in order to cancel the unstable modes in the presence of errors in state estimation, dynamics model, and control.
As traffic on the NRHO is expected to increase, autonomous GN\&C along this orbit is of critical importance, both to reduce reliance on ground-based navigation updates in nominal operations contexts, as well as to ensure a reliable onboard GN\&C capability in case of off-nominal scenarios such as communication failures. 
Today, with the lack of a dedicated cislunar position, navigation, and timing (PNT) architecture, horizon-based optical navigation (OPNAV)~\cite{ChristianRobinson2016, Christian2017, Christian2021Tutorial} is a suitable measurement source that may be acquired and processed onboard a spacecraft along the NRHO. 
A number of prior simulation-based works considered autonomous navigation \rreview{using images of the Moon through} horizon-based OPNAV on lunar libration point orbits (LPOs)~\cite{Franzese2019,Balossi2024}, as well as specifically on the NRHO~\cite{Yun2020, Qi2023-aw} \review{and across various perspectives with respect to the Moon~\cite{Machuca2026-rc}}, but without consideration of SK activities.
Several recent missions have successfully demonstrated OPNAV for orbit determination in cislunar flight: the \review{EQUULEUS mission~\cite{Oguri2020-wi}}, the Orion spacecraft during the Artemis I mission~\cite{Inman2024}, the LONEStar experiment during the Lunar Flashlight mission~\cite{Krause2024}, and the CAPSTONE mission~\cite{Givens2025}.
\review{LUMIO~\cite{Cervone2022-ax} is another upcoming small spacecraft mission planned to use its optical payload for horizon-based OPNAV, among other purposes.}

There has been significant focus on SK algorithms and their performance under orbit determination error.
Shirobokov et al.~\cite{Shirobokov2017} provides a survey of the various SK schemes and for LPOs in various dynamical regimes. 
Oguri~\cite{Oguri2024-of} demonstrates navigation with numerically simulated horizon-based OPNAV and SK based on an optimization-based pre-computed feedback policy.
In the context of SK for NRHOs, the $x$-\textit{axis crossing control}, where the $x$-axis corresponds to the Earth-Moon line, is known to perform favorably~\cite{Davis2017, Guzzetti2017, Newman2018, Davis2022}; in the recent CAPSTONE mission, this strategy was successfully deployed on the NRHO~\cite{Cheetham2021}, and has been chosen as the SK scheme for the Gateway~\cite{Davis2022}. 

In this work, we evaluate the feasibility and performance of autonomous GN\&C along NRHO by developing an end-to-end pipeline that consists of a navigation filter relying solely on measurements from processing realistic synthetic images of the Moon, and extensions to the $x$-axis crossing control algorithm that uses the recursively filtered state estimate to compute control actions.
\review{The contributions of this paper are as follows:}
\begin{itemize}
    \item We consider the image processing for generating measurements via horizon-based OPNAV~\cite{ChristianRobinson2016,Christian2017} in parallel with navigation, where we augment the fidelity of the measurement covariance expression of the horizon-based OPNAV algorithm by accounting for the attitude error of the camera.
    We conduct Monte Carlo experiments to report the levels of navigation accuracy that can be achieved by horizon-based OPNAV for various choices of camera, and quantify its limitations along the NRHO.
    \item We compare two variants of the $x$-axis crossing control, namely the traditional differential correction-based approach~\cite{Guzzetti2017, Davis2017, Davis2022} and the optimization-based approach~\cite{Elango2022SequentialLinear}. 
    In addition, we propose an unscented transform (UT)-based targeting scheme that uses the filter's state covariance information to improve the state prediction. 
    We also introduce a hysteresis \review{mechanism within the $x$-axis crossing control scheme, where a tunable margin is introduced to desensitize the control scheme from noise by setting a tighter \textit{targeting} tolerance than a \textit{triggering} tolerance condition that prompts the control maneuver.}
    \review{We compare variants of the $x$-axis crossing control through Monte Carlo experiments, and we highlight the significant performance difference that arises due to the short-sighted, single maneuver SK optimization as an approximation of the infinite-horizon cumulative SK cost.
    We also report the cost reduction achieved via UT-based targeting.}
    \item \review{The full GN\&C pipeline is used to evaluate the SK performance along NRHO navigating based on horizon-based OPNAV alone.
    We also study the sensitivity of the cumulative $\Delta V$ cost to the SK maneuver location, impacted by the periodic structure of the filter covariance achieved by horizon-based OPNAV.}
\end{itemize}

The remainder of this paper is organized as follows. Section~\ref{sec:background} introduces the full-ephemeris dynamics and the key dynamical properties of the NRHO. 
In Section~\ref{sec:autonomous_navigation}, the horizon-based OPNAV and the navigation filter implemented in this work are introduced.
Section~\ref{sec:stationkeeping_control} concerns the $x$-axis crossing controller, along with the modifications and augmentations that we propose. 
Section~\ref{sec:filter_experiments} provides results based on the navigation filter incorporating OPNAV measurements without any SK.
Next, in Section~\ref{sec:controller_experiments}, the SK schemes are compared with different levels of assumed state estimation errors. 
Section~\ref{sec:gncstack_experiments} provides end-to-end simulation results with both the OPNAV-based navigation filter and the SK algorithms. 
Finally, conclusions to this work are given in Section~\ref{sec:conclusion}.

\section{Preliminaries}
\label{sec:background}
We first introduce the high-fidelity ephemeris model in which the spacecraft's motion is studied. 
Then, we briefly discuss the NRHO, pointing out characteristics that pertain to OPNAV-based navigation and SK. 

\subsection{Equations of Motion}
We consider the motion of the spacecraft in cislunar space under the effect of gravitational forces of the Moon, Earth and Sun, along with J2 perturbation from the Moon and solar radiation pressure (SRP), in a Moon-centered inertial frame $\Frame_{\rm I}$. The equations of motion are given by
\begin{equation}
    \dot{\state} = 
    \begin{bmatrix}
        \dot{\rbold} \\ \dot{\vbold}
    \end{bmatrix}
    = 
    \fbold(t,\state) 
    =
    \begin{bmatrix}
        \boldsymbol{v}
        \\
        -\dfrac{\mu}{r^3} \boldsymbol{r} 
        + \sum_i \boldsymbol{a}_{N_i}(t,\rbold) 
        + \boldsymbol{a}_{\mathrm{SRP}}(t,\rbold)
        + \boldsymbol{a}_{\mathrm{SH}}(t,\rbold)
    \end{bmatrix}
    ,
    \label{eq:dyn_model} 
\end{equation}
where $\rbold \in \mathbb{R}^3$ is the position vector of the spacecraft with respect to an unforced particle at the origin of $\Frame_{\rm I}$, $\vbold \triangleq \dot{\boldsymbol{r}} \in \mathbb{R}^3$ is the rate of change of $\rbold$ in $\Frame_{\rm I}$, and $\state^T = [\rbold^T, \vbold^T] \in \mathbb{R}^6$ is the state vector. In the acceleration terms on the right-hand side, $r = \|\boldsymbol{r}\|$ is the distance from the center of the Moon to the spacecraft, $\mu$ is the standard gravitational parameter of the Moon,
$\boldsymbol{a}_{N_i}$ is the third-body perturbing acceleration of body $i$, $\boldsymbol{a}_{\mathrm{SRP}}$ is the perturbing acceleration due to SRP, and $\boldsymbol{a}_{\mathrm{SH}}$ is the perturbing acceleration due to the \review{spherical harmonics terms of the Moon}, 
In this work, we consider third-body perturbations due to the Earth and the Sun; \review{in $\abold_{\rm SH}$, we consider the term due to J2 coefficient.
While the J2 term is included as a representative perturbation due to SH, the Moon has SH coefficients with similar orders of magnitude as $J_2$ at up to degree and order 3, which is to be included for higher-fidelity analyses. Their inclusion is straightforward, and we avoided it here for simplicity.}
The perturbation terms are given by
\begin{align}
    \boldsymbol{a}_{N_i}(t,\rbold) &=  \mu_i \left( \dfrac{\dbold_i - \rbold}{\| \dbold_i -  \rbold \|^3} -\dfrac{\dbold_i}{d_i^3} \right),
    \\
    \boldsymbol{a}_{\mathrm{SRP}}(t,\rbold) &= P_{\Sun} \left( \dfrac{ \|\dbold_{\Earth} - \dbold_{\Sun}\|_2 }{r_{\Sun}} \right)^2 C_r (A/m) \dfrac{\rbold_{\Sun}}{r_{\Sun}},
    \label{eq:eom_SRP_term}
    \\
    \boldsymbol{a}_{\mathrm{SH}}(t,\rbold) &= 
    \Tbold^{\rm P}_{\rm I} 
    \left( -\dfrac{3 \mu J_2 R^2}{2r^5}
        \begin{bmatrix}
            \left( 1-5\frac{z_{\rm P}^2}{r^2} \right)x_{\rm P} \\
            \left( 1-5\frac{z_{\rm P}^2}{r^2} \right)y_{\rm P} \\
            \left( 3-5\frac{z_{\rm P}^2}{r^2} \right)z_{\rm P}
        \end{bmatrix}
    \right),
\end{align}
where $\Tbold^{\rm P}_{\rm I} \in \mathbb{R}^{3 \times 3}$ is the transformation matrix from the Moon's principal axes frame $\Frame_{\rm P}$ to $\Frame_{\rm I}$, $J_2$ is the oblateness coefficient of the Moon, $R$ is the equatorial radius of the Moon, and subscript $(\cdot)_{\rm P}$ denote position components in $\Frame_{\rm P}$; $\mu_i$ is the gravitational parameter of body $i$, $\dbold_i$ is the position of body $i$ with respect to the Moon, and $d_i = \| \dbold_i \|_2$; $P_{\rm Sun}$ is the SRP magnitude at \SI{1}{AU}, $C_r$ is the radiation pressure coefficient, $A/m$ is the area-to-mass ratio, $\rbold_{\rm Sun}$ is the position vector of the spacecraft with respect to the Sun, and $r_{\rm Sun} = \| \rbold_{\rm Sun} \|_2$.

The linearized solution to equation~\eqref{eq:dyn_model} in the vicinity of a reference state $\bar{\state}$ at time $t$ is given by 
\begin{equation}
    \delta \state(t) = 
    \dfrac{\partial \state(t)}{\partial \bar{\state}(t_0)} \delta \state(t_0)
    ,
\end{equation}
where $\delta \state(t_0)$ denotes the initial deviation from the initial reference state $\bar{\state}(t_0)$ at time $t_0$. The partial ${\partial \state(t)}/{\partial \bar{\state}(t_0)}$ is the state-transition matrix (STM) denoted by the shorthand $\boldsymbol{\Phi}(t,t_0)$, obtained through the nonlinear integration of the initial value problem
\begin{align}
\begin{cases}
      \dot{\boldsymbol{\Phi}}\left(t, t_0\right) = 
        \dfrac{\partial \fbold(t, \state)}{\partial \bar{\state}}
        \boldsymbol{\Phi}\left(t, t_0\right),
      \\ 
      \boldsymbol{\Phi}\left(t_0, t_0\right) = \boldsymbol{I}_{6}.
    \end{cases}
    \label{eq:stm_diffeqn}
\end{align}

\subsection{Near-Rectilinear Halo Orbit}
Libration point orbits (LPO), periodic motions revolving around libration points, offer the option for spacecraft to be placed in unique periodic regimes in cislunar space that are otherwise unattainable by classical Keplerian orbits about the Earth or the Moon. 
There exist multiple families of LPOs, each with varying characteristics in terms of their spatial location, energy, and/or stability, and the NRHOs are a subset of the so-called halo orbit families about the L1/L2 points with bounded linear stability. 
For details on the dynamic structure of NRHOs, see~\cite{Zimovan-Spreen2020, Zimovan-Spreen2022} and references therein. 
In this work, we focus on the 9:2 resonant NRHO of the Gateway, which completes 9 revolutions over 2 synodic periods of the Moon. 

While periodic motions may be numerically constructed in simplified models such as the circular restricted three-body problem, the analogous motion in the full-ephemeris model is only quasi-periodic and contains unstable modes, thus requiring SK corrections to maintain the spacecraft on the desired orbit. Integrating motion along the NRHO is further complicated by the relatively low perilunes of around \SI{3000}{km}, as numerical integration errors get amplified during the transits through perilunes, where the dominant acceleration, proportional to $1/r^2$, has high sensitivities. 
The compounding effects of the inherent dynamical, as well as numerical, instabilities, necessitate a numerical integrator that is efficient with tight error tolerances. In this work, we make use of the explicit embedded Runge-Kutta Prince-Dormand 9(8) from Verner~\cite{Verner2010} in the GNU scientific library (GSL)~\cite{gough2009gnu}.

\subsection{System Architecture for Optical Navigation and Station-Keeping}
\begin{figure}
    \centering
    \includegraphics[width=0.85\linewidth]{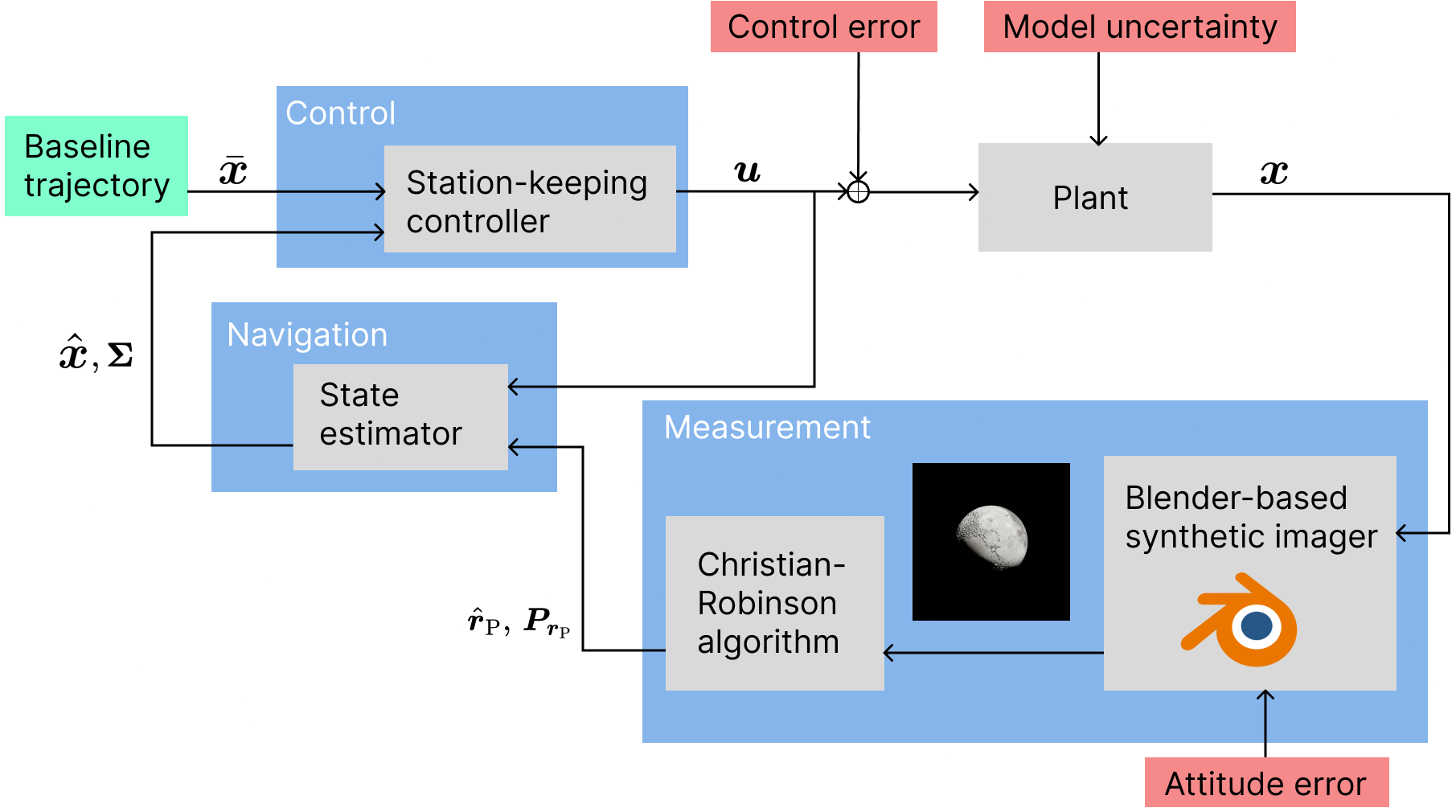}
    \caption{Overview of optical navigation-based station-keeping architecture and simulation environment}
    \label{fig:CONOPS}
\end{figure}

To validate autonomous navigation and SK on an NRHO, we develop a simulation environment that generates synthetic images of the Moon, processes the images to generate position measurements and measurement covariances, estimates the spacecraft state using a navigation filter, and computes orbital correction maneuvers using an SK controller. This framework is illustrated in Figure~\ref{fig:CONOPS}. 

Starting with the measurement block, synthetic images are generated using Blender; details of this implementation are provided in~\cite{Shimane2023oct}. 
The image is then processed using the Christian-Robinson algorithm, which is a horizon-based, non-iterative method for generating a position measurement $\boldsymbol{r}_{\mathrm{P}}$, along with an analytical expression for the measurement covariance $\boldsymbol{P}_{\boldsymbol{r}_{\mathrm{P}}}$. 
The measurement and measurement covariance are fed to the navigation filter; for simulation purposes, we keep track of both the filter's state estimate $\hat{\state}$ as well as the true state $\state$. 
The synthetic imaging block uses $\state$ to place the camera, while $\hat{\state}$ is used to point the camera toward the perceived location of the Moon. 
The control block on the bottom right is only given $\hat{\state}$ and $\boldsymbol{P}_{\boldsymbol{r}_{\mathrm{P}}}$ to compute a control command $\ubold$.
The state estimation assumes that $\ubold$ is executed, while in fact a corrupted $\ubold + \delta \ubold$ is executed on the true state $\state$.

\section{Autonomous Navigation with Optical Measurements}
\label{sec:autonomous_navigation}
We first provide an overview of the horizon-based OPNAV scheme. 
We adopt the non-iterative Christian-Robinson algorithm~\cite{ChristianRobinson2016} due to its superior performance over other iterative, ellipse-fitting-based variants~\cite{Christian2016Iterative}.
The original measurement covariance expression accounting for error associated with detecting the lit limb within the image~\cite{ChristianRobinson2016} is further developed to account for attitude uncertainty in addition to the error associated with detecting the lit limb within the image. 
Then, we briefly discuss the simulation environment developed in Blender in our previous work~\cite{Shimane2023oct}.
Finally, we describe the dynamic filter employed with the OPNAV measurements for simulating autonomous navigation.

\subsection{Horizon-Based Optical Navigation}
We consider the horizon-based OPNAV algorithm for computing the position vector of the spacecraft with respect to the imaged body assuming the camera attitude, defined by the camera frame~$\Frame_{\rm C}$, is known. 
This assumption is built on the fact that spacecraft attitude can be computed with high precision using star trackers. 
The small error on the spacecraft's attitude is incorporated into the image generation process, and the measurement covariance expression is updated from the expression provided in~\cite{ChristianRobinson2016} to reflect this source of error. 
The algorithm also requires ephemerides of the Moon and the Sun, as well as the radii of the Moon; these are assumed to be stored on the onboard computer.
\rreview{We first provide a brief overview of the Christian-Robinson algorithm, then further develop the measurement covariance model including attitude uncertainty.}

\subsubsection{Christian-Robinson Algorithm}
The Christian-Robinson algorithm~\cite{ChristianRobinson2016} provides both the position vector estimate as well as an analytical expression for the position measurement covariance \review{in the body's principal axes frame~$\Frame_{\rm P}$}. 
While a short overview is provided here for completeness, a full description is available for the Cholesky decomposition variant in~\cite{ChristianRobinson2016} and for the Singular Value Decomposition (SVD) variant in~\cite{Christian2017}. 
Starting with the $m$ pixel coordinates along the lit limb $\{u_i, v_i\}_{i=1}^m$ of the Moon \review{and assuming a pinhole camera model~\cite{Christian2021Tutorial}}, each point is projected onto a point on the image plane $\boldsymbol{s}_i$ by using the intrinsic camera calibration matrix $\boldsymbol{\kappa}$~\cite{Hartley2003, Christian2017, Christian2021Tutorial}
\begin{equation}
    \boldsymbol{s}_i = \boldsymbol{\kappa}^{-1} \begin{bmatrix}
        u_i \\ v_i \\ 1
    \end{bmatrix}
    ,
    \quad
    \boldsymbol{\kappa} = \begin{bmatrix}
        f / \mu_x & \alpha & u_p \\ 0 & f / \mu_y & v_p \\ 0 & 0 & 1
    \end{bmatrix}
    ,
\end{equation}
\review{where $f$ is the focal length, $\mu_x$ and $\mu_y$ are the horizontal and vertical pixel pitch, $[u_p, v_p]$ defines the principal point of the focal plane, and $\alpha$ is the skewness.
In this work, we assume the synthetic images contain no affine shear, i.e., $\alpha = 0$.
Here, we construct $\boldsymbol{\kappa}$ using the exact values for $f$, $\mu_x$, $\mu_y$, $u_p$, and $v_p$ used to generate the synthetic images of the Moon.
While in practice, errors in $\boldsymbol{\kappa}$ are inevitable, their impact is mitigated through on-orbit re-calibration of $\boldsymbol{\kappa}$ at the early stage of a mission~\cite{Christian2016-qp}.
In this work, we choose to assume a perfect $\boldsymbol{\kappa}$, thereby isolating the performance of the station-keeping pipeline from errors introduced by the optical hardware. If needed, these could be accounted for by appropriately increasing the measurement error covariance.}
Using SVD, the horizon-based OPNAV problem is transformed from being relative to a triaxial ellipsoid to being relative to a unit sphere via
\begin{equation}
    \bar{\boldsymbol{s}}_i = \boldsymbol{Q} \TCtoP \boldsymbol{s}_i,
\end{equation}
where $\boldsymbol{Q} = \operatorname{diag}(1/a, 1/b, 1/c)$ with $a$, $b$, $c$ the body's principal axes, and $\TCtoP \in \mathbb{R}^{3 \times 3}$ is the transformation matrix from $\Frame_{\rm C}$ to $\Frame_{\rm P}$. 
Finally, defining $\bar{\boldsymbol{s}}^{\prime}_i = \bar{\boldsymbol{s}}_i / \| \bar{\boldsymbol{s}}_i \|$, the position of the spacecraft in the camera frame is obtained by solving the linear least-squares problem
\begin{equation}
    \begin{bmatrix}
        {\bar{\boldsymbol{s}}^{\prime T}_1} \\ 
        {\bar{\boldsymbol{s}}^{\prime T}_2} \\ 
        \vdots \\ 
        {\bar{\boldsymbol{s}}^{\prime T}_m}
    \end{bmatrix}
    \boldsymbol{n} 
    =
    \boldsymbol{H} \boldsymbol{n}
    = \boldsymbol{1}_{m \times 1}
    ,
    \label{eq:opnav_leastsquares}
\end{equation}
\review{where $\boldsymbol{n}$ is a vector in the direction of $-\rbold$ with a length related to the distance between the camera and the center of the imaged body~\cite{Christian2021_TRN}.}
The matrix $\boldsymbol{H}$ is the vertical concatenation of ${\bar{\boldsymbol{s}}^{\prime T}_i}$ for $i =  1, \ldots, m$.
The position vector of the spacecraft in $\Frame_{\rm C}$ is given by 
\begin{equation}
    \hat{\rbold}_{\mathrm{C}} = -(\boldsymbol{n}^T \boldsymbol{n} - 1)^{-(1/2)} \TPtoC \boldsymbol{Q}^{-1} \boldsymbol{n}
    ,
\end{equation}
which is transformed to $\Frame_{\rm P}$ via
\begin{equation}
    \hat{\rbold}_{\mathrm{P}} = \TCtoP \hat{\rbold}_C
    = -(\boldsymbol{n}^T \boldsymbol{n} - 1)^{-(1/2)} \boldsymbol{Q}^{-1} \boldsymbol{n}
    ,
    \label{eq:opnavmeasurementsample}
\end{equation}
where $\TCtoP \in \mathbb{R}^{3\times 3}$ is the transformation matrix from $\Frame_{\rm P}$ to $\Frame_{\rm C}$.

\subsubsection{Analytical Measurement Covariance Under Attitude Uncertainty}
Following Christian and Robinson~\cite{ChristianRobinson2016}, we compute the covariance of the position measurement, in the presence of error associated with detecting the lit limb within the image, quantified in terms of the ``standard deviation of an observed horizon point in units of pixels'' $\sigma_{\mathrm{pix}}$. 
This model assumes perfect attitude knowledge, and as a consequence, perfect knowledge of the transformation matrix $\TPtoC$. 
However, in reality, the camera will have some unknown misalignment due to the imperfection of the estimated attitude, the mounting process of the hardware, and the vibration during launch. 
To this end, we develop an expression for the measurement covariance of the position in $\Frame_{\rm P}$, denoted by $\boldsymbol{P}_{\boldsymbol{r}_{\mathrm{P}}}$, that incorporates both the error introduced by the limb detection process, as well as the imperfect attitude knowledge.
\rreview{A similar modification to the measurement covariance was introduced in Franzese~\cite{Franzese2025-wb}, accounting for both pixel detection error and attitude error.
In~\cite{Franzese2025-wb}, the covariance is modeled as a pointing error expressed purely in the plane perpendicular to the boresight. 
In contrast, we model the attitude through a generic small perturbation $\delta \boldsymbol{\phi}$ acting on~$\Tbold^{\rm C}_{\rm P}$, which allows for the direct incorporation of attitude uncertainty information without having to project it onto $\mathcal{F}_{\rm C}$ a priori.}

\begin{figure}[t]
    \centering
    \begin{subfigure}[b]{0.8\textwidth}
        \centering
        \includegraphics[width=\textwidth]{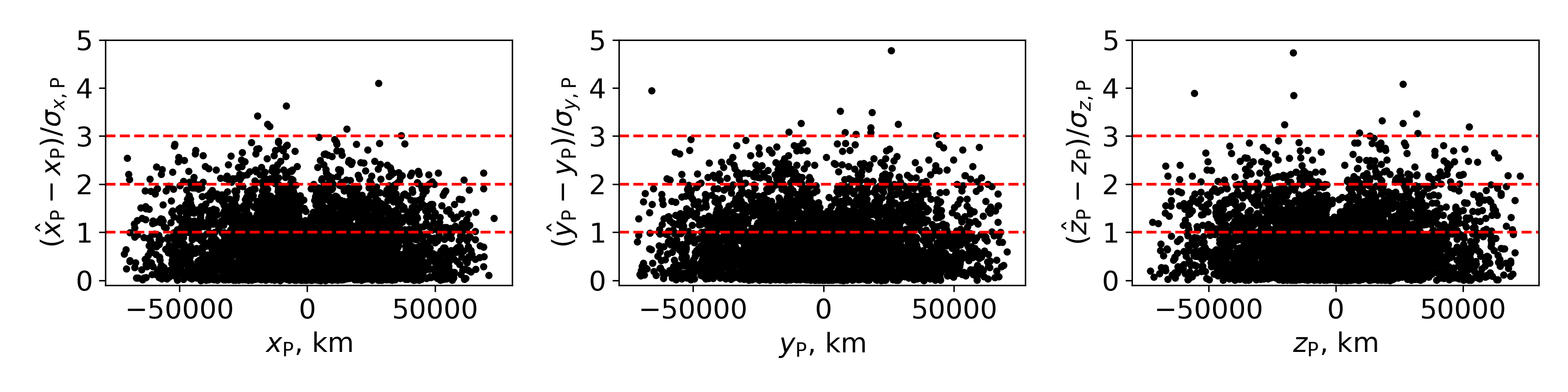}
        \caption{Ratio of measurement error over predicted $1$-$\sigma$ against true position vector component}
        \label{fig:meas_covar_exp_errsig_scatter}
    \end{subfigure}
    \\
    \begin{subfigure}[b]{0.8\textwidth}
        \centering
        \includegraphics[width=\textwidth]{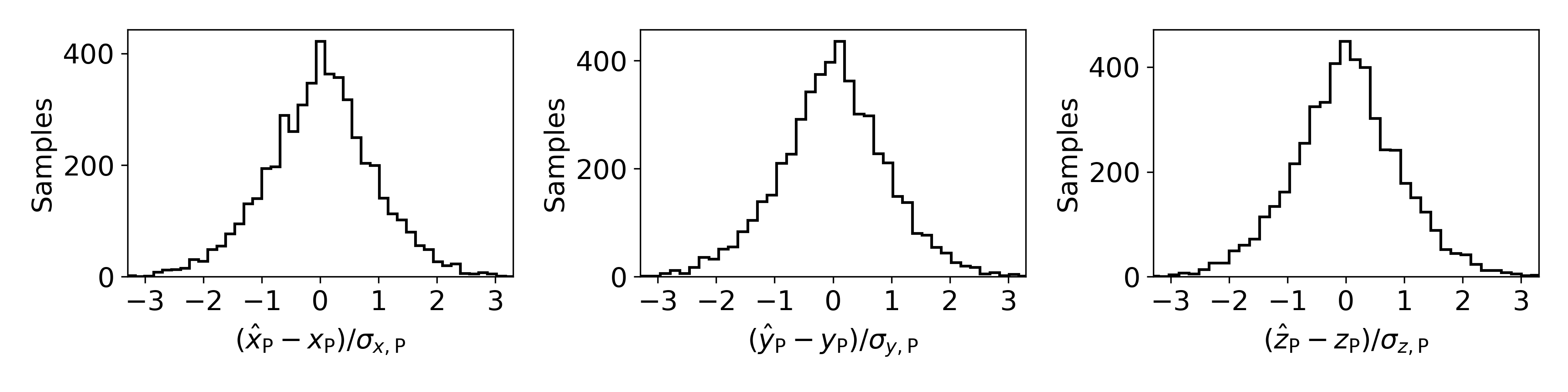}
        \caption{Distribution of ratio of measurement error over predicted $1$-$\sigma$}
        \label{fig:meas_covar_exp_errsig_hist}
    \end{subfigure}
    \caption{Component-wise ratio of measurement error over predicted $1$-$\sigma$ by measurement covariance model}
    \label{fig:meas_covar_attitude_error}
\end{figure}

\review{Let $\boldsymbol{\phi}$ denote the rotation vector corresponding to $\Tbold^{\rm C}_{\rm P}$.}
Taking the variation of~\eqref{eq:opnavmeasurementsample} with respect to both $\boldsymbol{n}$ and $\boldsymbol{\phi}$, 
\begin{equation}
    \delta \rbold_{\mathrm{P}} = \boldsymbol{F} \delta \boldsymbol{n} + \boldsymbol{G} \delta \boldsymbol{\phi}
    ,
\end{equation}
where $\boldsymbol{F}$ and $\boldsymbol{G}$ are given by
\begin{equation}
    \boldsymbol{F} = - \left(\boldsymbol{n}^T \boldsymbol{n}-1\right)^{-(1 / 2)} \boldsymbol{Q}^{-1}
    \left(\boldsymbol{I}_{3 \times 3}-\frac{\boldsymbol{n} \boldsymbol{n}^T}{\boldsymbol{n}^T \boldsymbol{n}-1}\right),
\end{equation}
\begin{equation}
    \boldsymbol{G} = \boldsymbol{T}^C_P [\rbold_{\mathcal{C}} \times ] 
    ,
\end{equation}
$[\, \cdot \,\times ]$ is the skew-symmetric matrix $[\boldsymbol{a} \times] \boldsymbol{b} = \boldsymbol{a} \times \boldsymbol{b}$, and $d_x$ is the pixel pitch in terms of pixels per radian. 
The covariance of $\hat\rbold_{\mathrm{P}}$ is given by 
\begin{equation}
    \boldsymbol{P}_{\boldsymbol{r}_{\mathrm{P}}} = \mathbb{E} [\delta \hat\rbold_{\mathrm{P}} \delta \hat\rbold_{\mathrm{P}}^T]
    = \boldsymbol{F} \boldsymbol{P}_n {\boldsymbol{F}}^T + \boldsymbol{G} \boldsymbol{P}_{\phi} \boldsymbol{G}^T
    \label{eq:measnoisecov}
    ,
\end{equation}
where $\boldsymbol{P}_n$ is the covariance of the vector $\boldsymbol{n}$, and $\boldsymbol{P}_{\phi}$ is the covariance of the camera attitude. 
The expression for $\boldsymbol{P}_{\phi}$ is given by
\begin{equation}
    \boldsymbol{P}_{\phi} = \sigma_{\phi}^2 \boldsymbol{I}_{3 \times 3}.
\end{equation}
An approximation for $\boldsymbol{P}_n$ developed in~\cite{ChristianRobinson2016} for Cholesky decomposition is adapted to the SVD via
\begin{equation}
    \boldsymbol{P}_n = \left( \boldsymbol{H}^T \boldsymbol{R}_y \boldsymbol{H} \right)^{-1}
    ,
\end{equation}
where $\boldsymbol{R}_y \in \mathbb{R}^{m \times m}$ is the covariance of the residuals of the least-squares problem~\eqref{eq:opnav_leastsquares}, given by
\begin{equation}
\begin{aligned}
    \boldsymbol{R}_y &= \operatorname{diag}(\sigma_{y_1}^2, \ldots, \sigma_{y_m}^2) 
    ,
    \\
    \sigma_{y_i} &= 
        \boldsymbol{J}_i 
        \boldsymbol{Q} \boldsymbol{T}^{\rm C}_{\rm P}
        \boldsymbol{R}_s
        \boldsymbol{T}^{\rm P}_{\rm C} \boldsymbol{Q}^T
        \boldsymbol{J}_i^T
    ,
\end{aligned}
\end{equation}
$\boldsymbol{R}_s$ is the covariance of the horizon measurements, approximated by
\begin{equation}
    \boldsymbol{R}_s \approx 
        \left( \dfrac{\sigma_{\mathrm{pix}}}{d_x}\right)^2
        \begin{bmatrix}
            1 & 0 & 0 \\ 0 & 1 & 0 \\ 0 & 0 & 0
        \end{bmatrix}
    ,
\end{equation}
$\sigma_{\mathrm{pix}}$ is the standard deviation of an observed horizon point in units of pixels~\cite{ChristianRobinson2016}, and $\boldsymbol{J}_i$ is given by
\begin{equation}
    \boldsymbol{J}_i = 
        \dfrac{1}{\| \bar{\boldsymbol{s}}_i \|} 
        \boldsymbol{n}^T ( \boldsymbol{I}_{3 \times 3} - {\bar{\boldsymbol{s}}_i}^{\prime} {\bar{\boldsymbol{s}}_i}^{\prime T} )
    .
\end{equation}
The measurement covariance expression~\eqref{eq:measnoisecov} is validated using a Monte Carlo experiment with the following steps:
\begin{enumerate}
    \item Using a camera with fixed $\boldsymbol{\kappa}$, generate random position vector in camera frame, $\rbold_{\rm C}$, and random transformation matrix $\boldsymbol{T}^{\rm C}_{\rm P}$, corresponding to the attitude \textit{estimate} of the camera with respect to $\mathcal{F}_{\rm P}$.
    \item Generate 100 limb pixel coordinates $\{u_i, v_i\}_{i=1}^{100}$ along the true limb of the body, located at equal angle intervals over a circular sector of $140^{\circ}$, perturbed by individually drawing a realization of $\sigma_{\rm pix}$
    \begin{equation}
        \{u_i, v_i\} 
        =
        \{u_i, v_i\}_{\mathrm{true}} + \{\delta u_i, \delta v_i \}
        \quad i = 1, \ldots, m
        ,
    \end{equation}
    where $\delta u_i, \delta v_i \sim \mathcal{N}(0, \sigma_{\rm pix})$.
    \item Construct the attitude perturbation matrix $\delta \boldsymbol{T} \in \mathbb{R}^{3 \times 3}$ using the Rodrigues' rotation formula
    \begin{equation}
        \delta \Tbold
        =
        \cos(\delta \phi) \boldsymbol{I}_3 + \sin(\delta \phi)
        \boldsymbol{i}^{\times} +
        [1 - \cos(\delta \phi)] \boldsymbol{i} \boldsymbol{i}^T
        ,\quad
        \boldsymbol{i}
        =
        \dfrac{1}{\sqrt{i_x^2 + i_y^2 + i_z^2}}
        \begin{bmatrix}
            i_x \\ i_y \\ i_z
        \end{bmatrix}
        ,
    \end{equation}
    where $\delta \phi \sim \mathcal{N}(0, \sigma_{\phi}^2)$, $i_x,i_y,i_z \sim \mathcal{U}(-1,1)$, and $\boldsymbol{i}^{\times}$ is the skew-symmetric form of $\boldsymbol{i}$. 
    \item Compute true position in the planet frame~$\mathcal{F}_{\rm P}$,
        \begin{equation}
            \rbold_{\mathrm{P}} = 
            \delta \boldsymbol{T} 
            \boldsymbol{T}^{\rm C}_{\rm P} \rbold_{\rm C}
            \rbold_{\rm C}
            .
        \end{equation}
    \item Compute position estimate $\hat{\rbold}_{\mathrm{C}}$ and covariance $\Pbold_{\rbold_{\mathrm{C}}}$ with Christian-Robinson algorithm and convert to $\Frame_{\rm P}$, $\hat{\rbold}_{\mathrm{P}}$ and $\Pbold_{\rbold_{\mathrm{P}}}$, using estimate transformation matrix $\boldsymbol{T}^{\rm C}_{\rm P}$.
    \item Compute error in~$\Frame_{\rm P}$ as $\hat{\rbold}_{\mathrm{P}} - \rbold_{\mathrm{P}}$.
\end{enumerate}
\rreview{Using $\sigma_{\rm pix} = 0.5$ pixel, $\sigma_{\phi} = 15$ $\mathrm{arcsec}$, and a camera with a field of view (FOV) of approximately $18^{\circ}$, we sample 10,000 points and repeat this procedure.
Note that, to ensure the Moon fits within the FOV, we uniformly sample $\rbold_{\rm C}$ such that $40000 \leq \| \rbold_{\rm C} \|_2 \leq 75000$, where the upper-bound is chosen based on the approximate furthest distance from the Moon along the NRHO.
As summarized in Table~\ref{tab:measurement_covariance_model}, we find the updated covariance model~\eqref{eq:measnoisecov} significantly improves the prediction of the standard deviation in the presence of attitude error.
This result is also illustrated in Figure~\ref{fig:meas_covar_attitude_error}, where the ratio $\hat{\rbold}_{\rm P} - \rbold_{\rm P}$ over the standard deviation from diagonal components of $\Pbold_{\rbold_{\rm P}}$ component-wise.
This ratio visualizes the prediction accuracy of $\Pbold_{\rbold_{\rm P}}$ against the empirical distribution from the Monte Carlo samples, }

\begin{table}[]
    \centering
    \caption{Monte Carlo validation results for measurement covariance model}
    \begin{tabular}{@{}llll@{}}
    \toprule
        Measurement covariance model $\boldsymbol{P}_{\boldsymbol{r}_{\mathrm{P}}}$ & Ratio of samples in $1$-$\sigma$ & Ratio of samples in $2$-$\sigma$ & Ratio of samples in $3$-$\sigma$ \\ \midrule
        $\boldsymbol{F} \boldsymbol{P}_n {\boldsymbol{F}}^T + \boldsymbol{G} \boldsymbol{P}_{\phi} \boldsymbol{G}^T$ (eqn~\eqref{eq:measnoisecov})&
        $0.6305$ & $0.9393$ & $0.9969$ \\
        $\boldsymbol{F} \boldsymbol{P}_n {\boldsymbol{F}}^T$ (without accounting for $\delta \boldsymbol{\phi}$) &
        $0.5976$ & $0.9169$ & $0.9899$ \\
        \bottomrule
    \end{tabular}
    \label{tab:measurement_covariance_model}
\end{table}

\subsection{Simulation Environment for Horizon-Based Optical Navigation}
\begin{figure}
    \centering
    \includegraphics[width=0.99\linewidth]{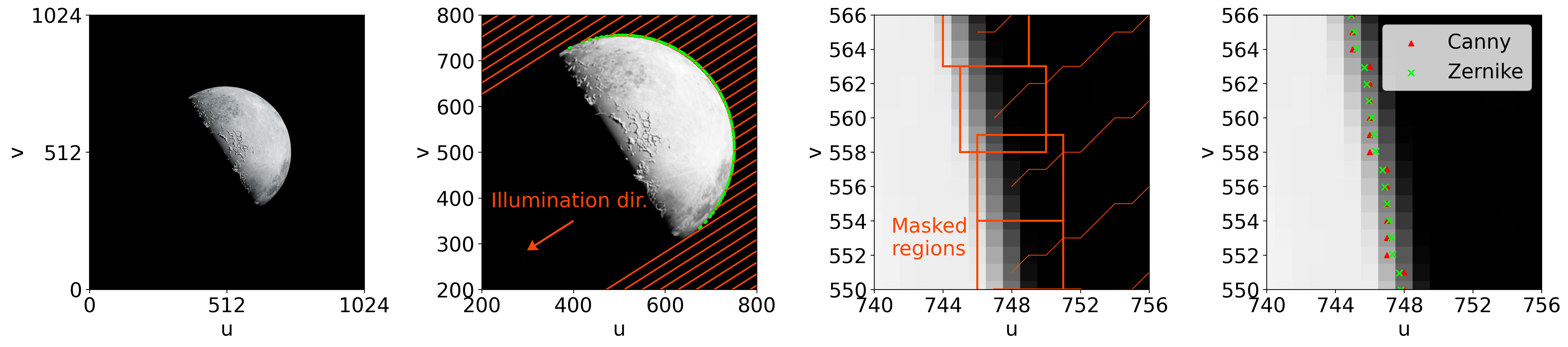}
    \caption{Synthetic image, illumination scan, masked regions, and subpixel-level edge detection}
    \label{fig:opnav_demo}
\end{figure}

The simulation environment is developed using Blender~\cite{Blender}, an open-source rendering engine. 
The development process of the simulation environment is detailed in our prior paper~\cite{Shimane2023oct}. 
Points on the image corresponding to the lit limb of the Moon are found through a three-step process following~\cite{Christian2017}, consisting of (i) creating a mask centered around approximate lit limb locations found by scanning the image along the projected direction of the illumination vector of the Sun, followed by (ii) applying a mask on edge points detected using the Canny edge detection algorithm, and finally (iii) applying a Zernike moment-based refinement scheme to correct the pixel-level point from (ii) down to subpixel-level locations using local mini patches centered around each pixel-level edge.
\review{Figure~\ref{fig:opnav_demo} shows, from left to right, the original synthetic image of the Moon, the illumination scan projected into the image, the masks created by the illumination scan, and the pixel-level and subpixel-level refined edges.}

\subsection{Navigation Filter}
The operation of a spacecraft requires a navigation filter to keep track of the best state estimate given measurements that may only include a subset of the state components. 
Due to the non-negligible nonlinearity of the dynamics, the traditional Kalman filter is insufficient for accurately predicting the state.
Thus, the extended Kalman filter (EKF) or the unscented Kalman filter (UKF) is more effective, providing a suitable trade-off between capturing the nonlinearity of the dynamics and the computational burdens. 

In the context of spaceflight, the EKF has been successfully applied in multiple missions~\cite{Carpenter2018}. 
Through preliminary studies, we found that EKF and UKF result in nearly identical navigation performance assuming a measurement update cadence of a few images per revolution. Thus, we adopt the EKF for navigation. 
Nevertheless, we also introduce the UT-based prediction, as its prediction scheme is leveraged for the design of the SK controller.

\subsubsection{Prediction in Extended Kalman Filter}
\review{The prediction step consists of propagating~\eqref{eq:dyn_model}, accounting for uncertainties in the SRP term~\eqref{eq:eom_SRP_term}. The filter assumes nominal SRP coefficients $\bar{C}_r$ and $\overline{A/m}$, while their true values are perturbed as
\begin{subequations}
\begin{align}
    C_r &= \bar{C}_r (1 + \delta_{C_r}),\quad 
        \delta_{C_r}\sim\mathcal{N}(0,\sigma_{C_r}^2)
    ,\quad
    \\
    A/m &= \overline{A/m} (1 + \delta_{A/m}),\quad  \delta_{A/m}\sim\mathcal{N}(0,\sigma_{A/m}^2)
    ,
\end{align}
\end{subequations}
at the beginning of each prediction.
The uncertainty is accounted for by an additive process noise~\cite{sarkka2023bayesian}, 
}
The prediction step of the EKF is given by
\begin{subequations}
    \begin{align}
        \hat{\state}_{j \mid j-1} &=
            \hat{\state}_{j-1 \mid j-1} + \int_{t_{j-1}}^{t_j} \fbold(\tau, \hat{\state}_{j-1}(\tau)) \mathrm{d} \tau
            \label{eq:meanpred_ekf}
        ,
        \\
        \boldsymbol{\Sigma}_{j \mid j-1}
            &=
            \boldsymbol{\Phi}\left(t_j, t_{j-1}\right) \boldsymbol{\Sigma}_{j-1 \mid j-1} \boldsymbol{\Phi}\left(t_j, t_{j-1}\right)^T + \boldsymbol{Q}_{j}
        .
    \end{align}
\end{subequations}
\review{$\boldsymbol{Q}_{j}$ is the additive process noise that accounts for the dynamics modeling error~\cite{sarkka2023bayesian},} modeled as an unbiased random process
\begin{align}
    \boldsymbol{Q}_j =
    \sigma_u^2
    \left[\begin{array}{cc}
        \frac{1}{3} h_j^3 \boldsymbol{I}_{3\times 3} & \frac{1}{2} h_j^2 \boldsymbol{I}_{3\times 3}
        \\[0.5em]
        \frac{1}{2} h_j^2 \boldsymbol{I}_{3\times 3} & h_j \boldsymbol{I}_{3\times 3}
    \end{array}\right],
    \label{eq:process_noise}
\end{align}
where $h_j=t_j-t_{j-1}$ is the duration of a sampling period and $\sigma_u^2$ is the diffusion coefficient to be adjusted to optimize the filter performance~\cite{Tapley_Schutz_Born_2004}.

\subsubsection{Prediction in Unscented Kalman Filter}
In the UKF, the prediction involves forming and propagating $2n+1$ sigma points, where $n$ is the dimension of the state vector $\xbold$, chosen as in~\cite{Sarkka2007},
\begin{subequations}
    \begin{align}
        \mathcal{X}_{j-1 \mid j-1}^{(0)} &= \hat{\state}_{j-1 \mid j-1}
        ,\\
        \mathcal{X}_{j-1 \mid j-1}^{(\ell)} &= \hat{\state}_{j-1 \mid j-1} 
            + \sqrt{n + \lambda} \left[\sqrt{\boldsymbol{\Sigma}_{j-1 \mid j-1}}\right]_{\ell}
            ,
        \quad \ell = 1, \ldots, n
        ,\\
        \mathcal{X}_{j-1 \mid j-1}^{(\ell+n)} &= \hat{\state}_{j-1 \mid j-1} 
            - \sqrt{n + \lambda} \left[\sqrt{\boldsymbol{\Sigma}_{j-1 \mid j-1}}\right]_{\ell}
            ,
        \quad \ell = 1, \ldots, n
        ,
    \end{align}
    \label{eq:ukf_sigmapoints}
\end{subequations}
where $\left[\sqrt{\boldsymbol{\Sigma}_{j-1 \mid j-1}}\right]_{\ell}$ corresponds to the $\ell^{\mathrm{th}}$ column of the matrix square root of ${\boldsymbol{\Sigma}_{j-1 \mid j-1}}$, and $\lambda$ is given by
\begin{equation}
    \lambda = \alpha^2 (n + \kappa) - n
    .
\end{equation}
The parameter $\alpha$ dictates the spread of the sigma points around the mean, usually chosen to be a small positive value, and $\kappa$ is a secondary scaling parameter.
\rreview{In this work, we use $\alpha=10^{-3}$ and $\kappa = 0$, the latter choice corresponding to the spherical cubature integration~\cite{sarkka2023bayesian}.}
Each sigma point is propagated using the nonlinear dynamics 
\begin{equation}
    \mathcal{X}_{j \mid j-1}^{(\ell)} = 
        \mathcal{X}_{j-1 \mid j-1}^{(\ell)} + \int_{t_{j-1}}^{t_j} 
        f\left(\tau, \mathcal{X}_{j-1 \mid j-1}^{(\ell)}(\tau)\right)
        \mathrm{d} \tau,
        \quad \ell = 0, \ldots, n
        .
\end{equation}
Finally, the time update of the mean and the covariance are given by
\begin{subequations}
    \begin{align}
         \hat{\state}_{j \mid j-1} &= \sum_{\ell=0}^{2n} W^{(\ell)}_{m} \mathcal{X}_{j \mid j-1}^{(\ell)}
         \label{eq:meanpred_ukf}
         ,
         \\
         \boldsymbol{\Sigma}_{j \mid j-1} &= \sum_{\ell=0}^{2n} W^{(\ell)}_{c}
            \left(\mathcal{X}_{j \mid j-1}^{(\ell)} - \hat{\state}_{j \mid j-1}\right)
            \left(\mathcal{X}_{j \mid j-1}^{(\ell)} - \hat{\state}_{j \mid j-1}\right)^T
            + \boldsymbol{Q}_j
        ,
    \end{align}
    \label{eq:pred_ukf}
\end{subequations}
where $\boldsymbol{Q}_j$ is the process noise given by expression~\eqref{eq:process_noise}, while $W^{(\ell)}_{m}$ and $W^{(\ell)}_{c}$ are weights given by
\begin{equation}
    \begin{aligned}
        W^{(0)}_{m} &= \dfrac{\lambda}{n + \lambda},
        \\
        W^{(0)}_{c} &= \dfrac{\lambda}{n + \lambda + (1-\alpha^2+\beta)},
        \\
        W^{{(\ell)}}_{m} &= \dfrac{1}{2(n+\lambda)}, \quad \ell = 1,\ldots,2n,
        \\
        W^{(\ell)}_{c} &= \dfrac{1}{2(n+\lambda)}, \quad \ell = 1,\ldots,2n,
    \end{aligned}
\end{equation}
and $\beta$ is an additional parameter to incorporate prior knowledge of $\hat{\state}$; $\beta = 2$ is optimal assuming the distribution of $\hat{\state}$ is Gaussian~\cite{sarkka2023bayesian}, and is used in this work.

\subsubsection{Measurement Update}
The OPNAV measurements are modeled as direct position measurements with additive Gaussian noise,
\begin{equation}
   \yvec_j = \boldsymbol{E}\state_j + \boldsymbol{w}_j, \;\;\boldsymbol{w}_j\sim\mathcal{N}(\boldsymbol{0},\boldsymbol{R}_j), \;\; \boldsymbol{E} = \begin{bmatrix}
        \boldsymbol{I}_{3\times 3} & \boldsymbol{0}_{3\times 3}
    \end{bmatrix}.
\end{equation}
The resulting linear update step of the filter is
\begin{subequations}
\begin{align}
    \boldsymbol{K}_j & =\boldsymbol{\Sigma}_{j \mid j-1} \boldsymbol{E}^T\left(\boldsymbol{E}\boldsymbol{\Sigma}_{j \mid j-1} \boldsymbol{E}^T+\boldsymbol{R}_j\right)^{-1}, \\
    \hat{\state}_{j \mid j} &=
        \hat{\state}_{j \mid j-1}+\boldsymbol{K}_j\left(\boldsymbol{y}_j - \boldsymbol{E} \hat{\state}_{j \mid j-1}\right),
    \\
    \boldsymbol{\Sigma}_{j \mid j} & =
    \left(\boldsymbol{I} - \boldsymbol{K}_j \boldsymbol{E}\right) \boldsymbol{\Sigma}_{j \mid j-1}
    \left(\boldsymbol{I} - \boldsymbol{K}_j \boldsymbol{E}\right)^T
    + \boldsymbol{K}_j \boldsymbol{R}_j \boldsymbol{K}_j^T
    ,
\end{align}    
\end{subequations}
where the measurement $\yvec_j$ and the measurement noise covariance $\boldsymbol{R}_j$ are obtained from~\eqref{eq:opnavmeasurementsample} and~\eqref{eq:measnoisecov}, respectively.

\subsection{Measurement Collection for Horizon-Based Optical Navigation on Near-Rectilinear Halo Orbit}
\begin{figure}
    \centering
    \includegraphics[width=0.55\linewidth]{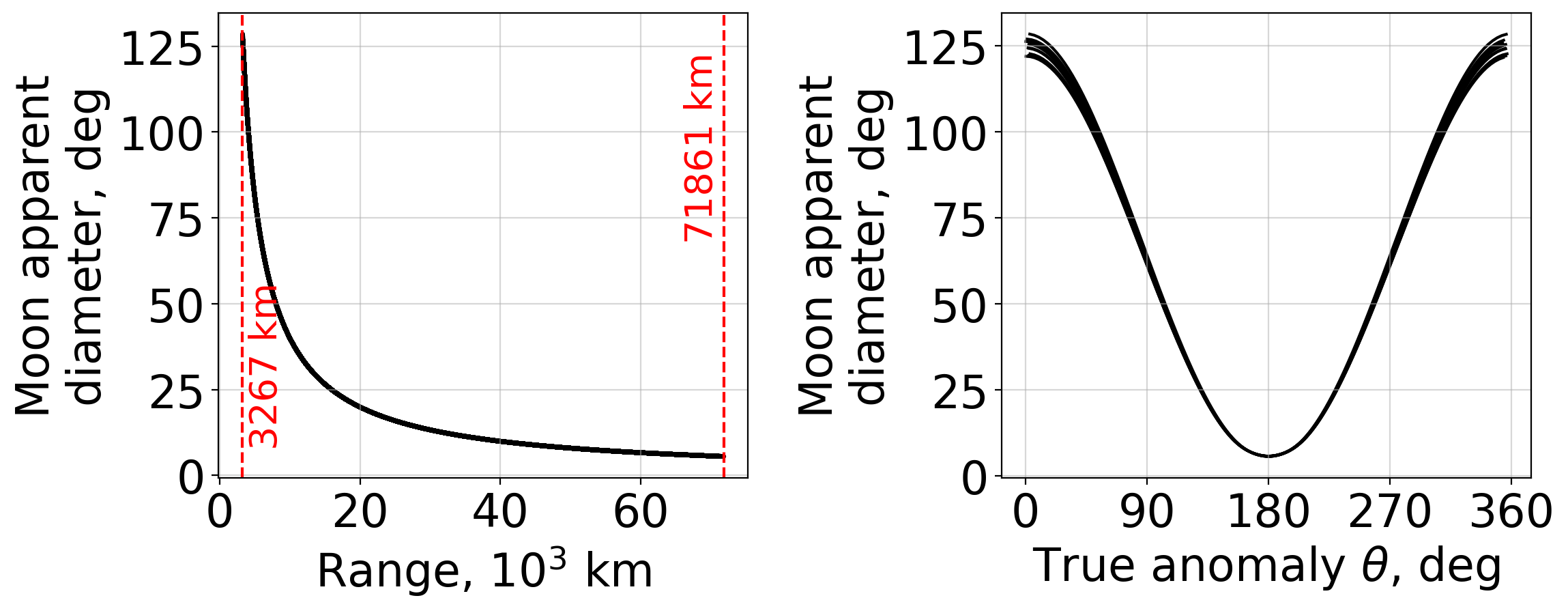}
    \caption{Range and apparent Moon diameter against osculating true anomaly along NRHO}
    \label{fig:ta_vs_range_fov}
\end{figure}

The choice of the onboard camera specifications directly impacts the number of limb points $m$ detected within an image, and therefore the quality of the position measurement itself. 
Furthermore, for a given position in space, if the FOV is too narrow and the limb of the Moon does not fit within the frame of the image, the measurement may be unusable. 
The choice of camera FOV is particularly important on the NRHO due to the large variety in distance from the Moon at perilune and apolune. 
Figure~\ref{fig:ta_vs_range_fov} shows the range and the apparent diameter of the Moon as seen along the NRHO across 10 superimposed revolutions.

\subsubsection{Effect of Field of View on Measurement}
The least-squares solution of~\eqref{eq:opnav_leastsquares} is affected by the number of detected limb points $m$.
One way to increase $m$ may be to use a camera with a larger number of pixels on its sensor; however, this requires larger onboard processing overhead. 
Another tuning knob is the camera's FOV. A camera with an FOV that exactly matches the apparent diameter of the Moon will contain the largest possible $m$, assuming that the Moon is exactly centered along the boresight. 
However, in practice, the camera is pointed to the Moon using the state estimate $\hat{\state}$.
Other operational considerations on attitude may further lead to scenarios where it is desirable to image the Moon without slewing the spacecraft too much.

\subsubsection{Choice of Camera for Horizon-Based OPNAV on NRHO}
\label{sec:choice_of_fov}
For a given range from the Moon and a given level of pointing margin, the narrowest possible FOV that fits the entire limb results in the largest $m$, and thus the best obtainable measurement $\hat{\rbold}$. 
However, choosing the optimal camera for a spacecraft on the NRHO is not trivial due to the large variability in range along the orbit, from $3 \times 10^3$~\SI{}{km} to over $7 \times 10^4$~\SI{}{km}: in the extreme case, choosing an FOV that fits the Moon within the image at perilune will result in significantly poorer measurements at other locations along the orbit, \rreview{where the spacecraft spends most of its time}, as the Moon will appear very small in the image.
Meanwhile, choosing an FOV that exactly fits the Moon at apolune will not allow for imaging the Moon at other locations, as the range is shorter at all other locations.
We can thus already hypothesize that the ``best'' camera will have an FOV that is suitable in some in-between ranges between perilune and apolune. 

If the navigation subsystem can accommodate more than a single camera, it would be ideal to have a near-range and a far-range camera, each with FOVs suitable for different locations along the orbit.
Within the scope of this work, we limit our study to a system with a single camera.
As we will show with numerical results, horizon-based OPNAV measurements using an appropriately chosen FOV and taken at appropriate locations along the orbit are remarkably reliable, thus resulting in satisfactory filter and station-keeping performance.

When choosing the FOV, it is important to keep in mind that for the purpose of autonomous navigation, we seek to have reliable filter estimates near apolune, where station-keeping control maneuvers, when necessary, are executed. 
On one hand, there is an inherent degradation of observability as the range increases. 
The worsened observability is manifested in the slow variation of the Moon's apparent diameter around apolune, as shown in Figure~\ref{fig:ta_vs_range_fov}; as the image is taken further away from the Moon, a difference in range $\Delta r$ results in a rapidly diminishing difference in the apparent diameter. 
The degradation of the measurement quality is reported in previous works, including Figure 9 in~\cite{ChristianRobinson2016}.
Meanwhile, imaging at perilune, where we have the best observability, also has its downsides: first, from an operational perspective, this may lead to pointing requirements over short periods of time to ensure the Moon is within the frame of the camera. Second, \rreview{measurement collected anywhere along the NRHO outside of perilune with a large FOV camera will be noticeably worse due to the smaller number of limb points $m$ that can be detected within the image.}
Thus, we choose a sufficiently narrow FOV sized to obtain reliable measurements outside of the vicinity of the NRHO's perilune; to account for attitude uncertainties, we ensure the FOV is at least twice as large as the Moon's apparent diameter from where the measurement is acquired.
\rreview{This FOV choice, leading to discarding of imaging opportunities around perilune, is a direct consequence of the highly eccentric nature of the NRHO\footnote{While the NRHO is a non-Keplerian periodic orbit, we borrow the term ``eccentric'' in a restricted two-body sense to refer to its high variability in range to the Moon over one revolution.}; a horizon-based OPNAV system along other LPOs, such as the 2:1 resonant L2 Halo orbit adopted in the proposed LUMIO mission, where the range varies only between $3\times10^4$~\SI{}{km} to $9\times 10^4$~\SI{}{km}, may reasonably adopt an FOV that can image the Moon over the entire revolution without significant degradation in measurement quality around apolune~\cite{Franzese2019}.}

\review{Note that even with a narrow FOV, the camera may still be used near perilune to provide measurements via alternative, terrain-relative OPNAV methods, such as feature-based~\cite{McCabe2020-ck}, crater-based navigation~\cite{Christian2021-cz}, or odometry~\cite{Christian2021-qi,De-Vries2022-rt,Cowan2025-qt}.
In this paper, we choose to focus on horizon-based measurements alone and thereby test its limits in navigating the spacecraft along the NRHO, while fusion with estimates from terrain-relative navigation is possibly left for future studies.
For considerations on sensor fusion for cislunar navigation, see e.g.~\cite{Vila2025-tl}.}

\section{Station-Keeping Control}
\label{sec:stationkeeping_control}
There exists a multitude of controllers proposed for SK on LPOs~\cite{Shirobokov2017,Dei_Tos2020-ml}. 
In the context of L1 and L2 halo orbits, a popular SK controller is the so-called $x$-axis crossing control, which has been shown to perform well both in numerical simulations~\cite{Davis2017, Guzzetti2017} and in-flight~\cite{Cheetham2021}.
A particularity about the $x$-axis crossing control scheme is that it has often been formulated as a multivariable root-finding problem, and rarely as a constrained minimization problem~\cite{Elango2022}.
In this work, we introduce both the root-finding-based and the minimization problem-based formulations, point out their distinctions, and compare their numerical performances.
Then, we propose a UT-based modification of the $x$-axis control that utilizes the filter's covariance estimate together with the state estimate to improve the prediction accuracy of the targeted state, and thereby reduce the cumulative $\Delta V$.

Recent advancements in station-keeping on the NRHO also focused on tackling the phase-deviation issue associated with $x$-axis crossing control~\cite{Davis2022,Shimane2025-by}.
To avoid the issue altogether, a model predictive control (MPC) scheme has also been proposed to provide full-state tracking under the same operational assumptions as $x$-axis crossing control~\cite{Shimane2025}, yielding similar cumulative $\Delta V$ and improved tracking performance.
In this work, the conducted numerical experiments do not exceed $\sim\!400$ \SI{}{days}, and thus are minimally impacted by phase deviation.
The UT-based modification proposed here is applicable to any control scheme, including $x$-axis crossing control schemes with phase-deviation remedies~\cite{Davis2022,Shimane2025-by} or the MPC in~\cite{Shimane2025}.

In the rest of this Section, we first provide an overview of the $x$-axis crossing control scheme, highlighting key findings from the literature that are relevant to our work. This is followed by a description of the root-solving-based formulation, solved using differential correction, and the constrained minimization-based formulation.
Then, the UT-based modification, applicable to either formulation, is introduced.
Finally, considerations pertaining to the control frequency and the targeting tolerance are discussed.

\subsection{Overview of $x$-axis Crossing Control}
The $x$-axis crossing control is a popular station-keeping control scheme that is widely regarded as among the most efficient, particularly for almost linearly stable periodic orbits such as the NRHO. 
We start with a generic optimization problem formulation that encompasses the broader family of targeting-based control and point out the specific variant that is commonly adopted in literature for the NRHO.

\subsubsection{Targeting Event}
Consider a spacecraft at $t_0$ and state estimate $\state_0 \in \mathbb{R}^6$. 
The controller must bring a subset of the propagated state at time $t_f$, $\psibold(t_f) \in \mathbb{R}^n$ where $n \leq 6$, to an $n$-dimensional ellipsoid from a reference target $\bar{\psibold}(\bar{t}_f) \in \mathbb{R}^n$ at time $\bar{t}_f$, with radii $\epsilon_{j,\mathrm{tol}}$ for $j = 0, \ldots, n-1$.
We denote by \textit{targeting event} the event that defines $\bar{t}_f$.
While the integration of the dynamics from time $t_0$ to $t_f$ is typically done in $\mathcal{F}_{\mathrm{I}}$, the targeted state $\psibold$ and $\bar{\psibold}$ may be components in another frame $\mathcal{F}_{\mathrm{B}}$. 

In $x$-axis crossing control, $\bar{t}_f \neq t_f$, as the targeting event is not defined based on propagation time.
The non-equivalence of $\bar{t}_f$ and $t_f$ is an important distinction of this control scheme compared to more conventional tracking controllers. 
The $x$-axis crossing controller gets its name from the fact that the targeting event is defined based on the time where the trajectory crosses the $xz$-plane near the Moon in the Earth-Moon rotating frame,~$\mathcal{F}_{\rm EM}$. 
While this does not exactly coincide with perilune, preliminary experiments have shown that the two events result in no apparent difference in the SK performance, and thus may be used interchangeably; in this work, the targeting event is defined by the perilune. 

The optimal impulsive control maneuver $\ubold \in \mathbb{R}^3$ that brings the spacecraft to the $n$-dimensional ellipsoid centered at $\bar{\psibold}(\bar{t}_f)$ is computed by solving the nonlinear program (NLP)
\begin{subequations}
    \label{eq:xzplane_general_NLP}
    \begin{align}
        \min_{\ubold} \quad & \| \ubold \|_2 
        ,
        \\
        \text{such that} \quad
        & \left| \psibold^{\mathrm{B}} (t_f) - \bar{\psibold}^{\mathrm{B}}(\bar{t}_f) \right| \leq v_{x,\mathrm{tol}} 
        ,
        \label{eq:xzplane_general_NLP_constraint}
    \end{align}
\end{subequations}
where the absolute value in constraint $\eqref{eq:xzplane_general_NLP_constraint}$ is applied component-wise. 
The final targeted state $\psibold^{\mathrm{B}} (t_f)$ is the subset of the propagated state at $t_f$, $\state^{\mathrm{B}} (t_f)$, transformed from the inertial frame to frame $\mathcal{F}_{\mathrm{B}}$, obtained by propagating the nonlinear dynamics with control $\ubold$,
\begin{equation}
    \state^{\mathrm{B}} (t_f)
    =
    T^{\mathrm{Inr}}_{\mathrm{B}} 
    \left( 
        \state_0 +
        \begin{bmatrix}
            \boldsymbol{0}_{3 \times 3} \\ \boldsymbol{I}_3
        \end{bmatrix}
        \ubold 
        +
        \int_{t_0}^{t_f} \fbold[\tau, \state] \mathrm{d}\tau
        \right)
    ,
    \label{eq:ivp_prediction_u0}
\end{equation}
where $T^{\mathrm{Inr}}_{\mathrm{B}} \in \mathbb{R}^{6 \times 6}$ is the transformation matrix from $\mathcal{F}_{\mathrm{I}}$ to $\mathcal{F}_{\mathrm{B}}$. 

Instead of targeting the immediate perilune, it is common to choose the $N^{\mathrm{th}}$ perilune downstream with $N > 1$ to improve the $\Delta V$ performance, subject to the reliability of the predicted state affected by estimation, dynamics, and control execution errors.
In this work, $N=7$ is adopted in accordance to previous results on the NRHO~\cite{Guzzetti2017, Davis2022}. 

Finally, the $n$ state components to target, the frame in which these state components are represented, and the tolerance on the tightness of the targeting, must be chosen. 
Previous works have successfully demonstrated that targeting the $x$-component velocity in $\mathcal{F}_{\mathrm{EM}}$, denoted hereafter by $v_x^{\mathrm{EM}}$, is sufficient for tracking a baseline NRHO trajectory under reasonable levels of errors and uncertainties~\cite{Davis2017, Guzzetti2017, Davis2022}.
Following their definition, $\mathcal{F}_{\mathrm{B}}$ is $\mathcal{F}_{\mathrm{EM}}$, and $\psibold^{\mathrm{B}} = v_x^{\mathrm{EM}}$. 
Then, constraint~\eqref{eq:xzplane_general_NLP_constraint} is actually
\begin{equation}
    \left| v_x^{\mathrm{EM}}(t_f) - \bar{v}_x^{\mathrm{EM}}(\bar{t}_f) \right| \leq v_{x,\mathrm{tol}}
    ,
    \label{eq:xzplane_vxtarget_NLP_constraint}
\end{equation}
where $v_x^{\mathrm{EM}} (t_f)$ is part of the final state $\state^{\mathrm{EM}} (t_f)$ obtained by transforming the solution to the initial value problem~\eqref{eq:ivp_prediction_u0}, and $\bar{v}_x^{\mathrm{EM}}(\bar{t}_f)$ is reference $x$-axis velocity in the instantaneous $\mathcal{F}_{\mathrm{EM}}$ at the targeted perilune. 
One may interpret the $x$-axis crossing controller using $v_x^{\mathrm{EM}}$ alone as tracking only the state components necessary for stabilizability.
The effect of the tolerance on $v_x^{\mathrm{EM}}$ has been investigated by Guzetti et al.~\cite{Guzzetti2017} using deep space network-based orbit determination, but it is revisited in this work to highlight the controller's interplay with horizon-based OPNAV.

\subsubsection{Control Location Selection}
\begin{figure}
    \centering
    \includegraphics[width=0.65\linewidth]{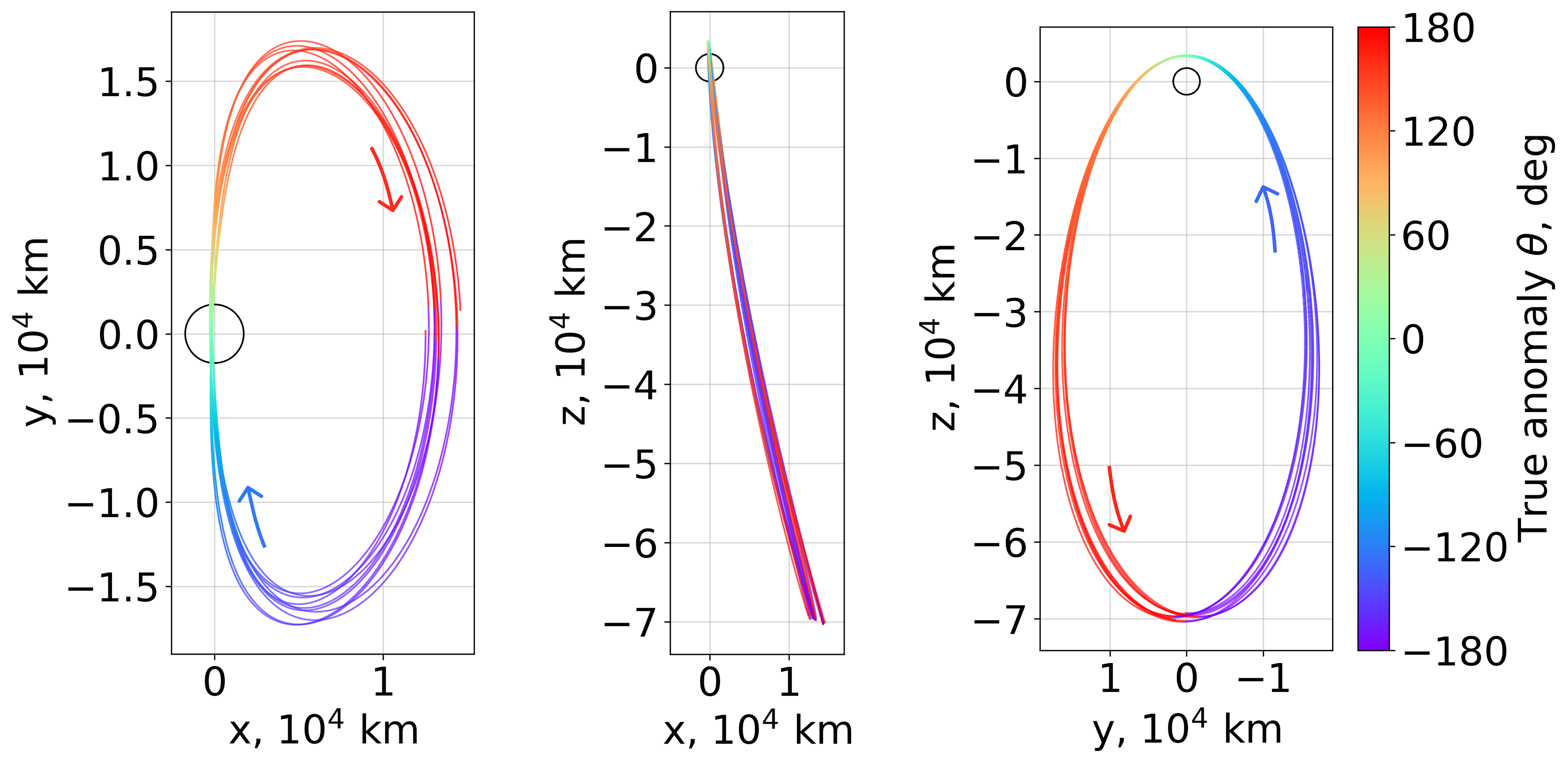}
    \caption{NRHO over 10 revolutions in Moon-centered Earth-Moon rotating frame}
    \label{fig:nrho_ta_diagram}
\end{figure}

The control location along the NRHO is parameterized in terms of the instantaneous true anomaly \review{$\theta$, calculated assuming a Moon-centered restricted two-body problem,
\begin{equation}
    \theta = \operatorname{atan2}{\left( h v_r, h^2/r - \mu \right)},
    \,\,\,
    h = \| \rbold \times \vbold\|_2, \,\,\, 
    v_r = \dfrac{\rbold \cdot \vbold}{r}
    ,
\end{equation}
as shown in Figure~\ref{fig:nrho_ta_diagram}.
In the Figure, the black arrows indicate the direciton of motion.}
The particularly high sensitivity of the dynamics near perilune on the NRHO makes this region sensitive to errors, and is usually considered unsuitable for station-keeping~\cite{Guzzetti2017}; thus, the control maneuver is to be located at or near the apolune, at a true anomaly of around $180^{\circ}$. 
Guzetti et al.~\cite{Guzzetti2017} previously provided results demonstrating a relatively low sensitivity of the maneuver placement at true anomalies between about $160^{\circ}$ and $200^{\circ}$ on the cumulative yearly station-keeping cost. For example, the Gateway's station-keeping strategy incorporates a single maneuver at $200^{\circ}$~\cite{Davis2022}, driven by the aforementioned reasons as well as operational considerations. 
We define the \textit{control action true anomaly} as the true anomaly about which the controller evaluates the need for control, and if needed, computes the required maneuver, to study its impact on the SK cost for a navigation based on OPNAV. 

\subsection{$x$-axis Crossing Control via Differential Correction}
A common solution approach for solving~\eqref{eq:xzplane_general_NLP} adopted in the literature~\cite{Davis2017, Guzzetti2017, Davis2022, Muralidharan2022} is differential correction (DC), which involves recasting~\eqref{eq:xzplane_general_NLP} as an under-determined root-finding problem with a shooting function $F: \mathbb{R}^3 \to \mathbb{R}$,
\begin{equation}
    F(\ubold) = v_x^{\mathrm{EM}} (t_f) - \bar{v}_x^{\mathrm{EM}}(\bar{t}_f)
    ,
    \label{eq:xzplane_vx_root_finding}
\end{equation}
seeking $F(\ubold) = 0$.
Employing an iterative scheme such as a generalized Newton-Raphson method, one may employ the minimum-norm update given at the $i^{\rm th}$ iteration
\begin{equation}
    \ubold^{(i+1)} = \ubold^{(i)} - \left(\dfrac{\partial F(\ubold^{(i)})}{\partial \ubold}\right)^T \left[ \left(\dfrac{\partial F(\ubold^{(i)})}{\partial \ubold}\right)^T \left(\dfrac{\partial F(\ubold^{(i)})}{\partial \ubold}\right) \right]^{-1} F(\ubold^{(i)})
    ,
    \label{eq:minimum_norm_update}
\end{equation}
where $\partial F / \partial \ubold$ is the Jacobian. Since $\ubold$ is essentially a perturbation on the initial velocity $\delta \vbold_0$, this Jacobian can be computed from the STM as
\begin{equation}
    \dfrac{\partial F}{\partial \ubold} =
    \left(
    T^{\mathrm{Inr}}_{\mathrm{EM}} 
    \begin{bmatrix}
        \Phi_{rv}(t_f, t_0) \\ \Phi_{vv}(t_f, t_0)
    \end{bmatrix} 
    \right)_{[4,:]}=
    \left(
    T^{\mathrm{Inr}}_{\mathrm{EM}} 
    \Phi_{[:,4:6]}(t_f,t_0)
    \right)_{[4,:]}
    ,
\end{equation}
where $\Phi_{[:,4:6]}$ is the last three columns of the STM and $(\cdot)_{[4,:]}$ denotes the fourth row of the matrix. 
While a DC-based approach does not directly minimize the control effort, it still yields adequate performance as the minimum-norm update~\eqref{eq:minimum_norm_update} promotes successive updates from $\ubold^{(i)}$ to $\ubold^{(i+1)}$ to be as small as possible.
Nevertheless, DC does not guarantee that the root-finding process will converge to $\ubold^*$ that satisfies~\eqref{eq:xzplane_vxtarget_NLP_constraint} with the smallest possible magnitude.
Instead, DC can result in ``over-compensating'' for the residual $F(\ubold)$ by returning a control that brings $F(\ubold) \to 0$ with its quadratic convergence nature.
The popularity of DC is attributed to its resemblance to the procedure used for trajectory design, with the only difference being the choice of the targeted state $\psi(t_f)$.
The overall DC process is shown in Algorithm~\ref{alg:xzplane_diffcorr} where the function $\operatorname{sxform}$ constructs the transformation matrix from $\mathcal{F}_{\rm I}$ to $\mathcal{F}_{\mathrm{EM}}$ at time $t_f$, and the function $\operatorname{solveIVPSTM}(t_0, t_f, \state_0 + \operatorname{vec}(\zerobold_{3 \times 1}, \bar{\ubold}))$ solves the IVP for the initial state $\state_0 + \operatorname{vec}(\zerobold_{3 \times 1}, \bar{\ubold})$ and the STM from time $t_0$ to $t_f$ by integrating the nonlinear dynamics and the matrix differential equation~\eqref{eq:stm_diffeqn}. 

\SetKwInOut{Input}{input}
\SetKwComment{Comment}{\# }{}
\begin{algorithm}
\caption{$x$-axis crossing control via differential correction}
\label{alg:xzplane_diffcorr}
\KwIn{$t_0$, $\state_0$, $t_f$, $\bar{v}_x^{\mathrm{EM}}$, $v_{x,\mathrm{tol}}$, $\mathrm{maxiter}$}
$T^{\mathrm{Inr}}_{\mathrm{EM}} \gets \operatorname{sxform}(t_f)$ \\
$\ubold \gets \zerobold_{3\times1}$ \comment*{\color{gray}Initialize cumulative control vector}
\For{$\mathrm{iter} = 1,\ldots,\mathrm{maxiter}$}{
    $\state_{f,0}, \Phi(t_f, t_0) \gets \operatorname{solveIVPSTM}(t_0, t_f, \state_0 + \operatorname{vec}(\zerobold_{3 \times 1}, \bar{\ubold}))$ \\

    \If{$\left| v_{x,0}^{\mathrm{EM}}(t_f) - \bar{v}_x^{\mathrm{EM}}\right| \leq v_{x,\mathrm{tol}}$}{
        \textbf{break}\\
    }

    $F \gets v_{x,0}^{\mathrm{EM}}(t_f) - \bar{v}_x^{\mathrm{EM}}$ \comment*{\color{gray}Compute residual of $v_x^{\mathrm{EM}}$}
    $\mathrm{D}F \gets \left( T^{\mathrm{Inr}}_{\mathrm{EM}} \Phi_{[:,4:6]}(t_f,t_0) \right)_{[4,:]}$ \comment*{\color{gray}Construct Jacobian}
    $\ubold \gets \ubold - \mathrm{D}F^T \left(\mathrm{D}F^T \mathrm{D}F \right)^{-1} F$ 
        \comment*{\color{gray}Minimum-norm update}
}
\textbf{return} $\ubold$ \\
\end{algorithm}




\subsection{$x$-axis Crossing Control as a Sequentially Linearized Minimization Problem}
Alternatively,~\eqref{eq:xzplane_general_NLP} can be solved through sequential linearization.
Specifically, we replace the IVP~\eqref{eq:ivp_prediction_u0} with linearization
\begin{equation}
    \state^{\mathrm{EM}} (t_f)
    =
    T^{\mathrm{Inr}}_{\mathrm{EM}} 
    \left(
        \int_{t_0}^{t_f} \fbold(\tau, \state_0) \mathrm{d}\tau
        + \begin{bmatrix}
            \Phi_{rv}(t_f, t_0) \\ \Phi_{vv}(t_f, t_0)
        \end{bmatrix} \ubold
    \right)
    =
    T^{\mathrm{Inr}}_{\mathrm{EM}} \state_{f,0} + B \ubold
    ,
    \label{eq:ivp_prediction_u0_linearized}
\end{equation}
where $\state_{f,0}$ is the final state at $t_f$ if no control is executed at $t_0$, and
\begin{equation}
    B = T^{\mathrm{Inr}}_{\mathrm{EM}} \begin{bmatrix}
        \Phi_{rv}(t_f, t_0) \\ \Phi_{vv}(t_f, t_0)
    \end{bmatrix} 
    = 
    T^{\mathrm{Inr}}_{\mathrm{EM}} \Phi_{[:,4:6]}(t_f,t_0)
    .
\end{equation}
Then, denoting $v_{x,0}^{\mathrm{EM}}(t_f) = \left( T^{\mathrm{Inr}}_{\mathrm{EM}} \state_{f,0} \right)_{[4]}$, the linearized problem is given by
\begin{subequations}
    \label{eq:xzplane_vxtarget_u0_seqlinear}
    \begin{align}
        \min_{\ubold} \quad & \| \ubold \|_2
        \\
        \text{such that} \quad
        &  \left| 
            v_{x,0}^{\mathrm{EM}}(t_f)
            + B_{[4,:]} \ubold
            - \bar{v}_x^{\mathrm{EM}} \right| \leq v_{x,\mathrm{tol}}^{\prime},
            \label{eq:xzplane_vxtarget_u0_seqlinear_constraint}
    \end{align}
\end{subequations}
where $(\cdot)_{[4]}$ denotes the fourth component of the vector.
The constraint~\eqref{eq:xzplane_vxtarget_u0_seqlinear_constraint} is bounded by $v_{x,\mathrm{tol}}^{\prime}$, which is given in terms of the original targeting tolerance $v_{x,\mathrm{tol}}$ by $v_{x,\mathrm{tol}}^{\prime} = s v_{x,\mathrm{tol}}$ where $0 < s < 1$ is a safety factor set to $s = 0.9$ in this work. 
Setting $v_{x,\mathrm{tol}}^{\prime} < v_{x,\mathrm{tol}}$ ensures the actual nonlinear constraint~\eqref{eq:xzplane_vxtarget_u0_seqlinear_constraint} is achieved by sequentially updating $\ubold$ with the solution of the linearized problem~\eqref{eq:xzplane_vxtarget_u0_seqlinear}. 

Solving the \textit{minimization} problem~\eqref{eq:xzplane_vxtarget_u0_seqlinear} is different from solving the root-finding problem~\eqref{eq:xzplane_vx_root_finding} in two regards.
First, the dynamics have been linearized using the STM in~\eqref{eq:ivp_prediction_u0_linearized}. Thus, the obtained solution $\ubold$ may not satisfy the constraint when propagated with the nonlinear IVP~\eqref{eq:ivp_prediction_u0}. 
In SK, the control that is sought is typically small in magnitude, and the linear approximation holds relatively well. Nevertheless, a few iterations (2 $\sim$ 3 in practice) solving~\eqref{eq:xzplane_vxtarget_u0_seqlinear} by updating $\state_{f,0}$ and $B$ may be necessary to incrementally obtain an aggregate control $\bar{\ubold}$ and satisfy the nonlinear constraint~\eqref{eq:xzplane_vxtarget_NLP_constraint}.
Specifically, after the $i^{\mathrm{th}}$ iteration solving the linearized problem, $\state_{f,0}$ is replaced by $\state_{f,0} \leftarrow \state_{f,0} + \ubold^{(i)}$ where $\ubold^{(i)}$ is the solution from the $i^{\mathrm{th}}$ iteration, and $B$ is reconstructed using the updated initial state for the $i+1^{\mathrm{th}}$ iteration. 
Second, because this is a minimization of $\|\ubold\|_2$, the resulting control $\ubold$ is the minimum feasible $\ubold$ that simultaneously ensures the final $v_x^{\mathrm{EM}}$ to be within $v_{x,\mathrm{tol}}^{\prime}$ of the targeted value $\bar{v}_x^{\mathrm{EM}}$ according to the linear approximated dynamics~\eqref{eq:ivp_prediction_u0_linearized}. 

Problem~\eqref{eq:xzplane_vxtarget_u0_seqlinear} is a SOCP, where the Euclidean norm in the objective may be formulated as a second-order cone constraint.
Its optimal solution may be obtained analytically by observing that it is equivalent to determining the projection of the origin to the intersection of two half-spaces~\cite{Elango2022SequentialLinear}. 
The control $\ubold$ satisfying~\eqref{eq:xzplane_vxtarget_u0_seqlinear_constraint} is a vector $\zbold$ that belongs to the intersection of two half-spaces given by
\begin{equation}
    \begin{aligned}
        C &= \left\{ 
            \zbold \, \left| \,
                B_{[4,:]} \zbold
                \leq v_{x,\mathrm{tol}}^{\prime} - v_{x,0}^{\mathrm{EM}}(t_f) + \bar{v}_x^{\mathrm{EM}}
            \right.
        \right\}
        \,\,\cap\,\,
        \left\{ 
            \zbold \, \left| \,
                -B_{[4,:]} \zbold
                \leq v_{x,\mathrm{tol}}^{\prime} + v_{x,0}^{\mathrm{EM}}(t_f) - \bar{v}_x^{\mathrm{EM}}
            \right.
        \right\}
        ,\\
        &= \left\{ 
            \zbold \, \left| \,
                \xibold_1^T \zbold \leq \eta_1
            \right.
        \right\}
        \,\,\cap\,\,
        \left\{ 
            \zbold \, \left| \,
                \xibold_2^T \zbold \leq \eta_2
            \right.
        \right\}
        ,
    \end{aligned}
\end{equation}
where
\begin{subequations}
\begin{align}
    \xibold_1^T &= B_{[4,:]}
    \label{eq:xzplane_seqlin_xi1_definition}
    ,\\
    \xibold_2^T &= - B_{[4,:]}
    \label{eq:xzplane_seqlin_xi2_definition}
    ,\\
    \eta_1 &= v_{x,\mathrm{tol}}^{\prime} - v_{x,0}^{\mathrm{EM}}(t_f) + \bar{v}_x^{\mathrm{EM}}
    \label{eq:xzplane_seqlin_eta1_definition}
    ,\\
    \eta_2 &= v_{x,\mathrm{tol}}^{\prime} + v_{x,0}^{\mathrm{EM}}(t_f) - \bar{v}_x^{\mathrm{EM}}
    .
    \label{eq:xzplane_seqlin_eta2_definition}
\end{align}
\end{subequations}
The expression for the projection of $\zbold$ onto $C$ is given by Proposition 28.19 in~\cite{BauschkeCombettes2010}
\begin{equation}
    P_C \zbold = \zbold - \nu_1 \xibold_1 - \nu_2 \xibold_2
    ,
    \label{eq:seqlinear_projection_Pcc}
\end{equation}
where exactly one of the following holds:
\begin{equation}
    \begin{bmatrix}
        \nu_1 \\ \nu_2
    \end{bmatrix}
    =
    \begin{cases}
        \begin{bmatrix}
            0 \\ 0
        \end{bmatrix}
        & 
        \xibold_1^T \zbold \leq \eta_1 \text{ and } \xibold_2^T \zbold \leq \eta_2
        ,\\[2.5em]
        \begin{bmatrix}
            \dfrac{\|\xibold_2\|^2(\zbold^T\xibold_1 - \eta_1) - \xibold_1^T\xibold_2(\zbold^T\xibold_2 - \eta_2)}{\|\xibold_1^2\|\|\xibold_2\|^2 - |\xibold_1^T \xibold_2|^2}
            \\[0.8em]
            \dfrac{\|\xibold_1\|^2 (\zbold^T \xibold_2 -\eta_2) - \xibold_1^T \xibold_2 (\zbold^T \xibold_1 - \eta_1)}{\|\xibold_1^2\|\|\xibold_2\|^2 - |\xibold_1^T \xibold_2|^2} 
        \end{bmatrix}
        & 
        \begin{array}{cc}
             \|\xibold_2\|^2 (\zbold^T \xibold_1 - \eta_1) >\xibold_1^T \xibold_2 (\zbold^T \xibold_2 - \eta_2))
        \text{ and } \\
             \|\xibold_1\|_2 (\zbold^T \xibold_2 - \eta_2) > \xibold_1^T\xibold_2 (\zbold^T \xibold_1 - \eta_1)
        ,\end{array}
        \\[2.5em]
        \begin{bmatrix}
            0 \\ \dfrac{\zbold^T \xibold_2 - \eta_2}{\|\xibold_2\|^2}
        \end{bmatrix}
        & 
        \begin{array}{cc}
             \zbold^T \xibold_2 > \eta_2 \text{ and } \\
             \|\xibold_2\|^2 (\zbold^T \xibold_1 - \eta_1) \leq \xibold_1^T \xibold_2 (\zbold^T\xibold_2 - \eta_2)
        ,\end{array}
        \\[2.5em]
        \begin{bmatrix}
            \dfrac{\zbold^T \xibold_1 - \eta_1}{\|\xibold_1\|^2} \\ 0
        \end{bmatrix}
        & 
        \begin{array}{cc}
             \zbold^T \xibold_1 > \eta_1 \\
             \text{ and } \|\xibold_1\|^2 (\zbold^T\xibold_2 - \eta_2) \leq \xibold_1^T\xibold_2 (\zbold^T\xibold_1 - \eta_1)
        .\end{array}
    \end{cases}
    \label{eq:xzplane_seqlin_nus_cases}
\end{equation}
Since we seek to minimize $\|\ubold\|_2$, choosing $\zbold = \zerobold_{3 \times 1}$ in equation~\eqref{eq:seqlinear_projection_Pcc} yields the optimal solution $\ubold^* = P_C \zerobold_{3 \times 1}$ to problem~\eqref{eq:xzplane_vxtarget_u0_seqlinear}. 

In summary, the sequential linearization approach solves the NLP~\eqref{eq:xzplane_general_NLP} by sequentially building the linearized problem~\eqref{eq:xzplane_vxtarget_u0_seqlinear} and computing its solution via~\eqref{eq:seqlinear_projection_Pcc}. 
The overall procedure is termed the ``sequentially linearized minimization problem'' (SLMP), and is summarized in Algorithm~\ref{alg:xzplane_seqlinear}, where the function $\operatorname{sxform}$ forms the transformation matrix at time $t_f$, and the function $\operatorname{originProjectionCoefficients}$ computes the coefficients $\nu_1$ and $\nu_2$ with $\zbold = \zerobold_{3\times1}$ using equation~\eqref{eq:xzplane_seqlin_nus_cases}. 
The algorithm returns the aggregated control vector $\bar{\ubold}$, which is the summation of the solution $\ubold$ of each sequential instance of the linearized problem~\eqref{eq:xzplane_vxtarget_u0_seqlinear}. 

\SetKwInOut{Input}{input}
\SetKwComment{Comment}{\# }{}
\begin{algorithm}   \DontPrintSemicolon
\caption{$x$-axis crossing control via sequentially linearized minimization problem}
\label{alg:xzplane_seqlinear}
\KwIn{$t_0$, $\state_0$, $t_f$, $\bar{v}_x^{\mathrm{EM}}$, $v_{x,\mathrm{tol}}$, $s$, $\mathrm{maxiter}$}
$v_{x,\mathrm{tol}}^{\prime} \gets sv_{x,\mathrm{tol}}$ \\
$T^{\mathrm{Inr}}_{\mathrm{EM}} \gets \operatorname{sxform}(t_f)$ \\
$\bar{\ubold} \gets \zerobold_{3\times1}$  \comment*{\color{gray} Initialize cumulative control vector} 
\For{$\mathrm{iter} = 1,\ldots,\mathrm{maxiter}$}{
    $\state_{f,0}, \Phi(t_f, t_0) \gets \operatorname{solveIVPSTM}(t_0, t_f, \state_0 + \operatorname{vec}(\zerobold_{3 \times 1}, \bar{\ubold}))$ \\

    \If{$\left| v_{x,0}^{\mathrm{EM}}(t_f) - \bar{v}_x^{\mathrm{EM}}\right| \leq v_{x,\mathrm{tol}}$}{
        \textbf{break}\\
    }
    
    $B \gets T^{\mathrm{Inr}}_{\mathrm{EM}} \Phi_{[:,4:6]}(t_f, t_0)$ \\
    $\eta_1, \xibold_1, \eta_2, \xibold_2 \gets $ equations~\eqref{eq:xzplane_seqlin_xi1_definition} $\sim$~\eqref{eq:xzplane_seqlin_eta2_definition} \\
    $\nu_1, \nu_2 \gets \operatorname{originProjectionCoefficients}(\eta_1, \xibold_1, \eta_2, \xibold_2)$ 
        \comment*{\color{gray}Via equation~\eqref{eq:xzplane_seqlin_nus_cases}}
    $\ubold \gets -\nu_1 \xibold_1 - \nu_2 \xibold_2$ 
        \comment*{\color{gray}Compute linearized control update}
    $\bar{\ubold} \gets \bar{\ubold} + \ubold$ 
        \comment*{\color{gray}Update cumulative control vector}
}
\textbf{return} $\ubold$ \\
\end{algorithm}


        

\subsection{Mean State Targeting via Unscented Transform}
Traditional $x$-axis crossing control introduced thus far involves targeting the state $\psibold(t_f)$, obtained by simply propagating the current state estimate. 
The resulting predicted state $\psibold(t_f)$ may be understood as the prediction output from the EKF, where the best estimate of a future state is obtained by propagating the current state estimate forward in time. 
While filter performance for navigation have resulted in no apparent difference between EKF and UKF, the state prediction involved in control is over a significantly longer time, across multiple revolutions along the periodic orbit. Thus, a UT-based prediction can better capture the nonlinearity involved.
Hence, we consider a control scheme which aims at steering the mean future state $\mathbb{E}\left[ \psibold^{\mathrm{B}} (t_f) \right]$ computed via the unscented transform. 
The resulting NLP is given by
\begin{subequations}
    \label{eq:xzplane_mean_targeting_NLP}
    \begin{align}
        \min_{\ubold} \quad & \| \ubold \|_2,
        \\
        \text{such that} \quad
        & \left| \mathbb{E}\left[ \psibold^{\mathrm{B}} (t_f) \right] - \bar{\psibold}^{\mathrm{B}}(\bar{t}_f) \right| \leq v_{x,\mathrm{tol}}
        .
        \label{eq:xzplane_mean_targeting_NLP_constraint}
    \end{align}
\end{subequations}
The mean targeted state $\mathbb{E}\left[ \psibold^{\mathrm{B}} (t_f) \right]$ under no control maneuver can be approximated using the unscented transform, as in the prediction step of the UKF~\eqref{eq:meanpred_ukf}.
The mean predicted state at time $t_f$ is approximated by
\begin{align}
    \mathbb{E}[ \psibold^{\mathrm{EM}}_{0}(t_f) ]
    &\approx 
    \left(
        \sum_{\ell=0}^{2n} W_m^{(\ell)} \mathcal{Y}^{(\ell)}
    \right)_{m}
    ,
    \label{eq:UnscentedTransform_mean}
\end{align}
where $\mathcal{Y}^{(\ell)}$ is \review{the propagation of the $\ell^{\rm th}$ sigma point $\mathcal{X}_{j-1|j-1}^{(\ell)}$ transformed to $\mathcal{F}_{\rm EM}$,}
\begin{equation}
    \mathcal{Y}^{(\ell)} =
         T^{\mathrm{Inr}}_{\mathrm{EM}} \mathcal{X}_{j \mid j-1}^{(\ell)}
         =
        T^{\mathrm{Inr}}_{\mathrm{EM}} \left(
        \mathcal{X}_{j-1 \mid j-1}^{(\ell)} + \int_{t_{j-1}}^{t_j} \fbold\left(\tau, \mathcal{X}_{j-1 \mid j-1}^{(\ell)}(\tau)\right) \mathrm{d} \tau
    \right),
    \quad \ell = 0, \ldots, n
    ,
    \label{eq:UnscentedTransform_propagation_controller}
\end{equation}
with sigma points computed by~\eqref{eq:ukf_sigmapoints}. 
By replacing $v_x^{\mathrm{EM}} (t_f)$ in~\eqref{eq:xzplane_vx_root_finding} or $v_{x,0}^{\mathrm{EM}} (t_f)$ in \eqref{eq:xzplane_vxtarget_u0_seqlinear_constraint} with $\mathbb{E}[ \psibold^{\mathrm{EM}}_{0}(t_f)]$, both the DC and SLMP approaches are modified to target the mean state, and are referred to as UT-DC and UT-SLMP, respectively. 

The intuition behind the expected performance improvement by solving problem~\eqref{eq:xzplane_mean_targeting_NLP} instead of~\eqref{eq:xzplane_general_NLP} is due to the improvement of the prediction accuracy that we obtain by capitalizing on the filter's covariance estimate.
The initial state estimate error, propagated over targeting time $t_f$ spanning multiple weeks, is more effectively countered by also propagating the sigma points from the initial covariance.

\subsection{Trigger Condition and Targeting Tolerance Tuning}
\begin{figure}
    \centering
    \includegraphics[width=0.75\linewidth]{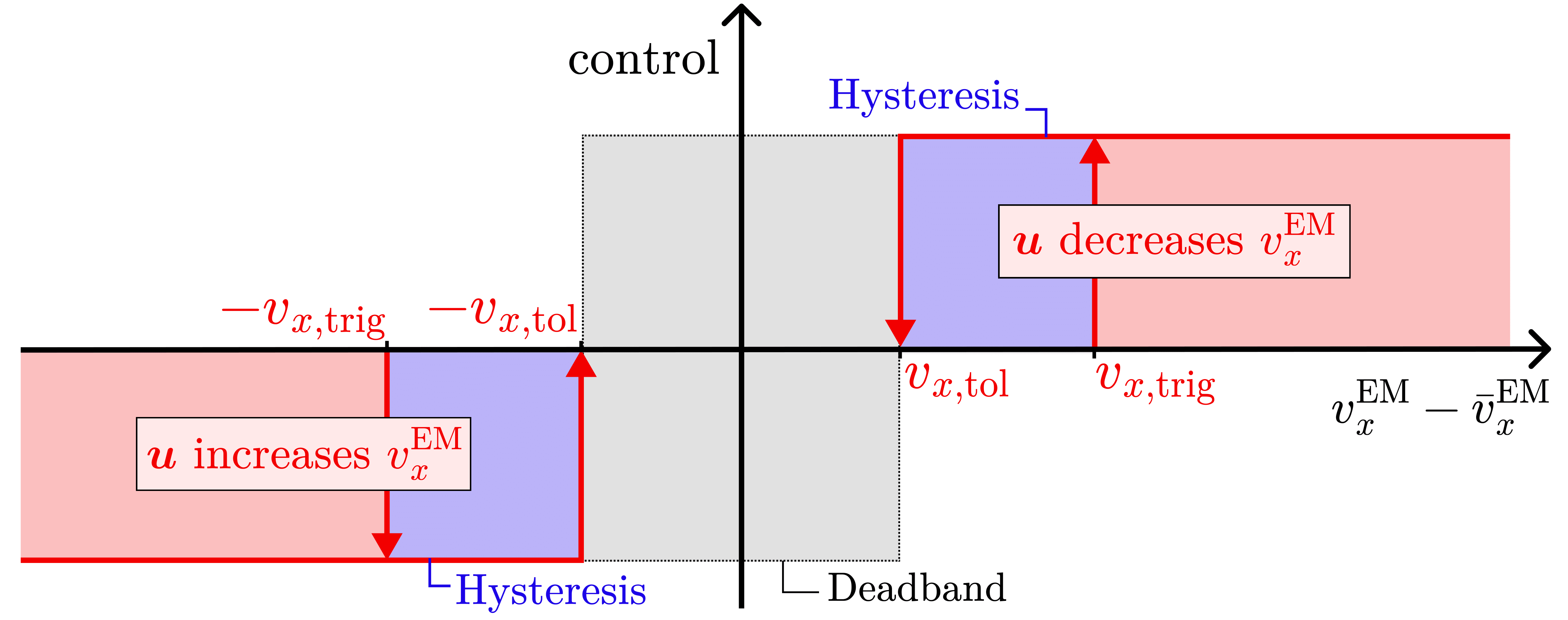}
    \caption{Hysteresis for $x$-axis crossing control}
    \label{fig:hysteresis}
\end{figure}

At each control action true anomaly, a trigger condition is used to determine whether a SK maneuver should be executed.
Such a condition necessitates an indicator that quantifies the difference between the state estimate and the baseline.
For the $x$-axis crossing control, a typical indicator is the difference in magnitude of $v_x^{\mathrm{EM}}$ at the $N^{\mathrm{th}}$ perilune with respect to the baseline. 
We define the threshold $v_{x,\mathrm{trig}}$ such that a control is triggered if $|v_x^{\mathrm{EM}} - \bar{v}_x^{\mathrm{EM}}| \geq v_{x,\mathrm{trig}}$.
\review{Setting $v_{x,\mathrm{trig}} > v_{x,\mathrm{tol}}$ builds in \textit{hysteresis} aimed at avoiding chattering across revolutions, where
\begin{enumerate}[label=(\roman*)]
    \item a control action is not engaged even if $|v_x^{\mathrm{EM}} - \bar{v}_x^{\mathrm{EM}}| > v_{x,\rm tol}$, until a larger tolerance $v_{x,\rm trig}$ is exceeded, and
    \item once engaged, the control action achieves $|v_x^{\mathrm{EM}} - \bar{v}_x^{\mathrm{EM}}| \leq v_{x,\rm tol}$, which is well below $v_{x,\rm trig}$.
\end{enumerate}
Figure~\ref{fig:hysteresis} illustrates hysteresis in the context of $x$-axis crossing control, where the horizontal axis corresponds to deviation in $v_x^{\rm EM}$, and the vertical axis indicates whether a control action is taken to decrease or increase $v_x^{\rm EM}$.
A control action is taken within the shaded regions, and hysteresis corresponds to the blue shaded regions.
With an appropriate tuning, the controller is effectively desensitized to the noise in the system, which is expected to lead to a reduction in cumulative $\Delta V$.}

When employing DC, the actual targeting violation $|v_x^{\mathrm{EM}} - \bar{v}_x^{\mathrm{EM}}|$ achieved is somewhat uncorrelated to $v_{x,\mathrm{tol}}$ due to the quadratic convergence property of the Newton-Raphson scheme; as such, the DC algorithm most often converges $|v_x^{\mathrm{EM}} - \bar{v}_x^{\mathrm{EM}}| \ll v_{x,\mathrm{tol}}$.
\review{Thus, even with $v_{x,\mathrm{trig}} = v_{x,\mathrm{tol}}$, the control action will drive $|v_x^{\mathrm{EM}} - \bar{v}_x^{\mathrm{EM}}|$ below $v_{x,\rm trig}$, which in turns results in a similar effect as (ii) due to hysteresis.}

In contrast, the SLMP constrained minimization problem consistently aims at achieving $|v_x^{\mathrm{EM}} - \bar{v}_x^{\mathrm{EM}}| = v_{x,\mathrm{tol}}$ to minimize the control action. 
As such, $v_{x,\mathrm{tol}}$ plays a concrete role in determining how well the controller tracks the baseline. \review{For a given maneuver instance, a tighter $v_{x,\mathrm{tol}}$ results in expending more $\Delta V$ to align the path closely to the baseline, while a relaxed $v_{x,\mathrm{tol}}$ results in the opposite behavior.
Meanwhile, the relationship between $v_{x,\mathrm{tol}}$ and the \textit{cumulative} $\Delta V$ over multiple SK maneuvers is non-trivial. A tighter $v_{x,\mathrm{tol}}$ pushes the controlled trajectory closer to the reference, potentially leading to long-term, cumulative $\Delta V$ cost reduction over successive maneuver instances.}

\section{Filter Experiments}
\label{sec:filter_experiments}
The first numerical experiment consists of running the synthetic image generation, processing it to generate position measurements, which are then fed to the recursive navigation filter. 
Results in this section do not include the station-keeping controller and focus on the performance of the filter alone. Without a controller, the trajectory naturally diverges after several revolutions due to the accumulation of numerical integration errors. 
As such, a simulation experiment without a controller can only last a limited number of revolutions.
Meanwhile, the filter's performance can be assessed within a few revolutions as the illumination conditions on an NRHO are almost periodic, and measurement qualities from one revolution to the next are nearly identical. 
Thus, in this section, for each considered filter configuration, 30 Monte Carlo samples each lasting 10 revolutions are taken. 
The initial baseline epoch and state are given in Table~\ref{tab:baseline_initial_state}. 
Across each Monte Carlo run, the initial state estimate is computed by adding a random initial injection error of $3\text{-}\sigma_{r} = 10$ \SI{}{km} and $3\text{-}\sigma_{v} = 10$ \SI{}{cm/s} to the baseline's initial state. 
At each measurement instance, the camera is pointed toward the Moon based on the current state estimate; due to the estimation error, the boresight never exactly aligns with the center of the Moon. Simulation parameters are summarized in Table~\ref{tab:exp_filter_errors}.

\rreview{We consider four measurement collection policies, summarized in Table~\ref{tab:filter_meas_config}. These policies span a range of FOV values within the tradeoff discussed in Section~\ref{sec:choice_of_fov}. A larger FOV allows measurements to be collected farther from apolune, while a narrower FOV improves the reliability of images acquired near apolune.
The measurement update frequency was selected based on preliminary experiments showing satisfactory navigation performance with three measurements per revolution. This choice allows us to evaluate the GN\&C pipeline under a scenario using minimal horizon-based OPNAV measurements.
More frequent imaging and measurement updates, if operationally feasible, would further improve navigation accuracy.
The measurement locations are selected as follows. The first and last measurements occur at the earliest and latest positions between successive perilunes where the Moon fits within the selected FOV, including margins. An intermediate measurement is then collected at a position that is $10^{\circ}$ in osculating true anomaly after the first measurement. This intermediate measurement is intended to improve navigation accuracy before the dead-reckoning phase that precedes the station-keeping maneuver.}
In all cases, we consider a square sensor of size $100 \times 100$ \SI{}{mm} and an image size $1024 \times 1024$ pixels for consistency.
While one may choose to further optimize both the choice of the FOVs and the locations where the measurements are collected, we focus on providing insights into the general sensitivity of the underlying trade-off.

\begin{table}[]
    \centering
    \caption{Baseline initial epoch and state in Moon-centered J2000}
    \begin{tabular}{@{}ll@{}}
    \toprule
    Baseline parameter & Value \\ \midrule
       Epoch,  seconds past J2000 & 946728069.183919 \\
       $x$, km & -100.3227942169551 \\
       $y$, km & 17287.240158966662 \\
       $z$, km & -68230.31701814539 \\
       $v_x$, km/s & -0.05947862362245673 \\
       $v_y$, km/s & 0.03798023721969298 \\
       $v_z$, km/s & 0.005508556661896624 \\
    \bottomrule
    \end{tabular}
    \label{tab:baseline_initial_state}
\end{table}

\begin{table}[]
    \centering
    \caption{Parameters and errors considered in filter experiments}
    \begin{tabular}{@{}ll@{}}
    \toprule
    Parameter and error & Value \\ \midrule
    Canonical distance $\rm DU$, km   & 3000.0                                \\ 
    Canonical time $\rm TU$, s      & 2346.711856601253                     \\
    Process noise diffusion coefficient $\sigma_u$ & 1.5       \\
    Initial 3-$\sigma$ position error, \SI{}{km}   & 10        \\
    Initial 3-$\sigma$ velocity error, \SI{}{cm/s} & 10        \\
    Attitude uncertainty $\sigma_{\phi}$, $\mathrm{arcsec}$  & 15       \\
    Pixel standard deviation $\sigma_{\mathrm{pix}}$ & 0.5    \\ 
    SRP $A/m$ relative uncertainty $\sigma_{\bar{A/m}}$ & 0.1           \\
    SRP $C_r$ relative uncertainty $\sigma_{\bar{C_r}}$ & 0.05          \\
    \bottomrule
    \end{tabular}
    \label{tab:exp_filter_errors}
\end{table}

\begin{table}[]
\centering
\caption{Camera specifications and measurement collection configurations}
\begin{tabular}{@{}llllll@{}}
\toprule
Focal length $f$, \SI{}{mm} &
Field of view, \SI{}{deg} &
Sensor size, \SI{}{mm} &
\begin{tabular}[c]{@{}l@{}}Image size,\\ \SI{}{pixels}\end{tabular} &
\begin{tabular}[c]{@{}l@{}}Number of \\ measurements\end{tabular} &
\begin{tabular}[c]{@{}l@{}}Measurement\\ true anomaly, \SI{}{deg} \end{tabular} \\ \midrule
300          & $18.92^{\circ}$ &
\multirow{4}{*}{$100 \times 100$} & 
\multirow{4}{*}{$1024 \times 1024$} & 
\multirow{4}{*}{3} & $140^{\circ}$, $150^{\circ}$, $220^{\circ}$ \\
360          & $15.81^{\circ}$ & & & & $145^{\circ}$, $155^{\circ}$, $215^{\circ}$ \\
450          & $12.68^{\circ}$ & & & & $150^{\circ}$, $160^{\circ}$, $210^{\circ}$ \\
550          & $10.39^{\circ}$ & & & & $155^{\circ}$, $165^{\circ}$, $205^{\circ}$ \\
\bottomrule
\end{tabular}
\label{tab:filter_meas_config}
\end{table}

\subsection{Trade-off on Measurement Collection}
\begin{figure}
     \centering
     \begin{subfigure}[b]{0.497\textwidth}
         \centering
         \includegraphics[width=\textwidth]{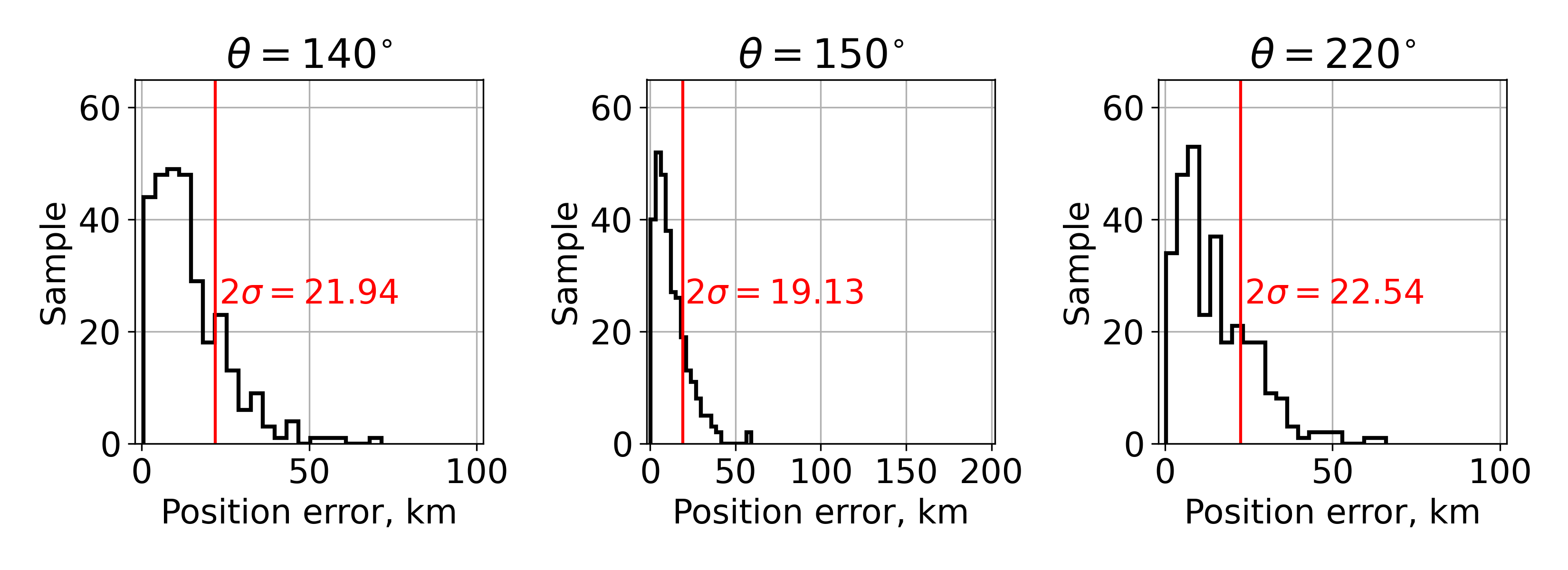}
         \caption{$f = 300$ \SI{}{mm}, $\mathrm{FOV} \approx 18.92^{\circ}$}
         \label{fig:meas_error_hist_ekf_nmeas3_f300_px1024}
     \end{subfigure}
     \hfill
     \begin{subfigure}[b]{0.497\textwidth}
         \centering
         \includegraphics[width=\textwidth]{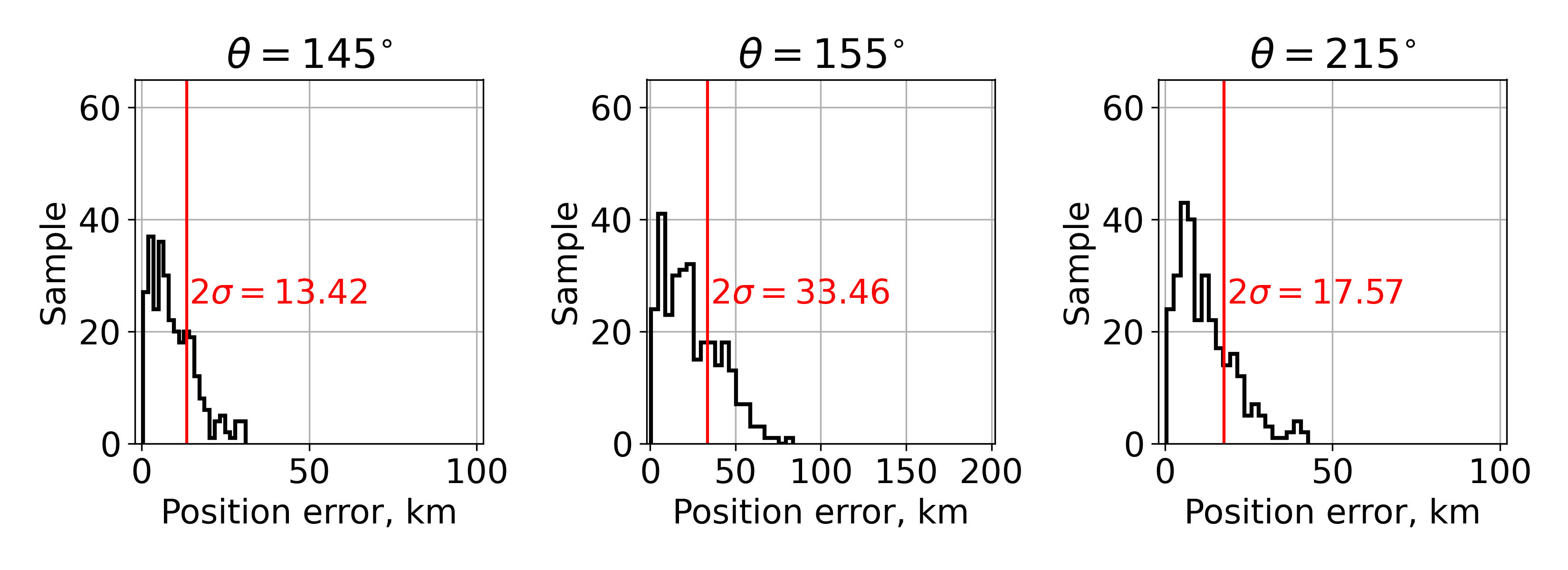}
         \caption{$f = 360$ \SI{}{mm}, $\mathrm{FOV} \approx 15.81^{\circ}$}
         \label{fig:meas_error_hist_ekf_nmeas3_f360_px1024}
     \end{subfigure}
     \\
     \begin{subfigure}[b]{0.497\textwidth}
         \centering
         \includegraphics[width=\textwidth]{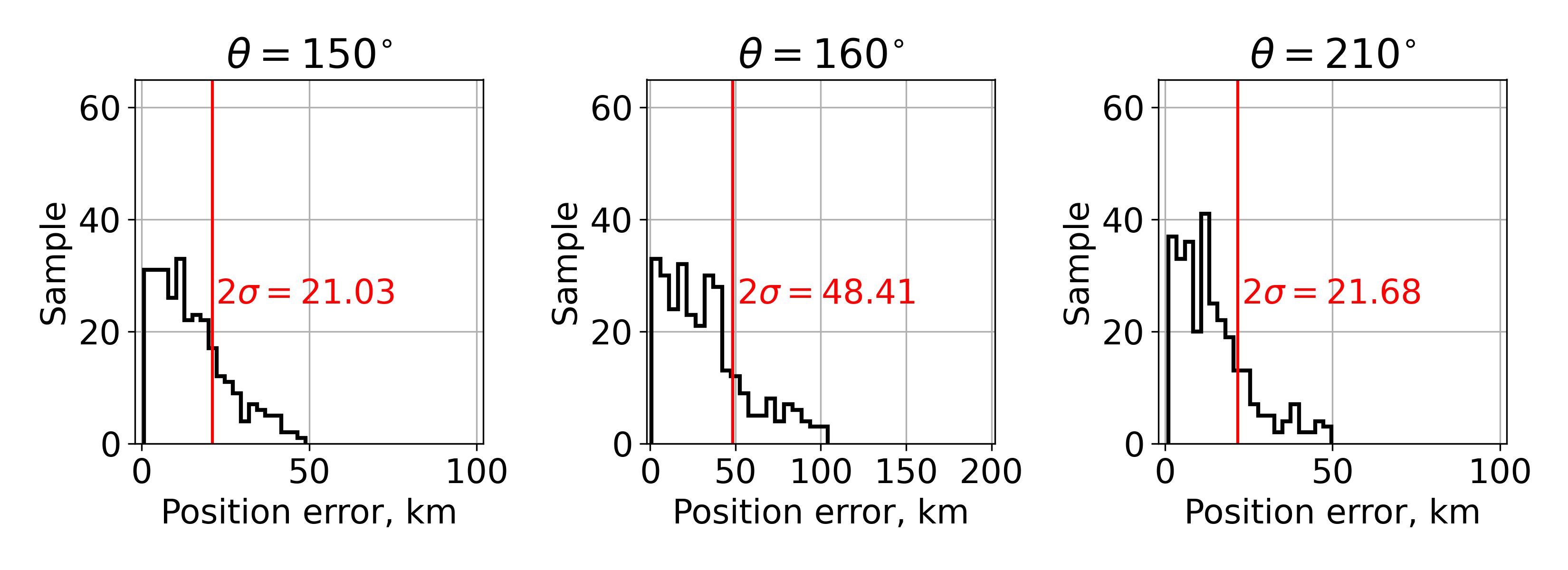}
         \caption{$f = 450$ \SI{}{mm}, $\mathrm{FOV} \approx 12.68^{\circ}$}
         \label{fig:meas_error_hist_ekf_nmeas3_f450_px1024}
     \end{subfigure}
     \hfill
     \begin{subfigure}[b]{0.497\textwidth}
         \centering
         \includegraphics[width=\textwidth]{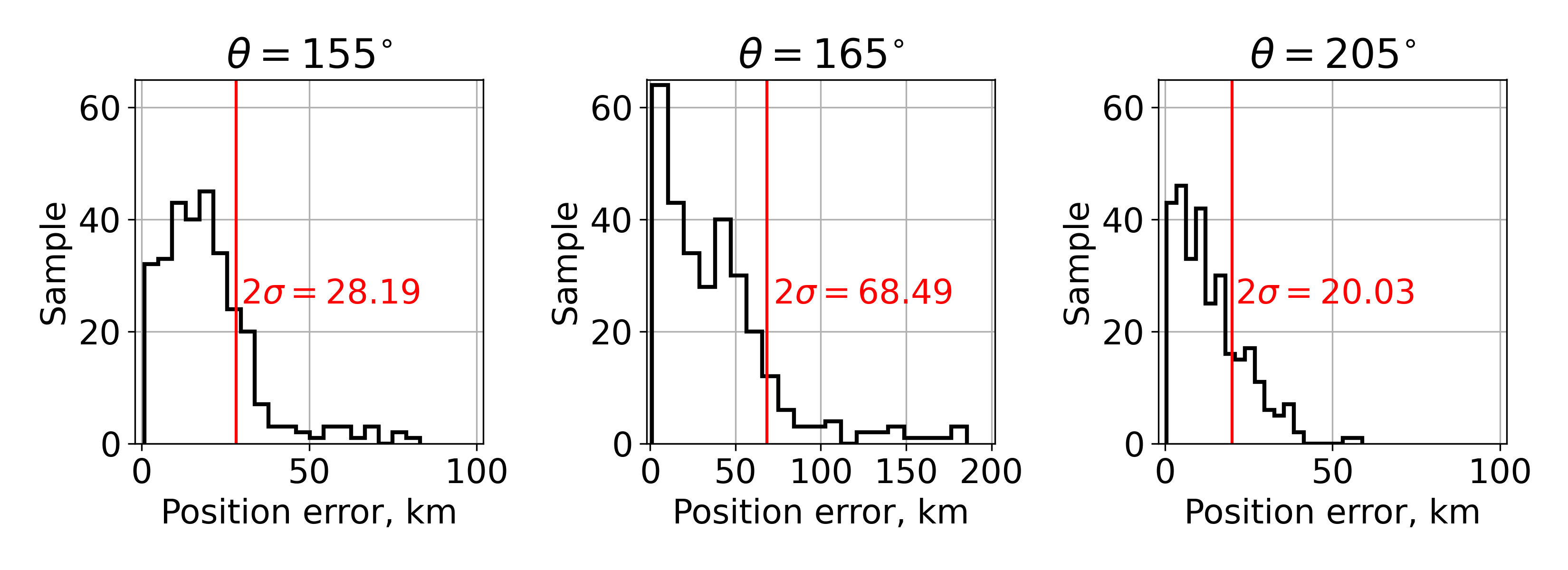}
         \caption{$f = 550$ \SI{}{mm}, $\mathrm{FOV} \approx 10.39^{\circ}$}
         \label{fig:meas_error_hist_ekf_nmeas3_f550_px1024}
     \end{subfigure}
     \caption{Measurement statistics for various camera FOV choices
     }
     \label{fig:measurement_statistics}
\end{figure}

Figure~\ref{fig:measurement_statistics} shows the distribution of measurement errors for each FOV choice, taken at the three respective osculating true anomaly positions along the NRHO. 
Each histogram contains 300 samples in total, coming from measurements collected over 10 revolutions in each of the 30 Monte Carlo runs. 
Due to the choice of the measurement collection strategy that is symmetric in true anomaly about the apolune, the first and third measurements share very similar quality, with similar distribution of measurement errors.

\rrreview{
In cases (b) through (d), there is a degradation of the empirical covariance of the second measurement; in contrast, case (a) results in an empirical covariance that is slightly better than the first and last measurements.
Multiple mechanisms are at play behind the measurement accuracies, including the degradation of observability along the boresight as the spacecraft approaches apolune ($\theta = 180^{\circ}$), the decrease in detected limb points as the Moon appears smaller within the image, as well as variation in illumination conditions and surface features along the limb imaged from different locations along the NRHO, as studied in detail in~\cite{Machuca2026-rc}.
All considered measurement configurations provide measurements of quality that are sufficient for the EKF to retain the state estimate in custody.
Further investigation would be required to elucidate an optimal measurement collection strategy, including both the measurement collection timings as well as the FOV for navigation along a given baseline, which is beyond the scope of this manuscript, aimed at establishing the feasibility of OPNAV-only navigation and station-keeping on the NRHO.
}

\begin{figure}
    \centering
    \includegraphics[width=0.49\linewidth]{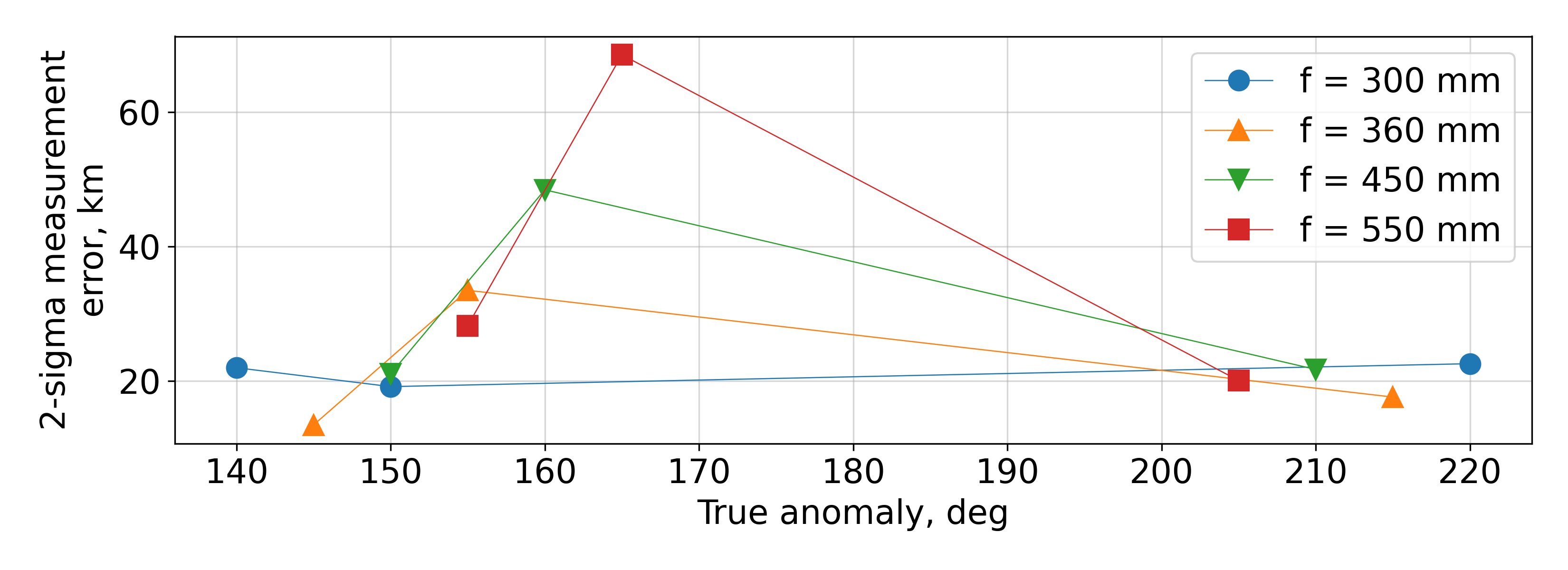}
    \caption{Empirical measurement covariance across various camera FOV choices}
    \label{fig:meas_stats_summary}
\end{figure}

\subsection{EKF Performance}
\begin{figure}
    \centering
    \includegraphics[width=0.99\linewidth]{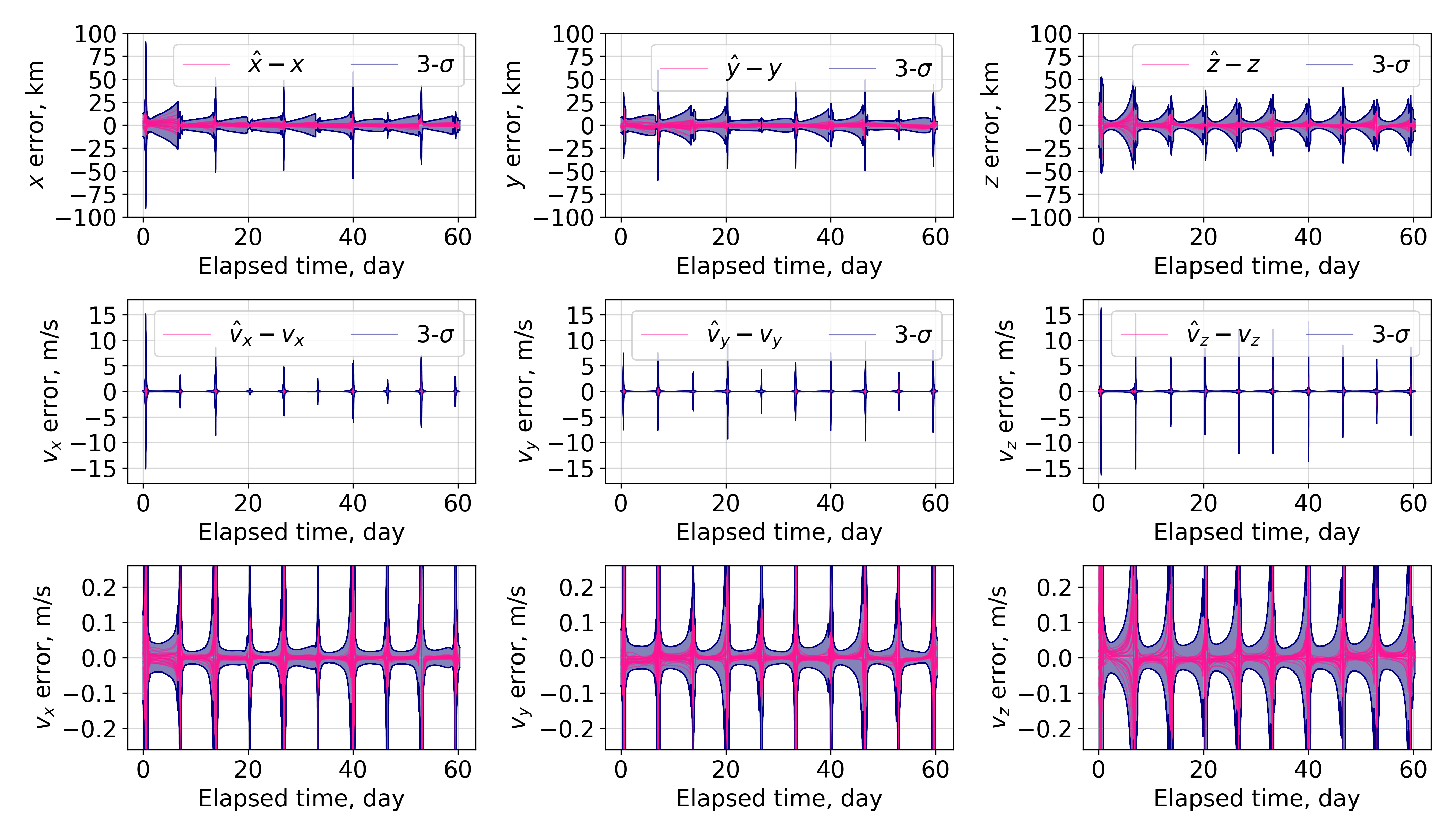}
    \caption{Monte Carlo runs of EKF with 3 measurements per revolution using $f = 360$ \SI{}{mm}}
    \label{fig:ekf_nmeas3_f360_px1024}
\end{figure}

With the measurement configurations in Table~\ref{tab:filter_meas_config}, the EKF is able to track the state estimate without diverging/wrongfully converging. 
As an illustrative example, the EKF performance using a $f = 360$ \SI{}{mm} camera is shown in Figure~\ref{fig:ekf_nmeas3_f360_px1024}.
\rrreview{In the Figure, the first row shows position estimate errors, while the second and third row show velocity estimate errors; the second row contains up to the maximum velocity estimate errors and $3$-$\sigma$ bounds, while the third row provides a zoomed-in view of the estimate error structure outside of the spikes corresponding to perilune passes.}
While each run is randomized in terms of initial state error, error on the SRP coefficients of the true dynamics, and attitude error, the measurement itself is taken at locations within tens of kilometers difference from one Monte Carlo run to another; thus, for any $i^{\mathrm{th}}$ measurement during a Monte Carlo run, the analytical measurement covariance from equation~\eqref{eq:measnoisecov} is nearly identical to that of the $i^{\mathrm{th}}$ measurement in any other Monte Carlo run.
We also note the spike in velocity errors, along with the corresponding covariances, at perilune passes, \rreview{consistent with other navigation results along the NRHO~\cite{Davis2017,Volle2018-ea,Parrish2020-sj}; the spike observed in this work is comparatively more pronounced as we assume no measurement is acquired during the perilune passes.}
The large error is due to the rapid rotation of the spacecraft's velocity vector around perilune, leading to along-track error growth. Following perilune, the error quickly reduces due to the dynamics as well as the incoming measurement.

\begin{figure}
     \centering
     \begin{subfigure}[b]{0.497\textwidth}
         \centering
         \includegraphics[width=\textwidth]{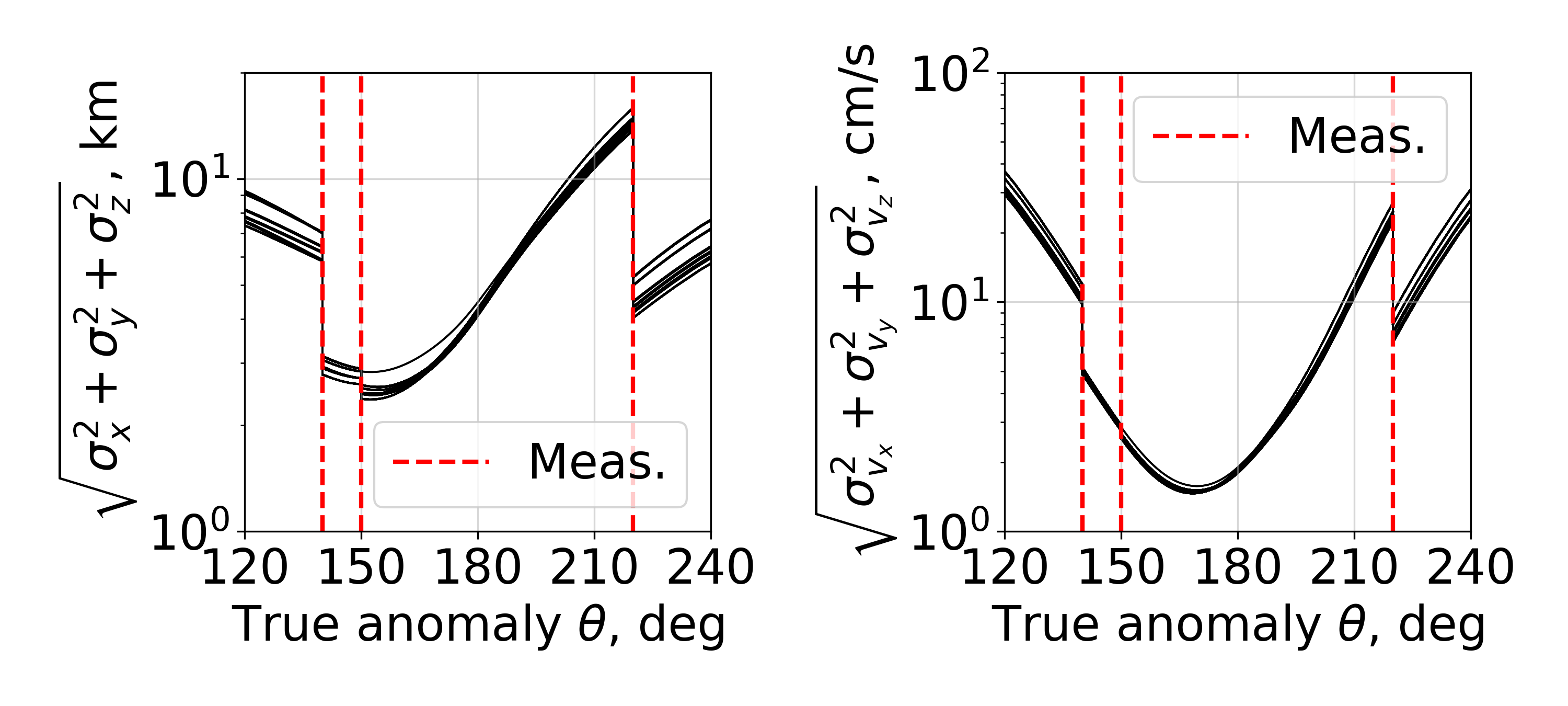}
         \caption{$f = 300$ \SI{}{mm}, $\mathrm{FOV} \approx 18.92^{\circ}$}
         \label{fig:rmssdv_vs_ta_ekf_nmeas3_f300_px1024_all_mc}
     \end{subfigure}
     \hfill
     \begin{subfigure}[b]{0.497\textwidth}
         \centering
         \includegraphics[width=\textwidth]{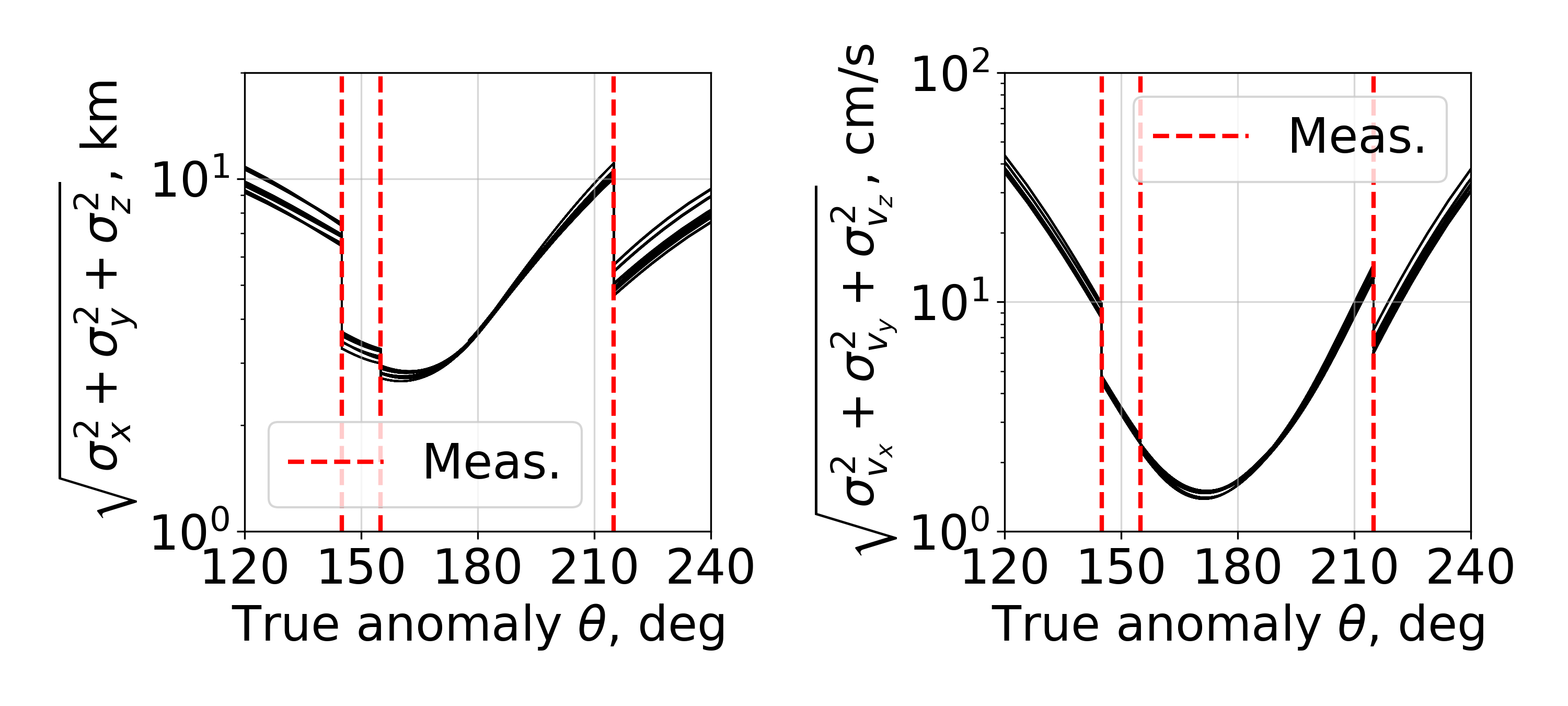}
         \caption{$f = 360$ \SI{}{mm}, $\mathrm{FOV} \approx 15.81^{\circ}$}
         \label{fig:rmssdv_vs_ta_ekf_nmeas3_f360_px1024_all_mc}
     \end{subfigure}
     \\
     \begin{subfigure}[b]{0.497\textwidth}
         \centering
         \includegraphics[width=\textwidth]{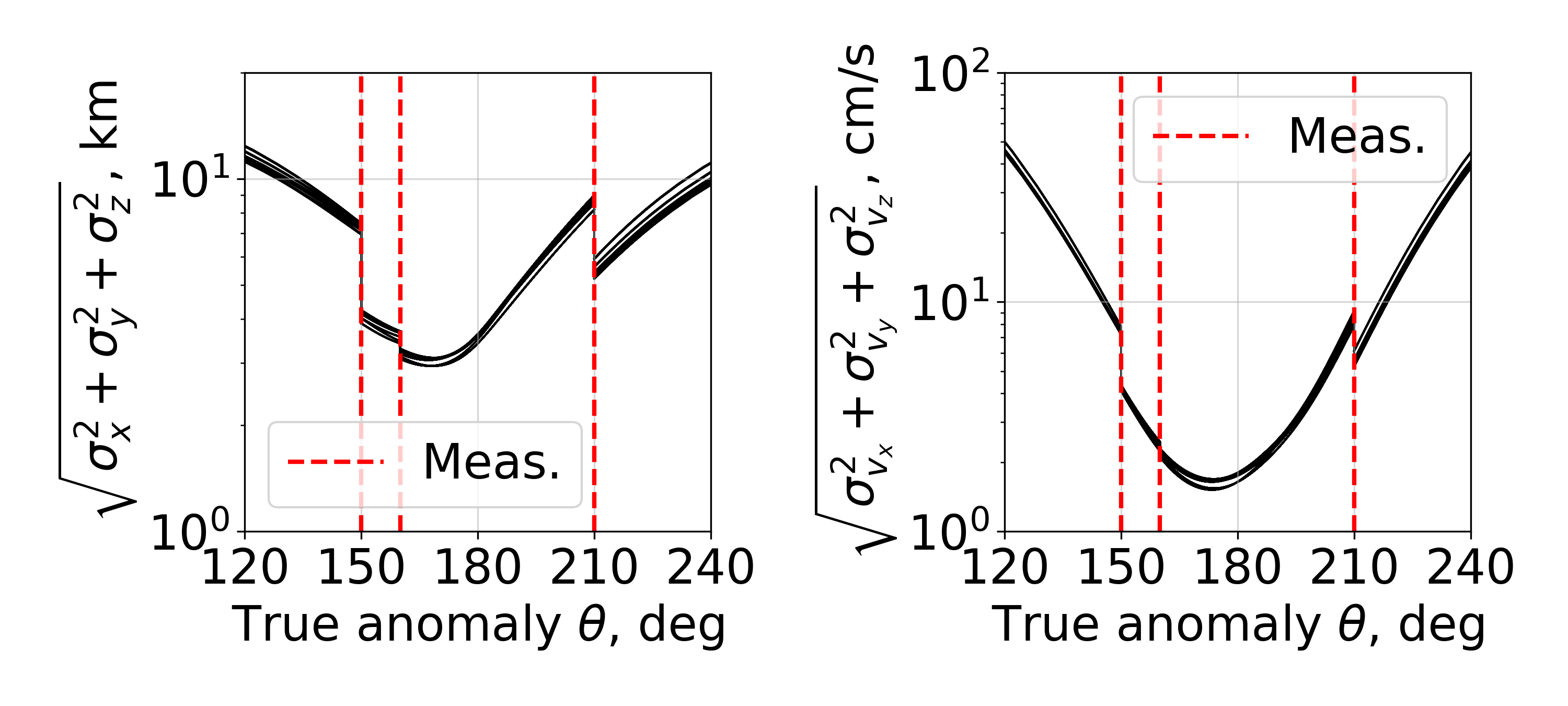}
         \caption{$f = 450$ \SI{}{mm}, $\mathrm{FOV} \approx 12.68^{\circ}$}
         \label{fig:rmssdv_vs_ta_ekf_nmeas3_f450_px1024_all_mc}
     \end{subfigure}
     \hfill
     \begin{subfigure}[b]{0.497\textwidth}
         \centering
         \includegraphics[width=\textwidth]{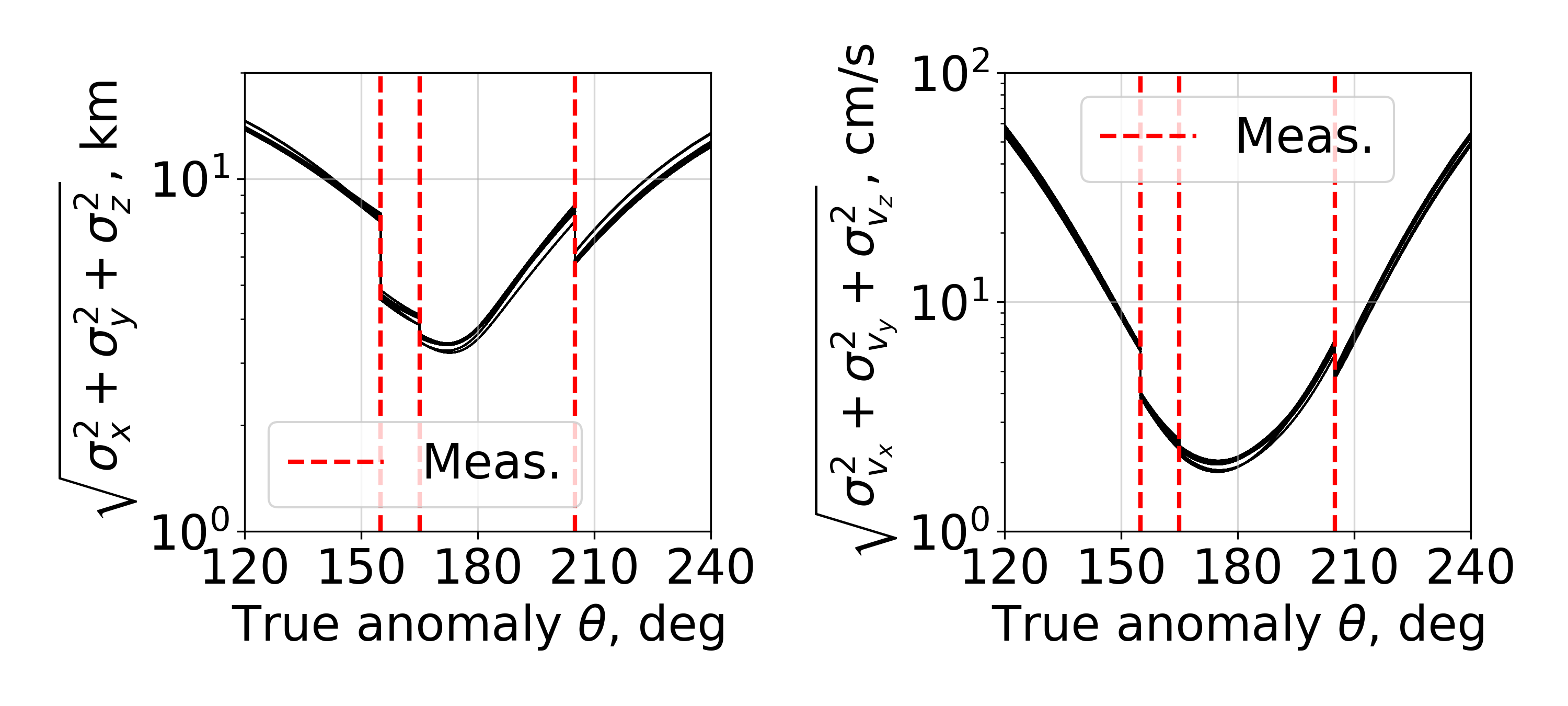}
         \caption{$f = 550$ \SI{}{mm}, $\mathrm{FOV} \approx 10.39^{\circ}$}
         \label{fig:rmssdv_vs_ta_ekf_nmeas3_f550_px1024_all_mc}
     \end{subfigure}
     \caption{Root mean squares of EKF standard deviation with 3 measurements per revolution}
     \label{fig:rmssdv_vs_ta}
\end{figure}

The difference in performance between each measurement configuration is visualized through the filter's standard deviation history: \review{Figure~\ref{fig:rmssdv_vs_ta} shows the root mean squares of the EKF's position and velocity standard deviation around apolune across 10 revolutions.}
The red vertical lines indicate measurements, where the standard deviation sees a discrete drop due to the Kalman update. 
The standard deviation in velocity, which is of particular interest for station-keeping, follows a local minima that falls in the vicinity of apolune where $\theta = 180^{\circ}$, but its exact location varies. 
Measurements with $f = 300$~\SI{}{mm} or $f = 360$~\SI{}{mm} result in lower velocity standard deviation; $f = 360$~\SI{}{mm} also sees a slight improvement in position standard deviations, and is thus chosen as the measurement collection configuration for the GN\&C pipeline experiments in Section~\ref{sec:gncstack_experiments}. 
While beyond the scope of this investigation, this experiment highlights the potential benefit of further optimizing and fine-tuning the camera FOV or measurement collection location and frequency for a specific orbit.

\section{Controller Experiments}
\label{sec:controller_experiments}
Prior to running the full GN\&C pipeline, a control-only experiment is run.
The purpose of experimenting only with the controller is to efficiently evaluate a large variety of strategies and controller tuning.
We study the use of DC versus the SLMP, mean state targeting via UT, and hysteresis via trigger condition and targeting tolerance tuning. 
The navigation error is drawn from a Gaussian distribution with position and velocity standard deviations of $3$-$\sigma_r = 5$ \SI{}{km} and $3$-$\sigma_v = 3$ \SI{}{cm/s}; these values are chosen as conservative estimations that can be achieved with OPNAV based on the filter experiment with $f = 360$ \SI{}{mm} and 3 measurements per revolution. 
To focus on the controller's tracking behavior, Monte Carlo experiments in this Section do not include any initial insertion error.
For each controller configuration, 100 Monte Carlo samples are taken, each lasting 60 NRHO revolutions, corresponding to about 393 days, using the same baseline NRHO as in Section~\ref{sec:filter_experiments}. 
In all cases considered within this experiment, the SK maneuver is executed at $\theta = 180^{\circ}$. 
The various sources of error considered along with their 3-$\sigma$ values are given in Table~\ref{tab:exp_control_errors}.
\review{In this work, we assume a ``quiet'' spacecraft station-keeping scenario~\cite{Davis2017}, where no random impulses are imparted due to attitude control.}

\begin{table}[]
    \centering
    \caption{Errors considered in controller experiments}
    \begin{tabular}{@{}ll@{}}
    \toprule
    Sources of error                  & 3-$\sigma$ values                       \\ \midrule
    Initial position error, \SI{}{km}   & 10                                    \\
    Initial velocity error, \SI{}{cm/s} & 10                                    \\
    Orbit determination position error, \SI{}{km} & 5                           \\ 
    Orbit determination velocity error, \SI{}{cm/s} & 5                         \\ 
    Control magnitude, relative       & 0.03                                   \\
    Control direction, deg            & 1.5                                     \\
    SRP $A/m$, relative               & 0.3                                     \\
    SRP $C_r$, relative               & 0.15                                    \\
    \bottomrule
    \end{tabular}
    \label{tab:exp_control_errors}
\end{table}

\begin{table}[]
\centering
\caption{Summary of 100 Monte Carlo samples targeting $v_x$ with 3-$\sigma_r = 5$ \SI{}{km} and 3-$\sigma_v = 3$ \SI{}{cm/s}}
\begin{tabular}{@{}lllllllll@{}}
\toprule
Case & Controller & \begin{tabular}[c]{@{}l@{}}$v_{x,\mathrm{trig}}$,\\ m/s\end{tabular} &
\begin{tabular}[c]{@{}l@{}}$v_{x,\mathrm{tol}}$,\\ m/s\end{tabular} & 
\begin{tabular}[c]{@{}l@{}}Success\\rate\end{tabular} & 
Iteration &
\begin{tabular}[c]{@{}l@{}}Mean yearly\\ cost, cm/s\end{tabular} & 
\begin{tabular}[c]{@{}l@{}}$95^{\mathrm{th}}$ percentile\\ yearly cost, cm/s\end{tabular} & \begin{tabular}[c]{@{}l@{}}Max yearly\\ cost, cm/s\end{tabular} \\ \midrule
1A   & DC         & 10 & 10 & 100\% & 1.89 & 83.8773 & 115.1519 & 130.4022 \\
     &            & 20 & 20 & 100\% & 1.62 & 83.8494 & 114.7258 & 129.0716 \\
     &            & 30 & 30 & 100\% & 1.42 & 85.4544 & 120.8226 & 130.8868 \\
\midrule
1B   & DC         & 10 & 1  & 100\% & 2.31 & \textbf{82.8235} & \textbf{107.5520} & \textbf{124.5990} \\
     &            & 20 & 1  & 100\% & 2.14 & 83.4813 & 109.2116 & 128.2814 \\
     &            & 30 & 1  & 100\% & 2.04 & 88.2740 & 125.7852 & 141.7324 \\
\midrule
1C   & SLMP       & 10 & 10 & 100\% & 2.17 & 105.5514 & 153.9997 & 187.8461 \\
     &            & 20 & 20 & 100\% & 2.11 & 149.8774 & 190.8868 & 200.1280 \\
     &            & 30 & 30 & 100\% & 2.07 & 194.1635 & 230.0571 & 247.6634 \\
\midrule
1D   & SLMP       & 10 & 1 &  100\% & 2.45 & 84.2067 & 117.9776 & 129.9781 \\
     &            & 20 & 1 &  100\% & 2.25 & 84.8724 & 114.3384 & 133.6056 \\
     &            & 30 & 1 &  100\% & 2.12 & 90.4282 & 121.6698 & 147.0017 \\
\bottomrule
\end{tabular}
\label{tab:control_experiments_results_target_vx}
\end{table}

\begin{table}[]
\centering
\caption{Summary of 100 Monte Carlo samples targeting UT-based $\mathbb{E}[v_x]$ with 3-$\sigma_r = 5$ \SI{}{km} and 3-$\sigma_v = 3$ \SI{}{cm/s}}
\begin{tabular}{@{}lllllllll@{}}
\toprule
Case & Controller & \begin{tabular}[c]{@{}l@{}}$v_{x,\mathrm{trig}}$,\\ m/s\end{tabular} &
\begin{tabular}[c]{@{}l@{}}$v_{x,\mathrm{tol}}$,\\ m/s\end{tabular} & 
\begin{tabular}[c]{@{}l@{}}Success\\rate\end{tabular} & 
Iteration &
\begin{tabular}[c]{@{}l@{}}Mean yearly\\ cost, cm/s\end{tabular} & 
\begin{tabular}[c]{@{}l@{}}$95^{\mathrm{th}}$ percentile\\ yearly cost, cm/s\end{tabular} & \begin{tabular}[c]{@{}l@{}}Max yearly\\ cost, cm/s\end{tabular} \\ \midrule
2A   & UT-DC    & 10 &  1 & 100\% & 2.54 & 77.4864 & 97.9576 & 104.2412 \\
     &            & 20 &  1 & 100\% & 2.29 & \textbf{76.7389} & \textbf{96.8750} & \textbf{106.5511} \\
     &            & 30 &  1 & 100\% & 2.06 & 82.4126 & 108.1957 & 138.7464 \\
\midrule
2B   & UT-SLMP  & 10 & 1 &  100\% & 3.18 & 77.3861 & 100.2806 & 119.0069 \\
     &            & 20 & 1 &  100\% & 2.84 & 77.7389 & 103.6512 & 116.1897 \\
     &            & 30 & 1 &  100\% & 2.57 & 81.8755 & 112.0431 & 123.8050 \\
\bottomrule
\end{tabular}
\label{tab:control_experiments_results_target_meanvx}
\end{table}

\subsection{Comparison of Differential Correction against SLMP}
\begin{figure}[t]
    \centering
    \begin{subfigure}[b]{0.66\textwidth}
        \centering
        \includegraphics[width=\textwidth]{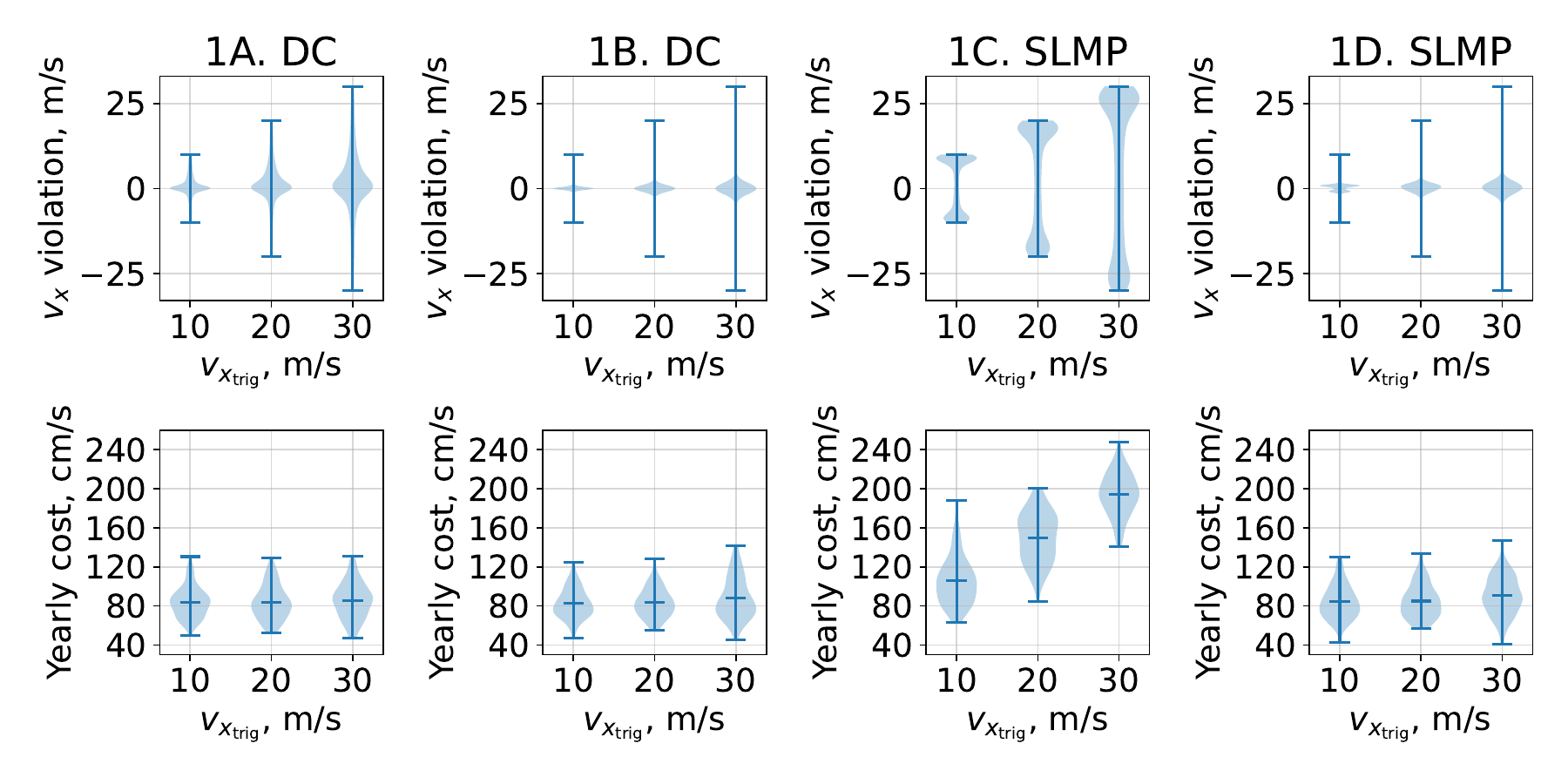}
        \caption{$v_x$-targeting schemes}
        \label{fig:sk_gncstack2_box_set1}
    \end{subfigure}
    \hfill
    \begin{subfigure}[b]{0.33\textwidth}
        \centering
        \includegraphics[width=\textwidth]{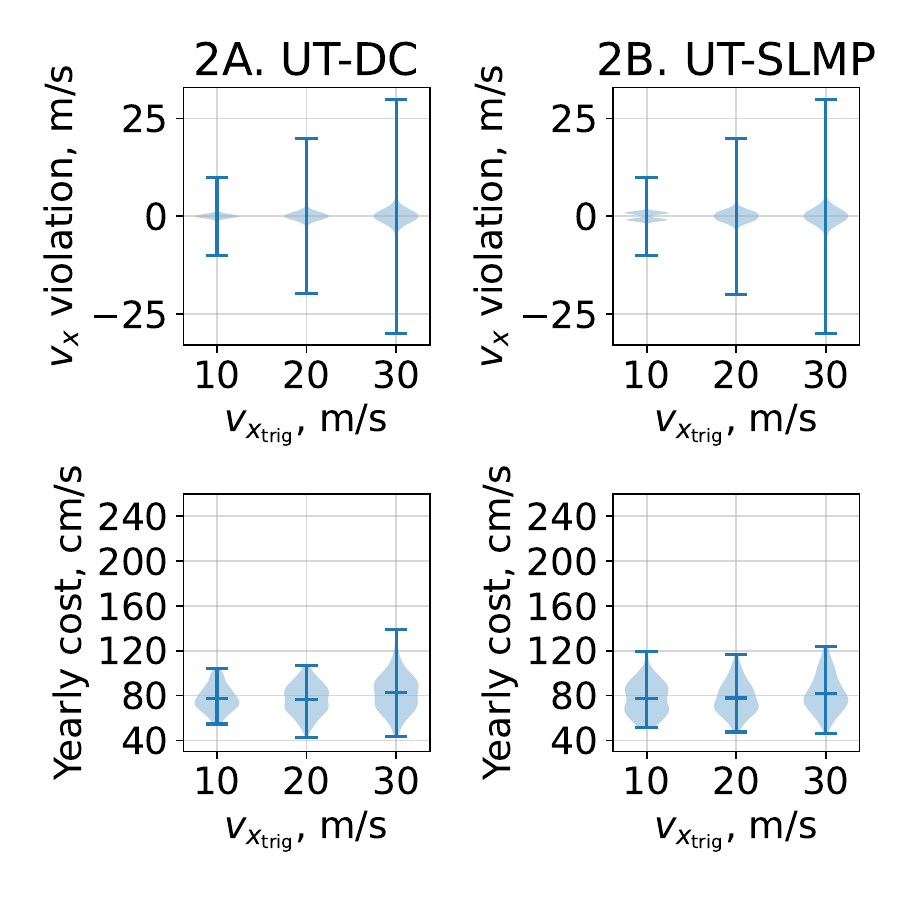}
        \caption{$\mathbb{E}[v_x]$-targeting schemes}
        \label{fig:sk_gncstack2_box_set2}
    \end{subfigure}
    \caption{Post-trigger/post-control $v_x$ violation and yearly cost}
    \label{fig:sk_gncstack2_box}
\end{figure}

We first focus on the difference in performance between DC and SLMP, without and with hysteresis. 
Table~\ref{tab:control_experiments_results_target_vx} summarizes the statistics from the Monte Carlo runs, and Figure~\ref{fig:sk_gncstack2_box_set1} shows the distribution of the achieved violation of $v_x$ after control on the top row, and the yearly cost on the bottom row. 

We observe by comparing cases 1A and 1B that the DC without or with hysteresis performs similarly; in contrast, looking at cases 1C and 1D, SLMP performs significantly worse than the other three cases when hysteresis is not used. 
The need for hysteresis is apparent by the top row in Figure~\ref{fig:sk_gncstack2_box_set1}, where the achieved $v_x$ violations between 1A and 1B are similar, and starkly different between 1C and 1D; in cases 1A, 1B and 1D a tight $v_x$ violation is achieved, while in 1C, the minimization problem aims to satisfy the targeting constraint at a higher tolerance, thus resulting in a higher cumulative cost.
\review{With SLMP, a tighter $v_{x,\rm tol}$ reduces the yearly station-keeping cost, highlighting that the long-term benefit of aligning the controlled trajectory closer to the reference outweighs the higher maneuver cost for a maneuver instance.
Such an effect is not pronounced for DC due to its quadratic convergence of $|v_x^{\rm EM} - \bar{v}_x^{\rm EM}| \ll v_{x,\rm tol}$.}
\review{In cases within 1A and 1B, we find similar performances with $v_{x,\rm trig} = 10$~\SI{}{m/s} and $v_{x,\rm trig} = 20$~\SI{}{m/s}, and a comparatively worse performance with $v_{x,\rm trig} = 30$~\SI{}{m/s}.
Such an increase in cumulative cost is attributed to the impact of an excessively high $v_{x,\rm trig}$, which results in less frequent, but cumulatively costly station-keeping controls.}

\begin{figure}
    \centering
    \includegraphics[width=0.563\linewidth]{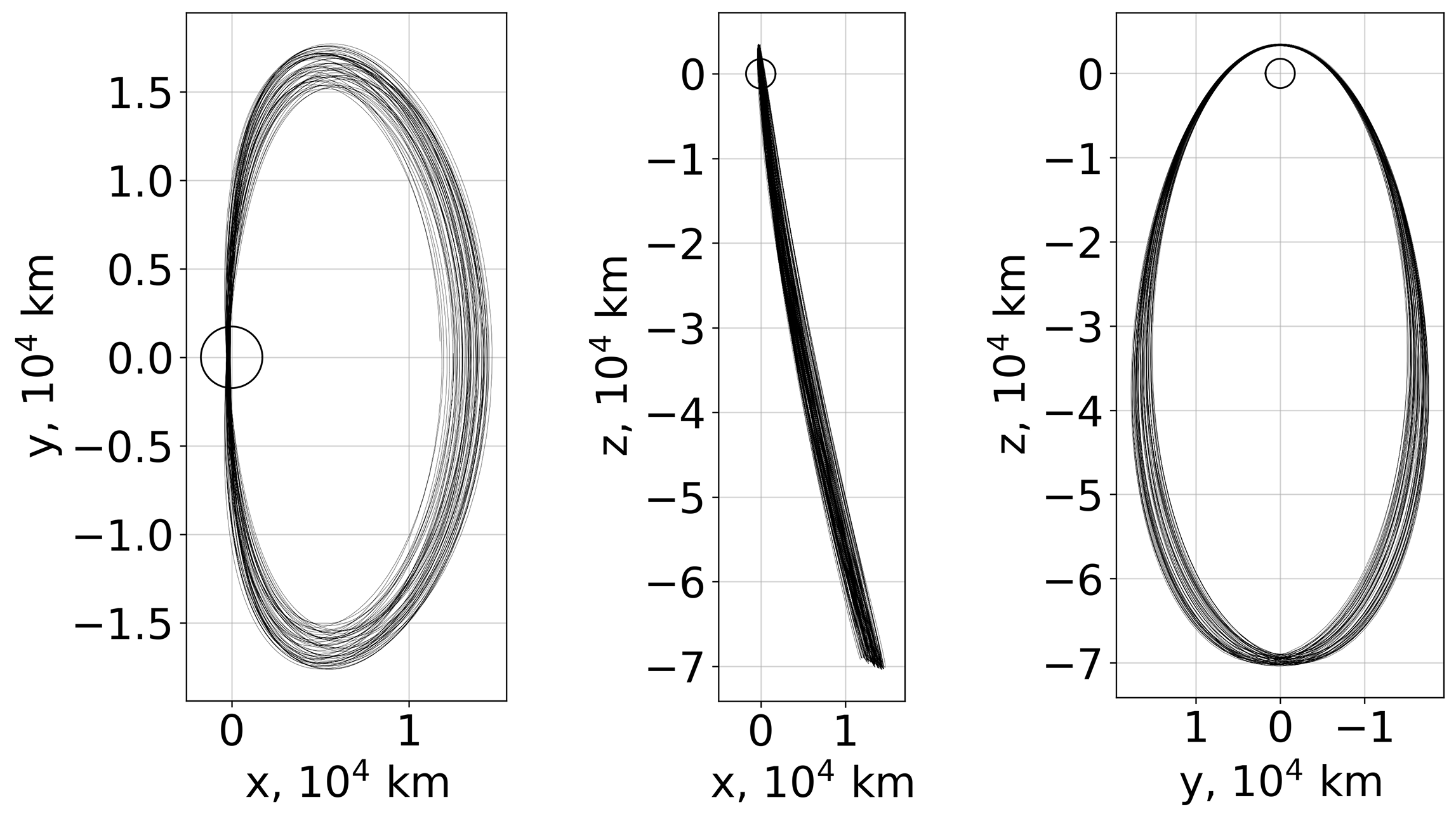}
    \caption{Example controlled trajectory over 60 revolutions from 1B}
    \label{fig:mc_sample_traj_xzplane_shooting_hysteresis_vxtol20_vxconstraint1}
\end{figure}

\begin{figure}
    \centering
    \includegraphics[width=0.49\linewidth]{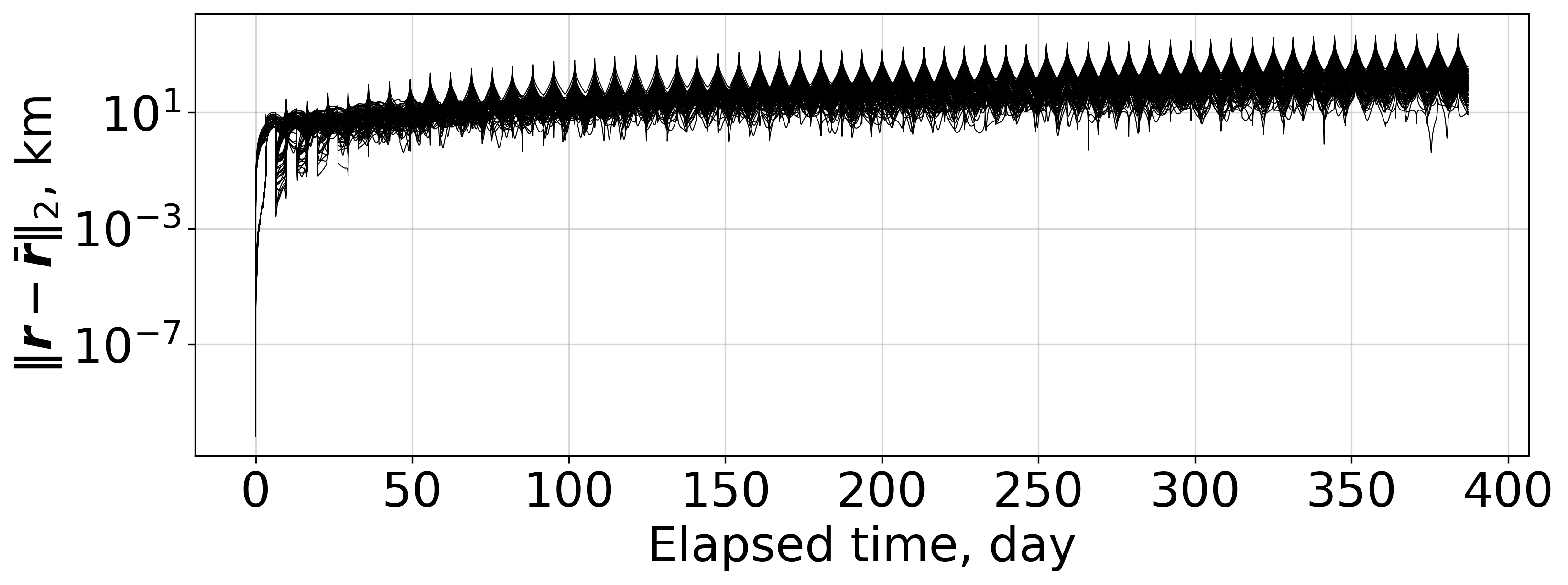}
    \caption{Position deviation histories from 1B}
    \label{fig:mc_traj_deviation_drnorm_xzplane_shooting_hysteresis_vxtol20_vxconstraint1}
\end{figure}

\review{We verify that the controlled trajectory follows the reference NRHO by looking at its position history.
Figure~\ref{fig:mc_sample_traj_xzplane_shooting_hysteresis_vxtol20_vxconstraint1} shows the controlled trajectory over 60 revolutions in~$\Frame_{\rm EM}$ for a single sample out of the Monte Carlo runs from 1B with $v_{x,\rm trig} = 20$~\SI{}{m/s}.
Controlled trajectories from other samples, as well as other controllers, closely resemble Figure~\ref{fig:mc_sample_traj_xzplane_shooting_hysteresis_vxtol20_vxconstraint1}.
We also show the norm of the position vector deviation over time, $\Delta r(t) = \| \rbold(t) - \bar{\rbold}(t) \|_2$, from 1B in Figure~\ref{fig:mc_traj_deviation_drnorm_xzplane_shooting_hysteresis_vxtol20_vxconstraint1}.
We observe a slow secular growth, but that remains under $100$~\SI{}{km} over the course of 60 revolutions.
Comparing with Figure~\ref{fig:mc_sample_traj_xzplane_shooting_hysteresis_vxtol20_vxconstraint1}, the growth in $\Delta r$ is primarily due to along-track deviation, as the qualitative geometry of the NRHO is maintained throughout the 60 revolutions.
The $\Delta r$ histories from other controllers also closely resemble Figure~\ref{fig:mc_traj_deviation_drnorm_xzplane_shooting_hysteresis_vxtol20_vxconstraint1}.}

\review{The secular growth in along-track deviation is a known side-effect of $x$-axis crossing control, and is related to its reliance on event-based targeting~\cite{Shimane2025-by}, which prioritizes maintaining the geometry of the orbit but neglects epoch deviation.
As a remedy, an additional infrequent (e.g., once a year) targeting maneuver may be executed to realign the trajectory to the reference~\cite{Davis2022}, or the $x$-axis crossing control may be replaced by a controller that recursively tracks the full-state~\cite{Shimane2025}.}

\subsection{Efficacy of UT-Mean State Targeting}
We now compare the use of the unscented transform. Table~\ref{tab:control_experiments_results_target_meanvx} summarizes the statistics from the Monte Carlo runs.
Figure~\ref{fig:sk_gncstack2_box_set2} shows the distribution of the achieved violation of $v_x$ after control on the top row, and the yearly cost on the bottom row. 
Comparing 2A against 1B and 2B against 1D, we see that the use of UT improves the performance for both the DC and SLMP approach. 
This improvement validates the motivation of the UT-based approach: the effect of the errors on the initial state estimate is attenuated by incorporating sigma points to predict the final targeted state. 
The use of UT results in an increase in the average number of iterations required, as the problem is capturing the nonlinearity better, and both DC and SLMP are based on linearized correction schemes or dynamics, respectively.
\review{The controlled trajectories and trend in $\Delta r$ histories resembles those shown in Figures~\ref{fig:mc_sample_traj_xzplane_shooting_hysteresis_vxtol20_vxconstraint1} and~\ref{fig:mc_traj_deviation_drnorm_xzplane_shooting_hysteresis_vxtol20_vxconstraint1}, and are omitted for conciseness.}

\section{GN\&C Pipeline Validation}
\label{sec:gncstack_experiments}
The final set of numerical experiments consists of validating the entire pipeline, with online generation of synthetic images, simulation of onboard navigation filtering, and station-keeping. 
Building upon the insights from Section~\ref{sec:filter_experiments}, we select a camera with $f = 360$ \SI{}{mm}, taking measurements at true anomalies of $145^{\circ}$, $155^{\circ}$, and $215^{\circ}$.
We test controller configurations 1B and 2A, which were the cases with the best performance from the experiments in Section~\ref{sec:controller_experiments}, without and with the UT-based targeting, respectively. 
In both cases, we select $v_{x, \mathrm{trig}} = 20$ \SI{}{m/s} and $v_{x, \mathrm{tol}} = 1$ \SI{}{m/s}.
\review{In each experiment in this Section, we consider 100 Monte Carlo runs, each lasting 60 revolutions, or approximately $388$ days. 
As with Monte Carlo runs from previous Sections, we re-introduce a random initial insertion error, with standard deviations $3\text{-}\sigma_{r} = 10$~\SI{}{km} and $3\text{-}\sigma_{v} = 10$~\SI{}{cm/s}.}

\subsection{Comparison against Filter-Only and Controller-Only Experiments}
\label{sec:results_control_only}
\begin{figure}
    \centering
    \includegraphics[width=0.49\linewidth]{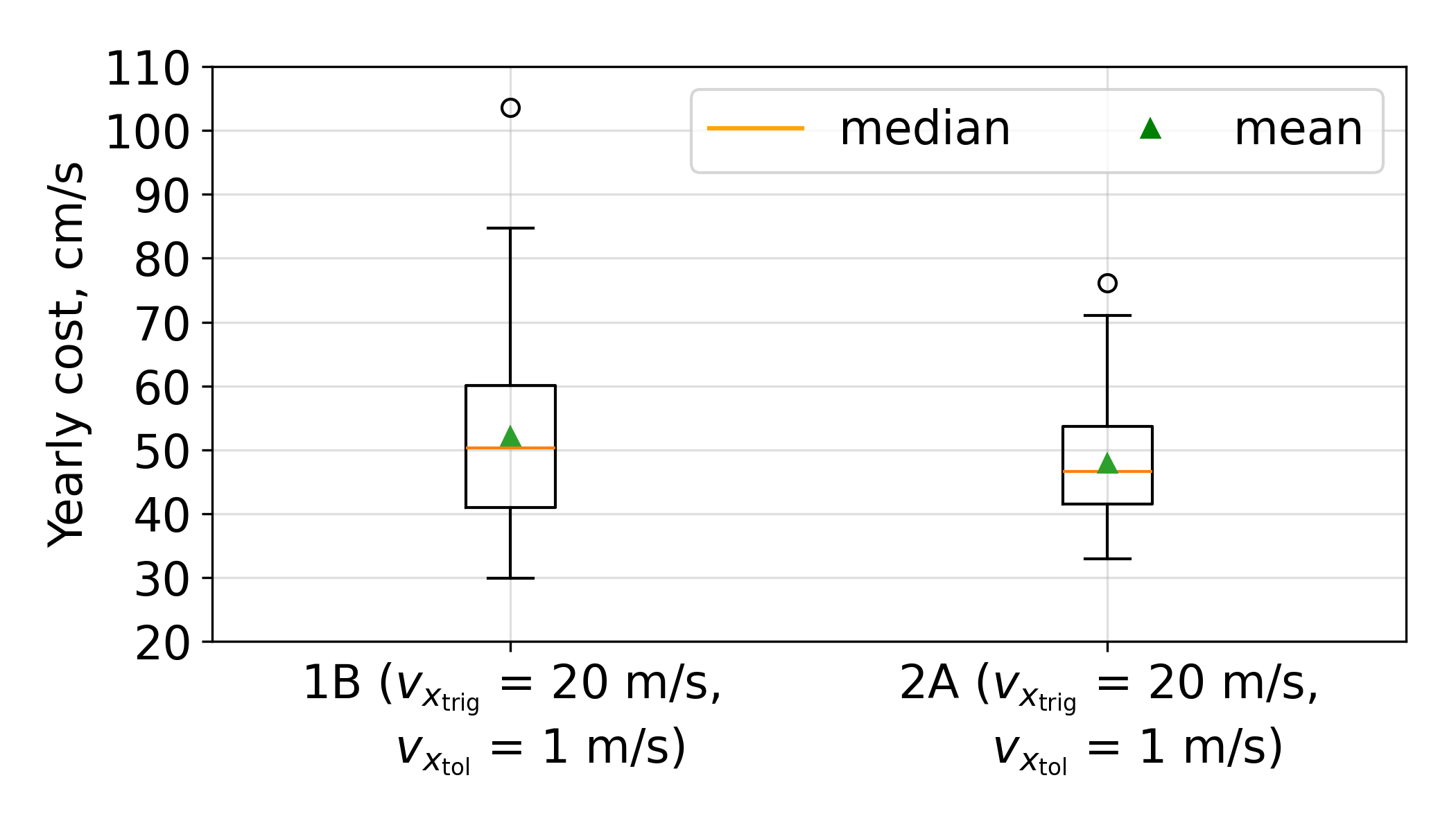}
    \caption{Distribution of yearly cost with full autonomous GN\&C pipeline}
    \label{fig:gnc_stack_yearly_1B2A}
\end{figure}

\begin{figure}
    \centering
    \begin{subfigure}[b]{0.48\textwidth}
        \centering
        \includegraphics[width=\textwidth]{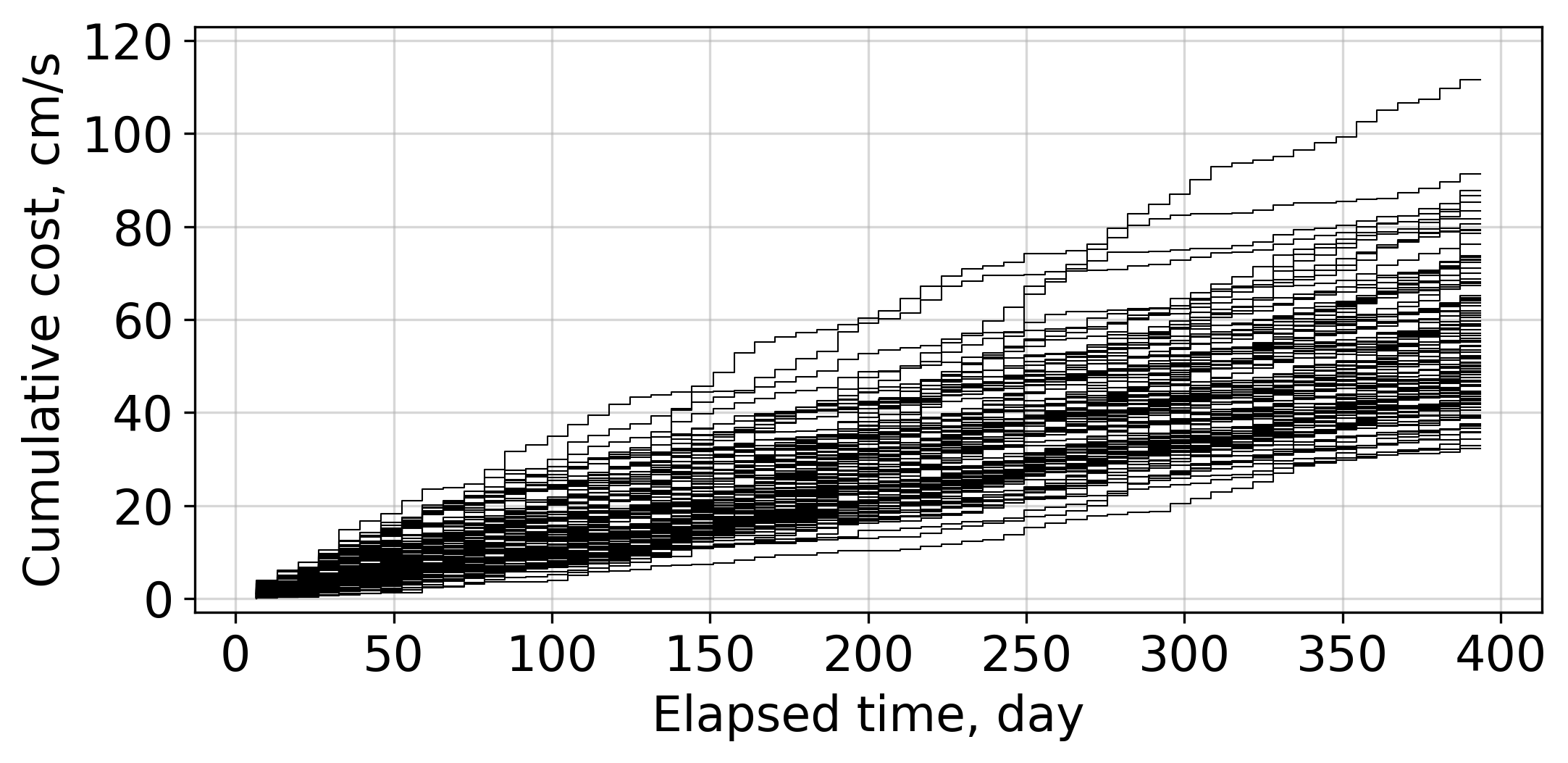}
        \caption{1B ($v_{x,\rm trig} = 20$~\SI{}{m/s}, $v_{x,\rm tol} = 1$~\SI{}{m/s})}
        \label{fig:gnc_stack_yearly_1B_cumulative}
    \end{subfigure}
    \hfill
    \begin{subfigure}[b]{0.48\textwidth}
        \centering
        \includegraphics[width=\textwidth]{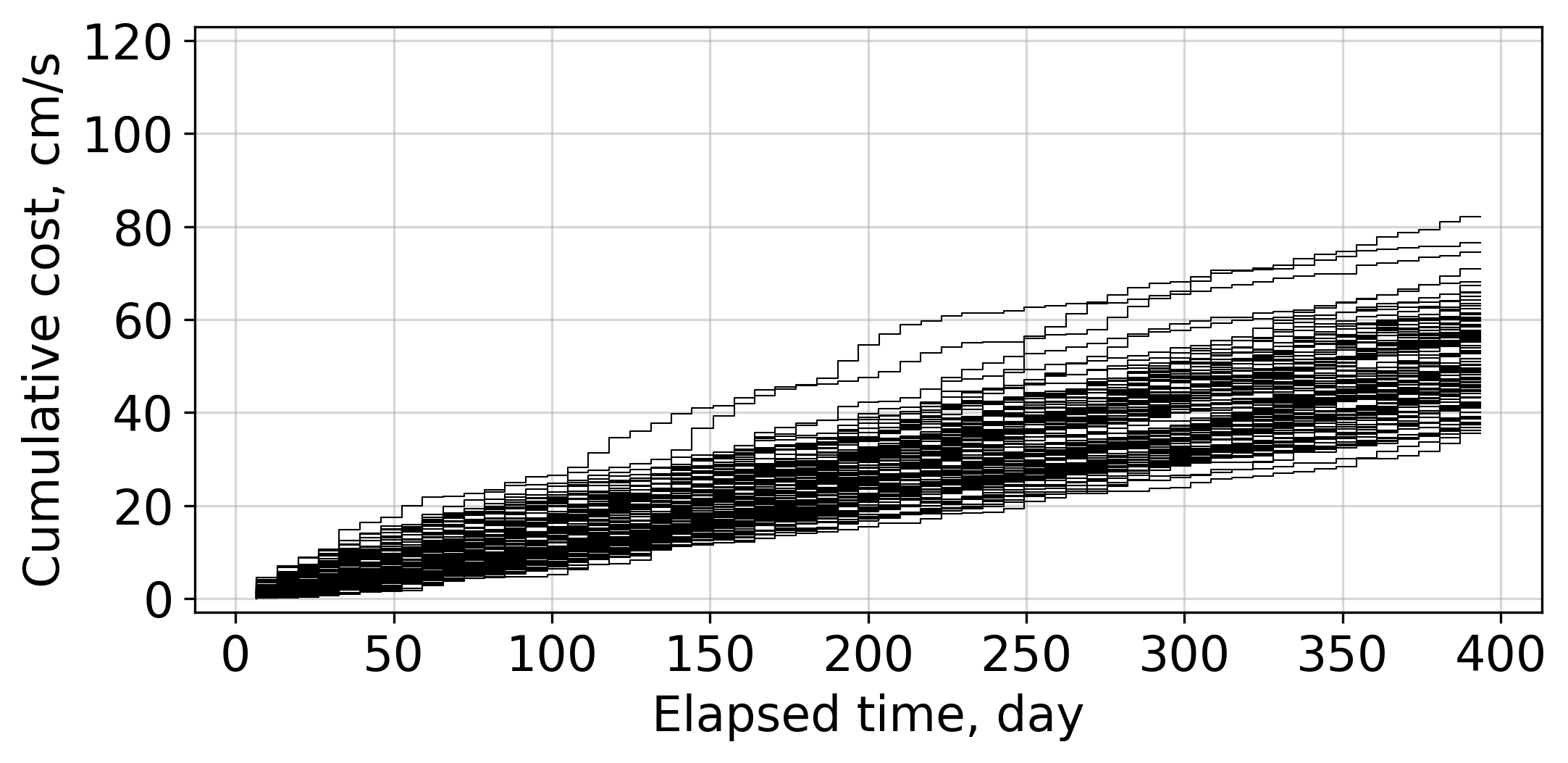}
        \caption{2A ($v_{x,\rm trig} = 20$~\SI{}{m/s}, $v_{x,\rm tol} = 1$~\SI{}{m/s})}
        \label{fig:gnc_stack_yearly_2A_cumulative}
    \end{subfigure}
    \caption{Cumulative station-keeping cost with full autonomous GN\&C pipeline}
    \label{fig:gnc_stack_yearly_1B2A_cumulative}
\end{figure}

\begin{figure}
    \centering
    \includegraphics[width=0.87\linewidth]{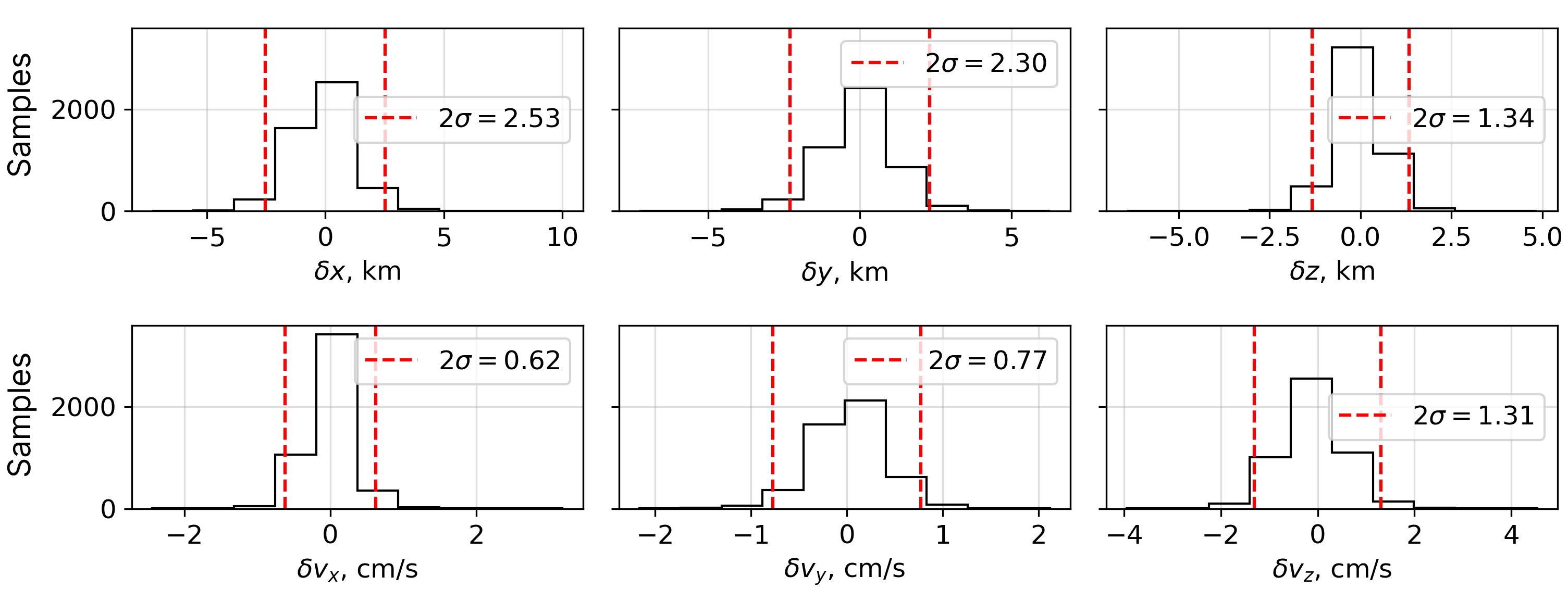}
    \caption{Distribution of filter's state estimate error immediately before maneuver at $\theta = 180^{\circ}$}
    \label{fig:naverror_pre_control_2A}
\end{figure}

\begin{figure}
    \centering
    \includegraphics[width=0.87\linewidth]{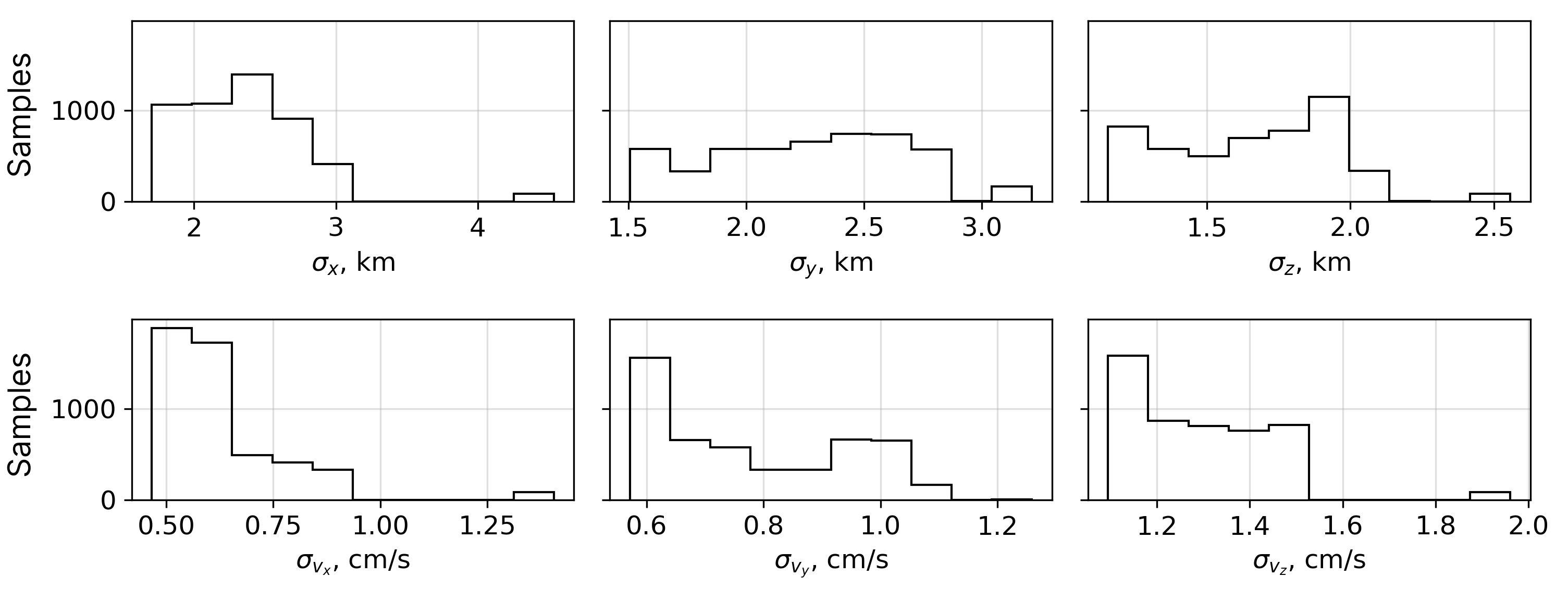}
    \caption{Navigation filter's standard deviation distribution immediately before maneuver at $\theta = 180^{\circ}$}
    \label{fig:navcovar_pre_control_2A}
\end{figure}

Table~\ref{tab:gncstack3_cost_summary} summarizes the yearly $\Delta V$, Figure~\ref{fig:gnc_stack_yearly_1B2A} shows the distribution of the cumulative cost, and the corresponding cumulative cost histories are shown in Figure~\ref{fig:gnc_stack_yearly_1B2A_cumulative}. 
The resulting $\Delta V$ are lower than those reported from the controller experiments in Tables~\ref{tab:control_experiments_results_target_vx} and~\ref{tab:control_experiments_results_target_meanvx}, since the previously assumed $3$-$\sigma_r = 5$ \SI{}{km} and $3$-$\sigma_v = 3$ \SI{}{cm/s} were conservative. Figure~\ref{fig:naverror_pre_control_2A} shows that the state estimate errors immediately before the maneuver are at lower values. Figure~\ref{fig:navcovar_pre_control_2A} shows the diagonal components of the covariance immediately before the maneuver, which are consistent with the empirical errors in Figure~\ref{fig:naverror_pre_control_2A}. 

The relative performance between 1A and 2B is consistent both from Section~\ref{sec:controller_experiments} and the GN\&C pipeline results presented here, where we see a substantial benefit from UT-based targeting.
From Table~\ref{tab:gncstack3_cost_summary}, the advantage of UT is particularly pronounced in the $95^{\mathrm{th}}$ percentile and max cost from the Monte Carlo experiments; as shown in Figure~\ref{fig:gnc_stack_yearly_1B2A_cumulative}, out of the 100 Monte Carlo samples, controller 2A has a smaller standard deviation, and there are few samples where the error realizations result in higher cumulative cost with controller 1B.
While the same realizations are also apparent for controller 2A, the cumulative cost does not accumulate as fast as in 1B. 

\begin{table}[]
    \centering
    \caption{Monte Carlo summary from DC and UT-DC, using EKF with 3 measurements per revolution, $f = 360$ \SI{}{mm} camera, $v_{x,\mathrm{trig}} = 20$ \SI{}{m/s} and $v_{x,\mathrm{tol}} = 1$ \SI{}{m/s}, maneuvering at $\theta = 180^{\circ}$}
    \begin{tabular}{@{}llllll@{}}
    \toprule
    Case &
    Controller &
    \begin{tabular}[c]{@{}l@{}}Success\\rate\end{tabular} & 
    \begin{tabular}[c]{@{}l@{}}Mean yearly\\ $\Delta V$, cm/s\end{tabular} & 
    \begin{tabular}[c]{@{}l@{}}$95^{\mathrm{th}}$ percentile\\ yearly $\Delta V$, cm/s\end{tabular} &
    \begin{tabular}[c]{@{}l@{}}Max yearly\\ $\Delta V$, cm/s\end{tabular} \\
    \midrule
    1B   & DC      & 100\% & 52.1766 & 77.4933 & 103.5663\\
    2A   & UT-DC & 100\% & 47.9092 & 62.5284 & 76.1859 \\ \bottomrule
    \end{tabular}
    \label{tab:gncstack3_cost_summary}
\end{table}

\begin{figure}
    \centering
    \includegraphics[width=0.99\linewidth]{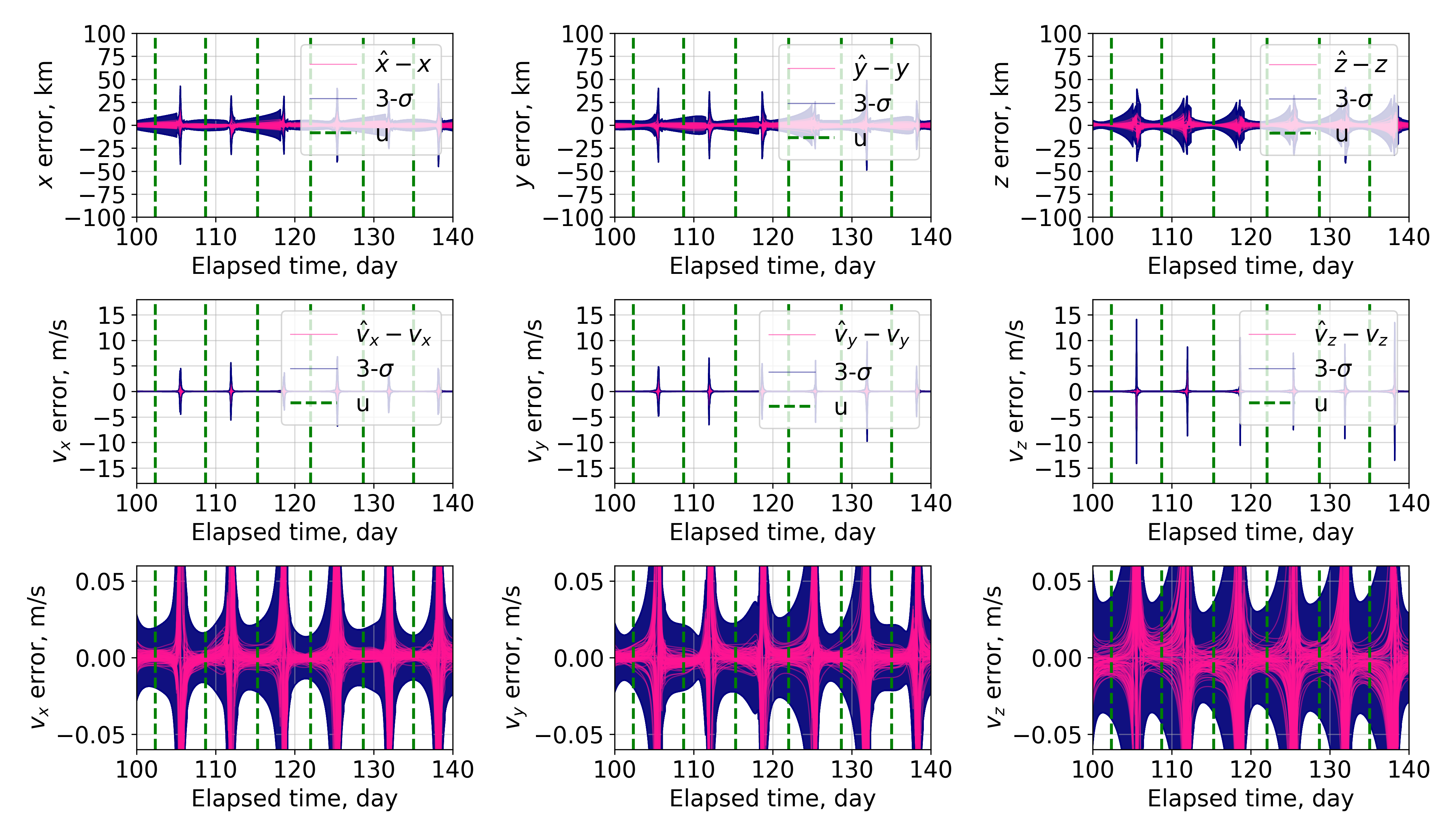}
    \caption{Snapshot of Monte Carlo runs of EKF with station-keeping at $\theta = 170^{\circ}$}
    \label{fig:gncstack_filter_zoomed_ta180}
\end{figure}

\rrreview{Figure~\ref{fig:gncstack_filter_zoomed_ta180} shows the EKF estimation error history over a representative 40-day time span with configuration 2A.
In the Figure, the green vertical lines denote maneuvers, where the filter's velocity estimate is updated by the uncorrupted control. For simplicity, the estimate covariance is kept constant.
Due to magnitude of each maneuver being well under~$0.01$~\SI{}{m/s} on average, we find the filter's estimate covariance to contain the estimate error from all Monte Carlo samples even immediately after maneuvers.
Similar EKF behavior has been observed across the entire year-long simulation, with configuration 1A as well as with other maneuver locations, and their visualizations are omitted for brevity.}

\subsection{Sensitivity on Control Location}
Guzzetti et al.~\cite{Guzzetti2017} show that a constant covariance, for instance based on DSN measurements, results in a flat trend of total cumulative station-keeping cost against maneuver location.
Using the full GN\&C pipeline consisting of the synthetic images and simulation of the navigation filters, we provide insight into the effect of the maneuver location with higher realism than assuming a fixed level of navigation error, as is common in control-only MC experiments that assume fixed navigation errors~\cite{Davis2017,Guzzetti2017,Davis2022}.
For this experiment, we adopt the UT-DC controller 2B, which was found to provide the best cost performance from results in~\ref{sec:results_control_only}.

\begin{figure}
    \centering
    \includegraphics[width=0.49\linewidth]{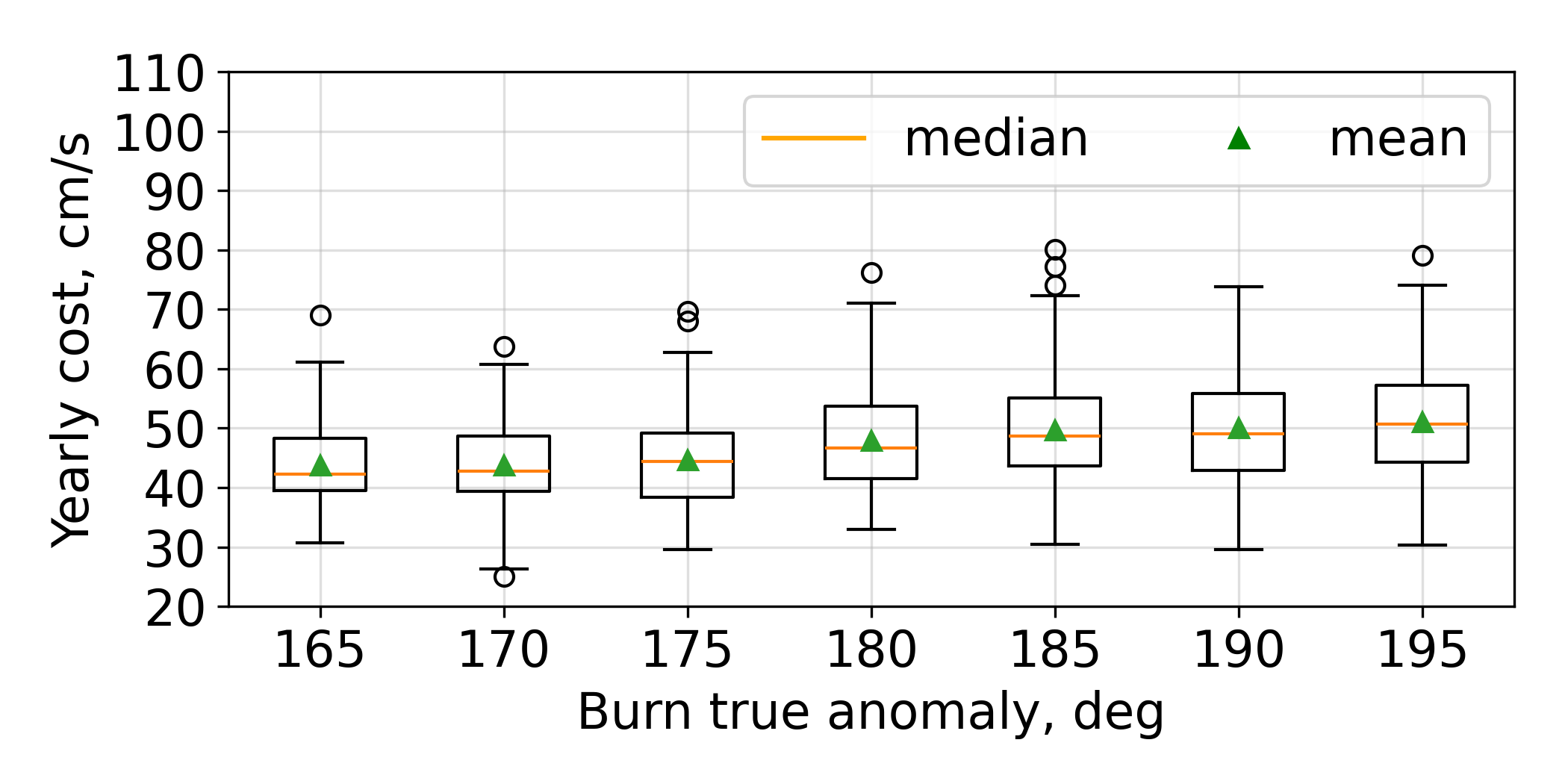}
    \caption{Distribution of yearly cost with UT-DC controller across varying maneuver true anomaly}
    \label{fig:navcovar_pre_control_2A_vs_trueanom}
\end{figure}

Figure~\ref{fig:navcovar_pre_control_2A_vs_trueanom} shows the results from 100 Monte Carlo runs with 2B, where the control action true anomaly is varied within a $\pm 15^{\circ}$ window about the apolune, with corresponding cost statistics summarized in Table~\ref{tab:gncstack3_cost_vs_trueanom}. 
While the average and median cost differences are below \SI{10}{cm/s}, a larger discrepancy in the worst-case performance, exceeding a difference of \SI{10}{cm/s}, is observed.
Specifically, with respect to the considered NRHO, camera properties, and measurement frequency, maneuvering at a true anomaly of $170^{\circ}$ yields the best performance, both in terms of the mean/median, and worst-case cumulative costs. 
We can interpret the small, albeit non-negligible difference in these yearly cost distributions by looking at the filter standard deviations from Figure~\ref{fig:rmssdv_vs_ta_ekf_nmeas3_f360_px1024_all_mc}.
The minimum in $\sigma_{v_z}$ occurs at around $170^{\circ}$, which coincides with the best-performing maneuver location in Figure~\ref{fig:navcovar_pre_control_2A_vs_trueanom}.
Such an observation is made possible by the realistic assessment that couples the navigation filter performance with station-keeping, especially in cases where a constant navigation covariance is not applicable. 

\begin{table}[]
    \centering
    \caption{Station-keeping cost with UT-DC controller for varying maneuver true anomaly}
    \begin{tabular}{@{}lllll@{}}
    \toprule
    Burn true anomaly, deg &
    \begin{tabular}[c]{@{}l@{}}Success\\rate\end{tabular} & 
    \begin{tabular}[c]{@{}l@{}}Mean yearly\\ cost, cm/s\end{tabular} & 
    \begin{tabular}[c]{@{}l@{}}$95^{\mathrm{th}}$ percentile\\ yearly cost, cm/s\end{tabular} &
    \begin{tabular}[c]{@{}l@{}}Max yearly\\ cost, cm/s\end{tabular} \\
    \midrule
    $165^{\circ}$ & 100\% & 43.7387 & 56.2019 & 69.0637\\
    $170^{\circ}$ & 100\% & 43.7115 & 58.0249 & 63.6931\\
    $175^{\circ}$ & 100\% & 44.6368 & 59.8371 & 69.7114\\
    $180^{\circ}$ & 100\% & 47.9092 & 62.5284 & 76.1859\\
    $185^{\circ}$ & 100\% & 49.7340 & 68.6846 & 80.0424\\
    $190^{\circ}$ & 100\% & 50.0258 & 68.6872 & 73.8191\\
    $195^{\circ}$ & 100\% & 51.0217 & 64.5657 & 79.0357\\
    \bottomrule
    \end{tabular}
    \label{tab:gncstack3_cost_vs_trueanom}
\end{table}

\section{Conclusion}
\label{sec:conclusion}
Several key insights have been obtained from the development and evaluation of the guidance, navigation, and control pipeline for operations in a Near Rectilinear Halo Orbit with horizon-based OPNAV.
From a navigation standpoint, the simulations demonstrate that only a limited number of horizon-based measurements are necessary for the filter to maintain sufficient state estimation accuracy on the NRHO. An investigation of the camera field of view and measurement locations along the orbit reveals notable variations in estimation error. Across all tested configurations, the filter covariance exhibits a persistent and well-defined periodic structure.
The structure is attributed to the inherent degradation of horizon-based measurement precision with increasing true anomaly, driven primarily by the rapid reduction in the apparent diameter of the Moon.
\review{With the EKF using 3 measurements collected per revolution, we find a consistent global minimum in standard deviations at a true anomaly of around $170^{\circ}$.}

On the control side, the quasi-Newton convergence properties of differential correction provide an advantage in reducing cumulative station-keeping cost compared to an explicit minimization-based strategy. The lower performance of the latter, arises from its short-sighted per-maneuver cost objective that only satisfies the targeting constraint to a specified tolerance, which does not directly minimize the overall cumulative cost.
The introduction of hysteresis improves the performance for both differential correction and minimization-based approaches, as choosing a smaller targeting tolerance than the triggering tolerance enhances robustness against maneuver execution error and model uncertainty.
Additionally, incorporating an unscented transform-based targeting framework substantially reduces the $95^{\mathrm{th}}$-percentile and worst-case station-keeping costs observed in Monte Carlo simulations. By leveraging the filter covariance rather than relying solely on the state estimate, the unscented transform-based framework mitigates sensitivity to estimation errors, which vary considerably across realizations.
Finally, the adoption of OPNAV-based navigation introduces a periodic structure in the filter covariance, which directly influences station-keeping performance, as the cumulative station-keeping cost becomes dependent on the maneuver location along the NRHO.
\review{Using the UT-DC controller with maneuvers executed at the global minimum of the EKF standard deviation, we successfully maintain the NRHO at a yearly station-keeping cost of about \SI{60}{cm/s} at the $95^{\rm th}$ percentile.}

The OPNAV-based GN\&C pipeline developed in this work can be extended to a broader class of libration point orbits for which horizon-based OPNAV is applicable. Beyond horizon-based techniques, alternative OPNAV methods such as crater-based navigation and visual odometry may be considered to further enhance autonomous onboard navigation and reduce reliance on ground-based updates.



\section*{Acknowledgments}
The authors acknowledge Dr. Pedro Miraldo for his help with setting up the synthetic image generation, as well as Dr. Marcus Greiff and Dr. Purnanand Elango for their advice on the implementation of the navigation filter and the minimization-based control schemes, respectively. 

\bibliography{gnc_references.bib,references.bib}

\end{document}